\documentclass{article}

\usepackage[nonatbib,preprint]{neurips_2022}

\usepackage[utf8]{inputenc} 
\usepackage{hyperref}       
\usepackage{url}            
\usepackage{booktabs}       
\usepackage{amsfonts}       
\usepackage{nicefrac}       
\usepackage{microtype}      
\usepackage{xcolor}         
\usepackage{multirow}       
\usepackage{amsmath, amssymb, bm}
\usepackage{tikz}
\usepackage{tabularx}
\newcolumntype{Y}{>{\centering\arraybackslash}X}

\newcommand{\kd}{\boldsymbol{k}_{\hspace*{-0.1mm}d}}

\newcommand{\korm}{\boldsymbol{k}_{\hspace*{-0.1mm}\mathrm{orm}}}
\newcommand{\makeredbox}[1]
{{\begin{tikzpicture}
    \node[anchor=south west,inner sep=0] (image) at (0,0) {\includegraphics[width=0.2\textwidth]{#1}};
    \begin{scope}[x={(image.south east)},y={(image.north west)}]
    \draw[cyan,very thick] (0.45,0.55) rectangle (0.99,0.8);
    \end{scope}
\end{tikzpicture}}}

\newcommand{\makecropbox}[6]
{{\begin{tikzpicture}
			\node[anchor=south west,inner sep=0] (image) at (0,0) {\includegraphics[height=#2]{#1}};
			\begin{scope}[x={(image.south east)},y={(image.north west)}]
				\draw[cyan,very thick] (#3, #4) rectangle (#5, #6);
			\end{scope}
\end{tikzpicture}}}

\newcommand{\makecropboxroller}[6]
{{\begin{tikzpicture}
			\node[anchor=south west,inner sep=0] (image) at (0,0) {\includegraphics[width=#2]{#1}};
			\begin{scope}[x={(image.south east)},y={(image.north west)}]
				\draw[cyan,very thick] (#3, #4) rectangle (#5, #6);
			\end{scope}
\end{tikzpicture}}}

\newlength{\myfigsize}

\newlength{\secpre}
\newlength{\secpost}
\newlength{\parpre}
\title{Shape, Light, and Material Decomposition from Images using Monte Carlo Rendering and Denoising}

\author{
  Jon Hasselgren \\
  NVIDIA\\
  \And
  Nikolai Hofmann \\
  NVIDIA \\
  \And
  Jacob Munkberg \\
  NVIDIA \\
}

\begin{document}


\newcommand{\figRoller}{
	\begin{figure}[tbh]
		\centering
		\setlength{\tabcolsep}{1pt}
		\begin{tabular}{cccc}    
			\multicolumn{4}{c}{
				\setlength{\tabcolsep}{0mm}
				\begin{tabular}{cccc}
					\makecropboxroller{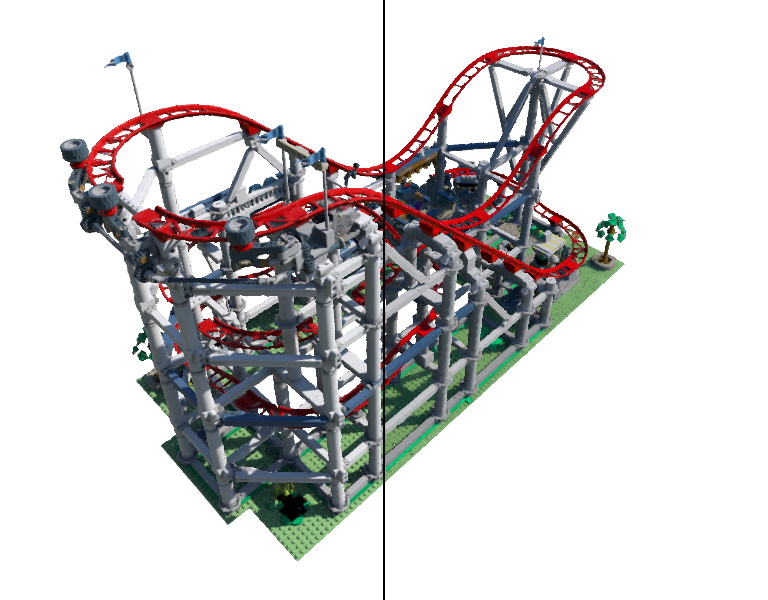}{0.24\columnwidth}{0.416666}{0.22}{0.583333}{0.386666} &
					\makecropboxroller{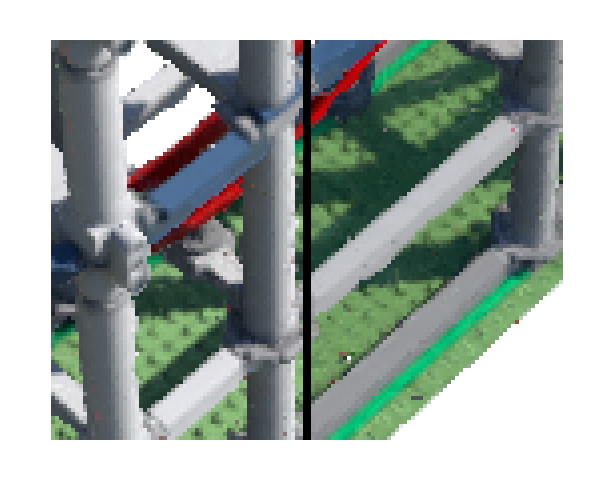}{0.24\columnwidth}{0.08}{0.08}{0.92}{0.92} &
					\makecropboxroller{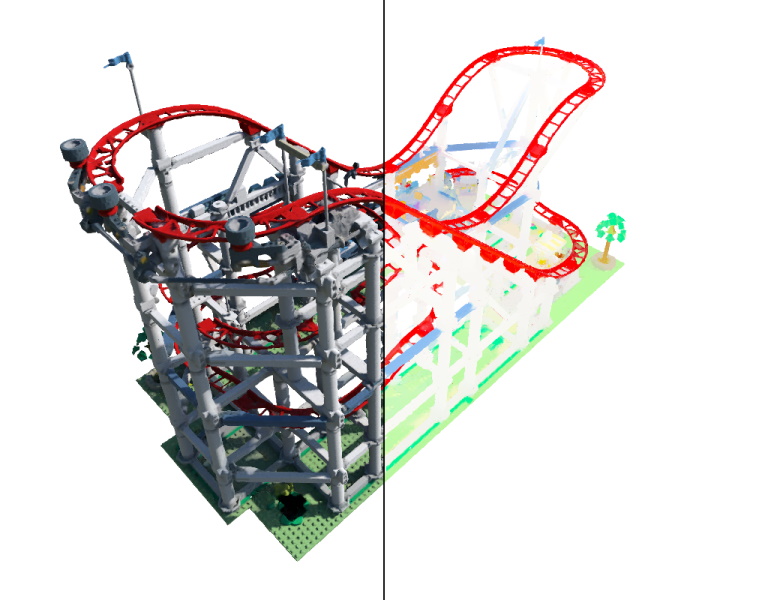}{0.24\columnwidth}{0.416666}{0.22}{0.583333}{0.386666} &
					\makecropboxroller{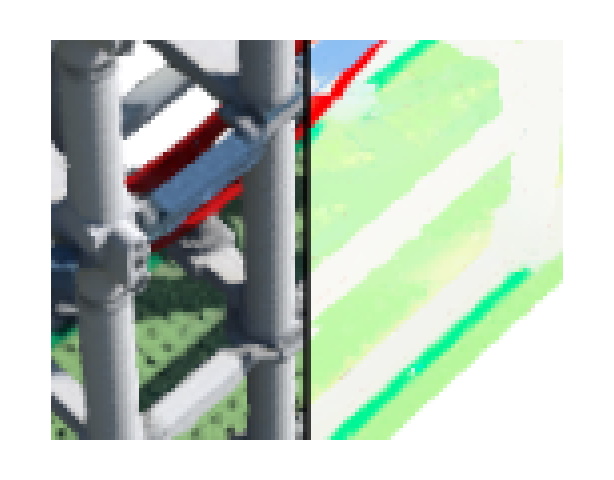}{0.24\columnwidth}{0.08}{0.08}{0.92}{0.92} \\
					\multicolumn{2}{c}{\textsc{nvdiffrec}} & \multicolumn{2}{c}{Our}
				\end{tabular}
			} \\
			\multicolumn{4}{c}{
				\setlength{\tabcolsep}{0mm}
				\begin{tabular}{ccc} 
					\includegraphics[width=0.33\columnwidth]{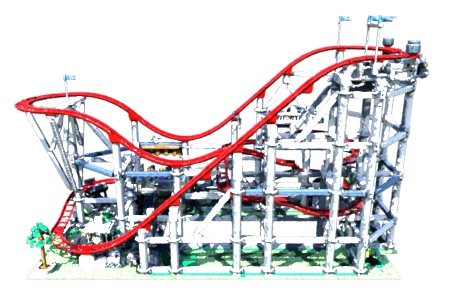} &
					\includegraphics[width=0.33\columnwidth]{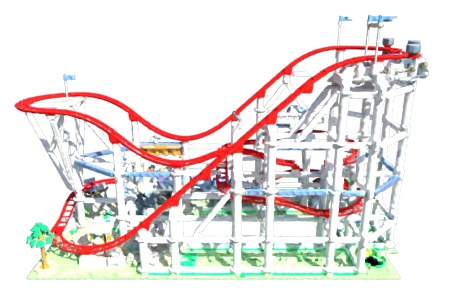} &
					\includegraphics[width=0.33\columnwidth]{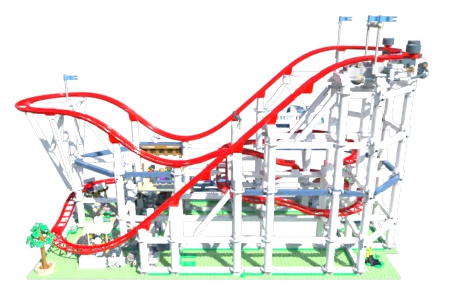} \\
					\textsc{nvdiffrec} & Our & Reference
				\end{tabular}
			} \vspace{1mm} \\
		\end{tabular}
		\caption{\protect \textsc{nvdiffrec}~\cite{munkberg2021nvdiffrec} successfully reconstructs complex geometry  from multi-view images, 
			but struggles with the material \& light separation.
			In the top row, we visualize split-screens of the rendered reconstruction and the diffuse albedo texture. Note that 
			\textsc{nvdiffrec} bakes most of the lighting in the albedo texture, which hurts quality in relighting scenarios (shown in the bottom row).
			In contrast, by leveraging a more advanced renderer, we successfully disentangle material and lighting (note the lack of shading in the 
			albedo texture), and improve relighting quality. The dataset consists of 200 views of the Rollercoaster from 
			LDraw \protect resources~\cite{Lasser2022} (CC BY-2.0).}
		\label{fig:roller}
	\end{figure}
}


\newcommand{\figSystem}{
    \begin{figure}
        \includegraphics[width=\textwidth]{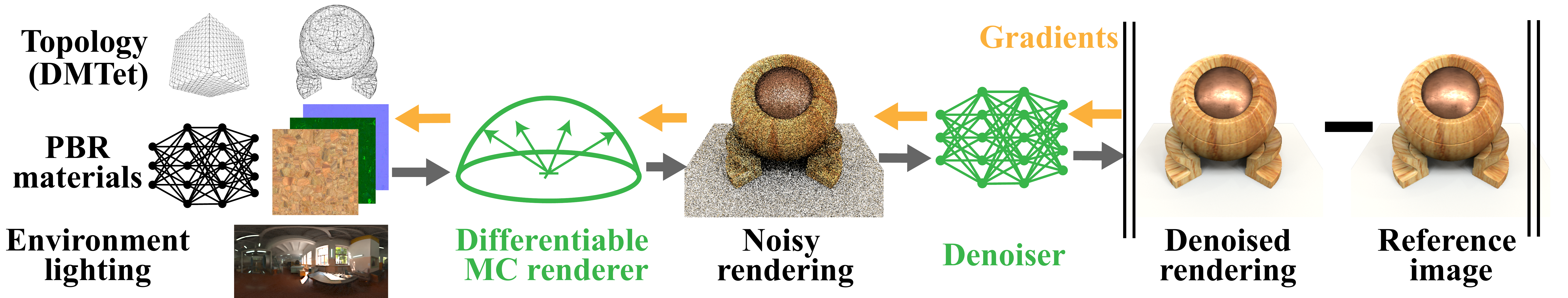}
        \caption{\protect 
            We extend \textsc{nvdiffrec}~\cite{munkberg2021nvdiffrec} with 
			a differentiable Monte Carlo renderer for direct illumination.
            Additionally, to reduce variance, we add a differentiable denoiser.
			These novel steps are highlighted in green. Following \textsc{nvdiffrec}, 
            the topology is parameterized using an SDF, and a triangular surface mesh is extracted in each iteration using DMTet~\cite{Shen2021},
			combined with spatially-varying PBR materials and HDR environment lighting. 
			The system is supervised using only photometric loss on the rendered, denoised image compared to a reference, 
			and gradients are back-propagated to the denoiser, shape, materials, and lighting parameters.
			All parameters are optimized jointly.
        }
        \label{fig:system}
    \end{figure}
}


\newcommand{\figTraining}{
\begin{figure}[tb]
    \centering
    \setlength{\tabcolsep}{1pt}
    \begin{tabular}{cccccc}    
        \includegraphics[width=0.16\columnwidth]{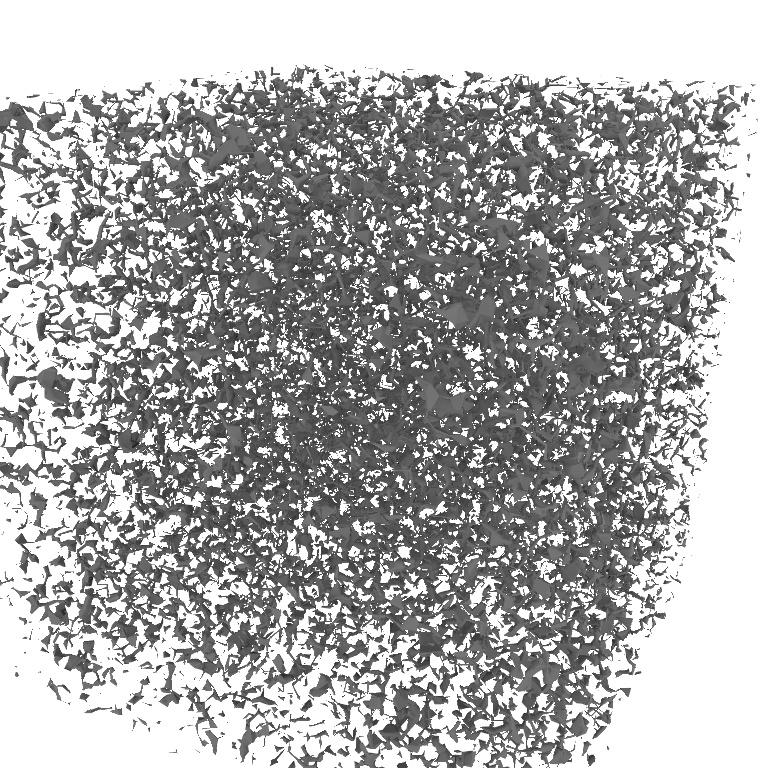} &
        \includegraphics[width=0.16\columnwidth]{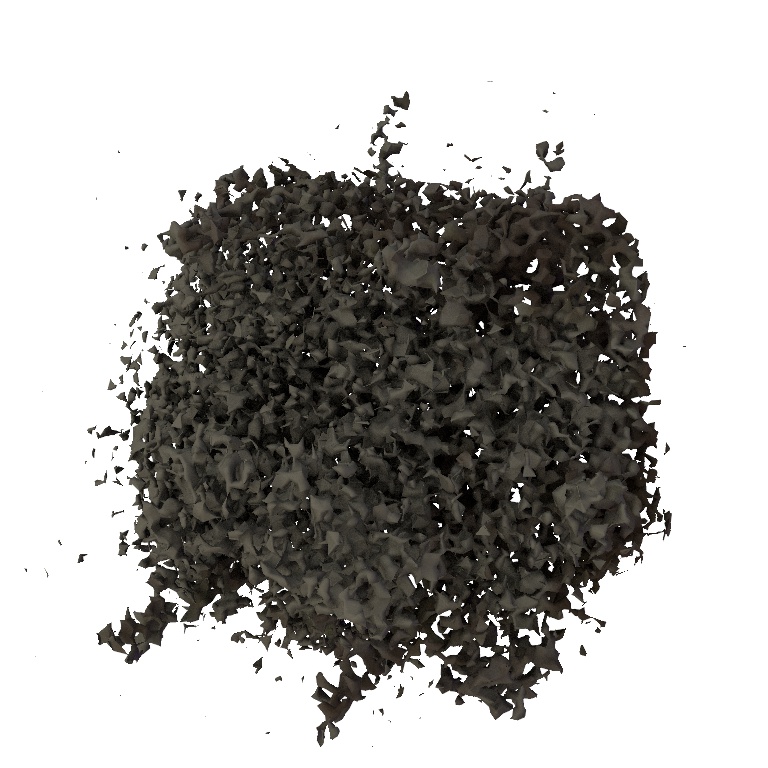} &
        \includegraphics[width=0.16\columnwidth]{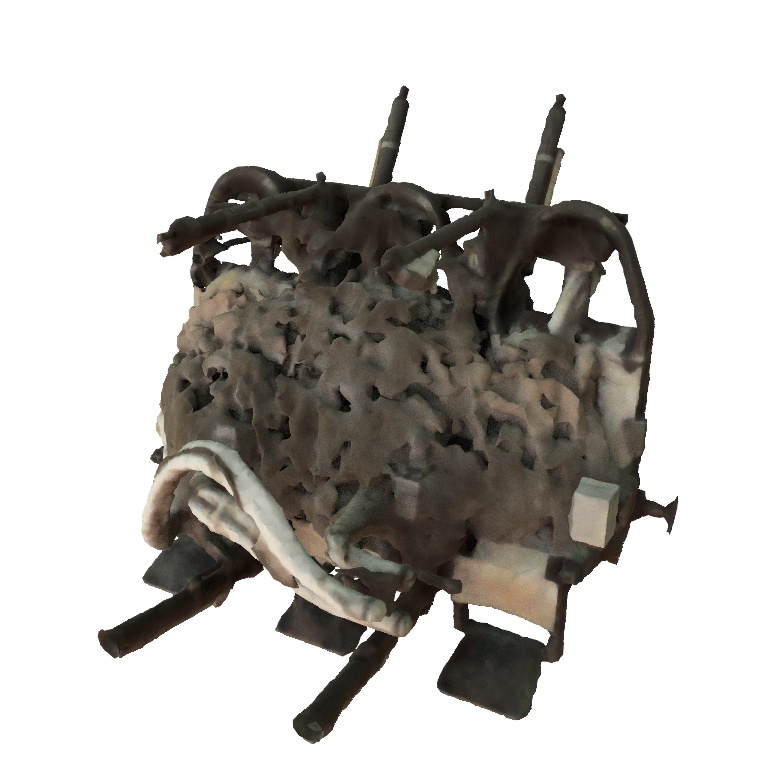} &
        \includegraphics[width=0.16\columnwidth]{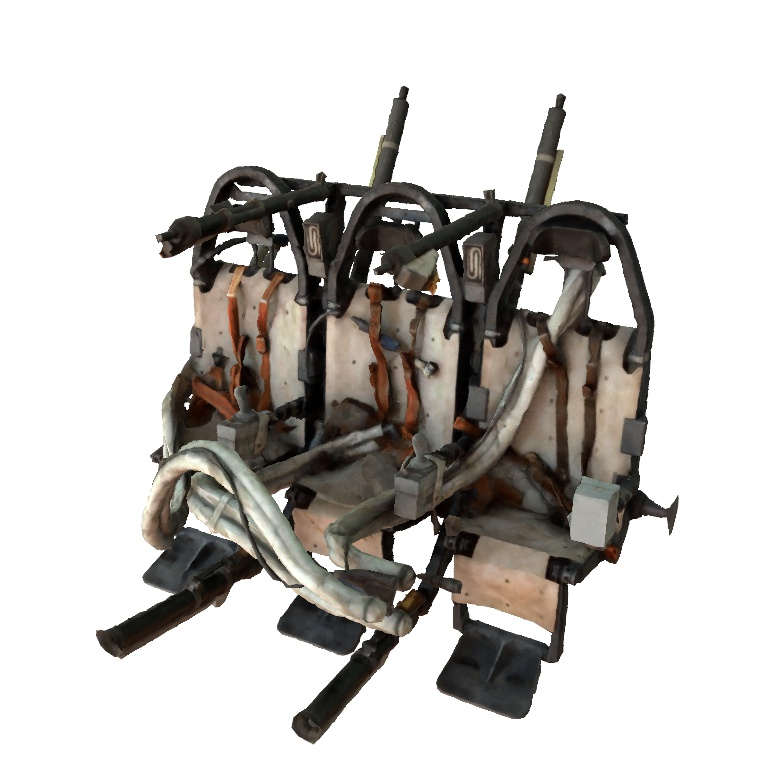} &
		\includegraphics[width=0.16\columnwidth]{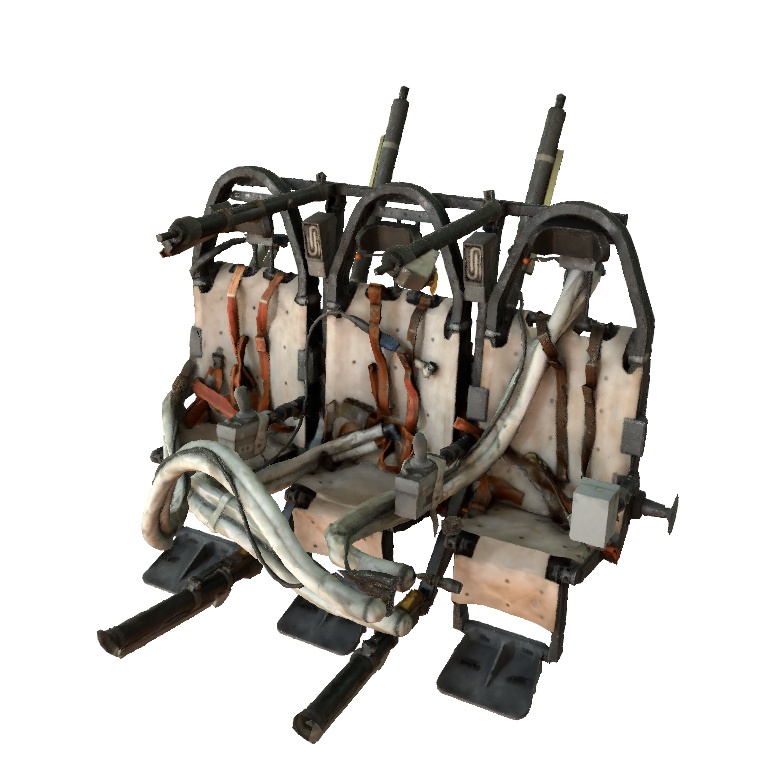} &
		\includegraphics[width=0.16\columnwidth]{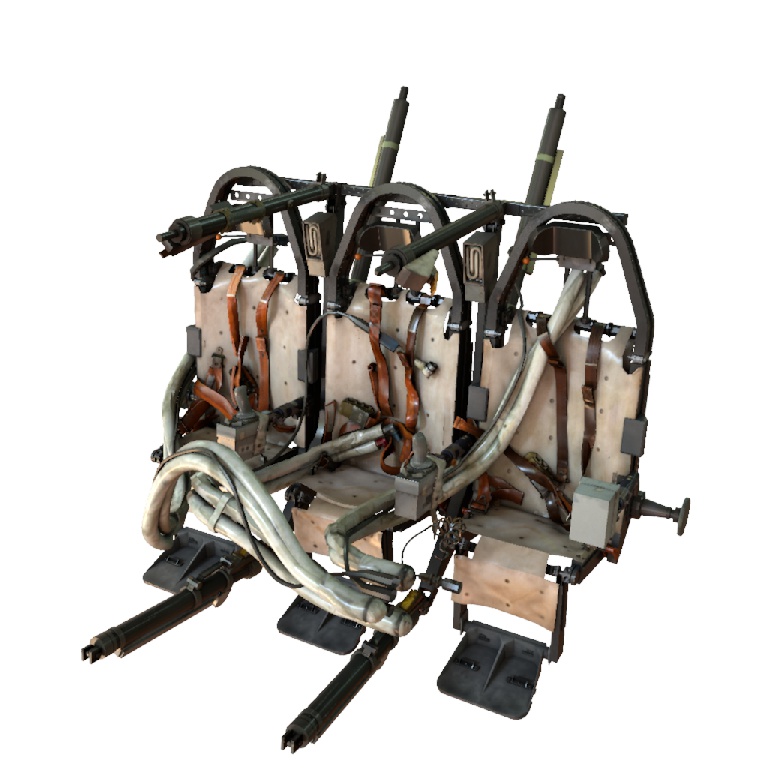}	\\
        \small{1} & \small{10} & \small{100} & \small{1000} & \small{5000} & \small{Reference}
    \end{tabular}
    \caption{Visualization of the optimization process. Note that the initial guess for
	topology are randomized SDF values on the grid. After 1000 iterations, we already have
	a high quality topology and plausible materials and lighting for this complicated asset. 
	Synthetic dataset with 200 frames, generated from a part of the Apollo capsule, 
	courtesy of the \protect Smithsonian~\cite{Smithsonian2020} (CC0-1.0).}
    \label{fig:training}
\end{figure}
}


\newcommand{\figSeparation}{
	\begin{figure}
		\centering
		\includegraphics[width=\textwidth]{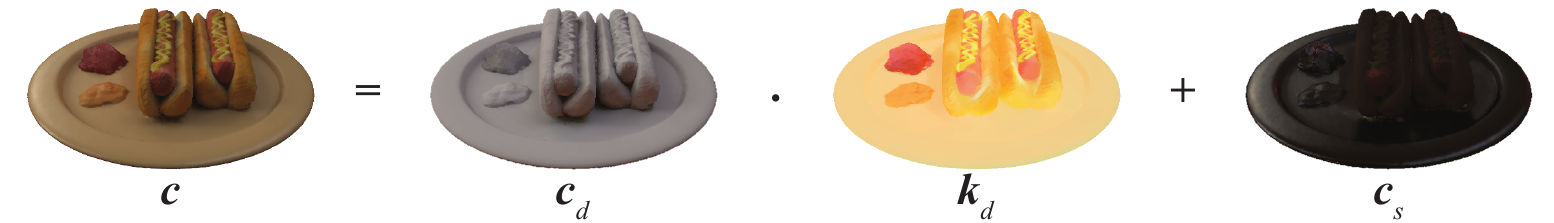}
		\caption{We separate lighting into diffuse lighting, $\mathbf{c}_d$, diffuse reflectance, $\mathbf{k}_d$, 
			and specular lighting, $\mathbf{c}_s$. This enables fine-grained regularization
			and denoising without smearing texture detail.}
		\label{fig:light_sep}
	\end{figure}
}


\newcommand{\figAblationDenoisePlots}{
	\begin{figure}
		\centering
		\includegraphics[width=\columnwidth]{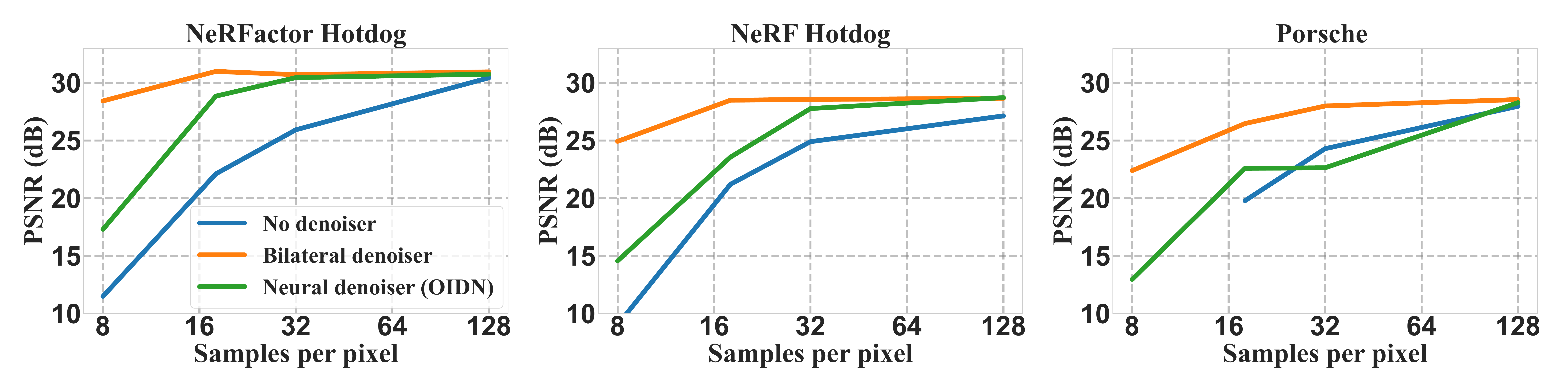}
		\caption{
			Ablation study on the effect of using different denoising algorithms during optimization at low sample counts on three different scenes of
			increasing complexity (from left to right). We plot averaged PSNR scores over 200 novel views,
			rendered without denoising, using high sample counts. In this experiment, we used decorrelated samples in the backward pass to 
			highlight the effect of denoising. The most complex scene (Porsche) failed to converge at 8~spp without denoising.
		}
		\label{fig:ablation-denoise-plots}
	\end{figure}
}


\newcommand{\figRelight}{
	\begin{figure}[tb]
		\centering
		\setlength{\myfigsize}{0.12\textwidth}	
		\setlength{\tabcolsep}{0.0mm}
		\begin{tabular}{ccccc@{\hskip 2.0mm}cccc}
			
			\rotatebox[origin=c]{90}{Ref}  &
			\raisebox{-0.5\height}{\includegraphics[trim={0cm 1.25cm 0cm 2cm},clip, width=\myfigsize]{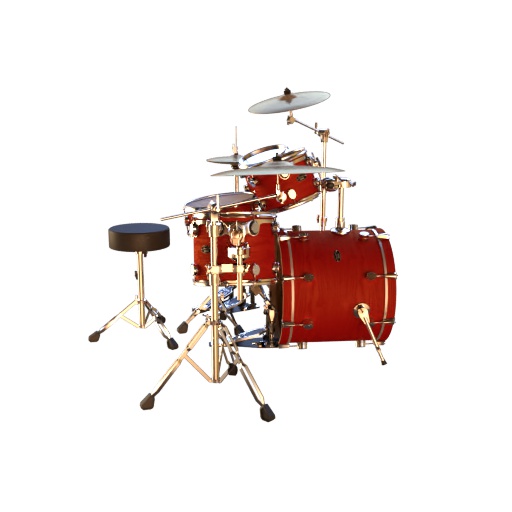}} &
			\raisebox{-0.5\height}{\includegraphics[trim={0cm 1.25cm 0cm 2cm},clip, width=\myfigsize]{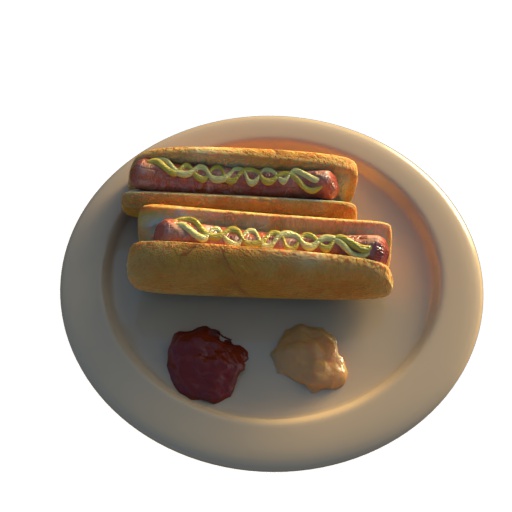}} &			
			\raisebox{-0.5\height}{\includegraphics[trim={0cm 1.25cm 0cm 2cm},clip, width=\myfigsize]{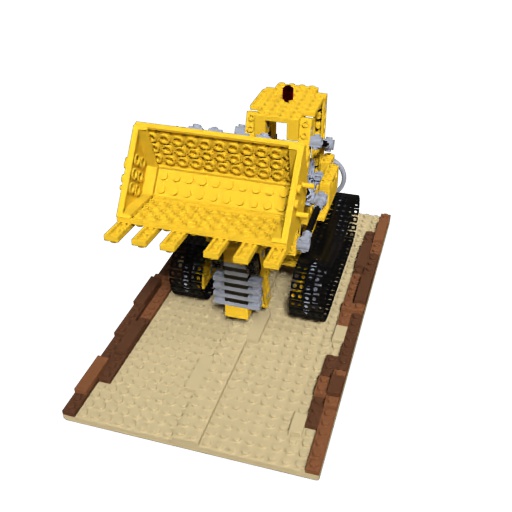}} &
			\raisebox{-0.5\height}{\includegraphics[trim={0cm 1.25cm 0cm 2cm},clip, width=\myfigsize]{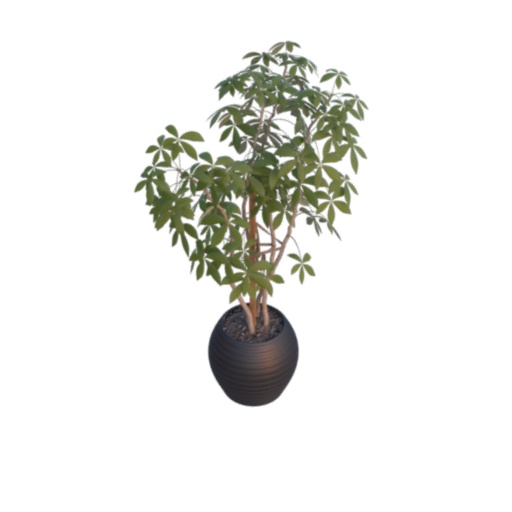}} &
			\raisebox{-0.5\height}{\includegraphics[trim={0cm 1.25cm 0cm 2cm},clip, width=\myfigsize]{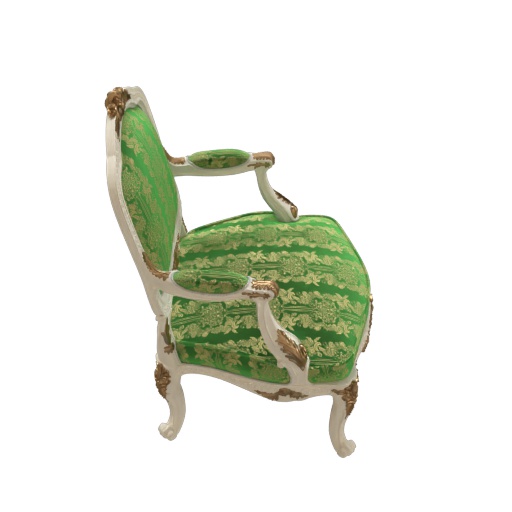}} &
			\raisebox{-0.5\height}{\includegraphics[trim={0cm 1.25cm 0cm 2cm},clip, width=\myfigsize]{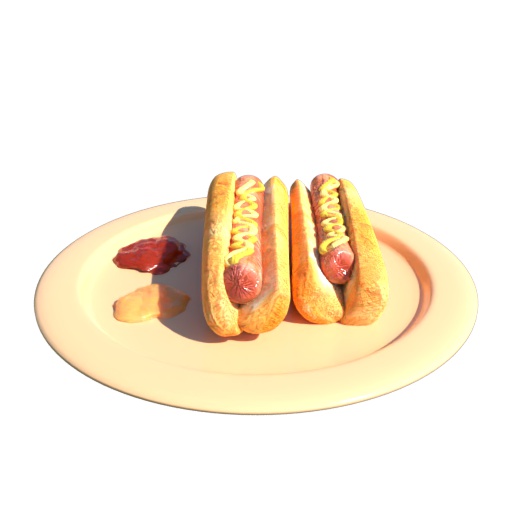}} &			
			\raisebox{-0.5\height}{\includegraphics[trim={0cm 1.25cm 0cm 2cm},clip, width=\myfigsize]{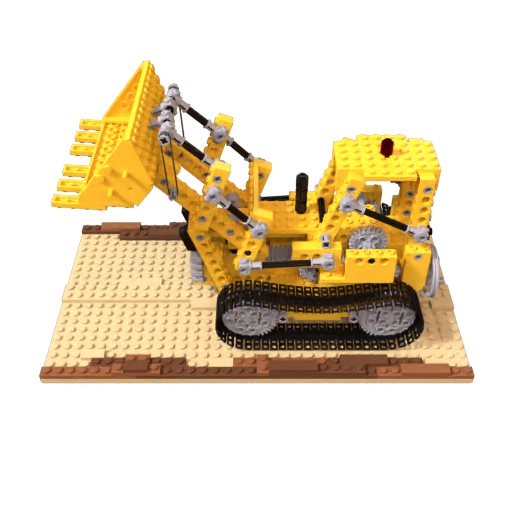}} &
			\raisebox{-0.5\height}{\includegraphics[trim={0cm 1.25cm 0cm 2cm},clip, width=\myfigsize]{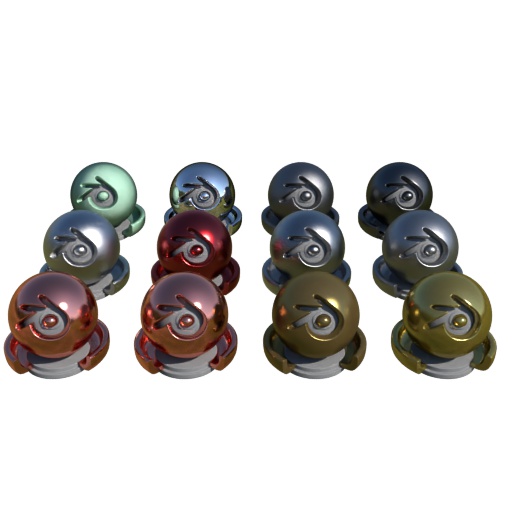}} \\
			
			\rotatebox[origin=c]{90}{\textsc{nvdiffrec}}  &
			\raisebox{-0.5\height}{\includegraphics[trim={0cm 1.25cm 0cm 2cm},clip, width=\myfigsize]{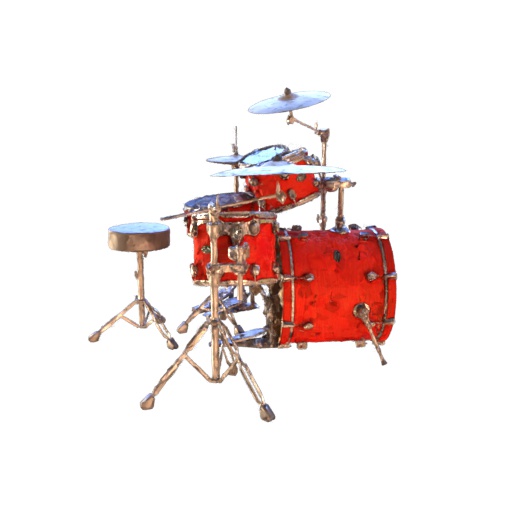}} &
			\raisebox{-0.5\height}{\includegraphics[trim={0cm 1.25cm 0cm 2cm},clip, width=\myfigsize]{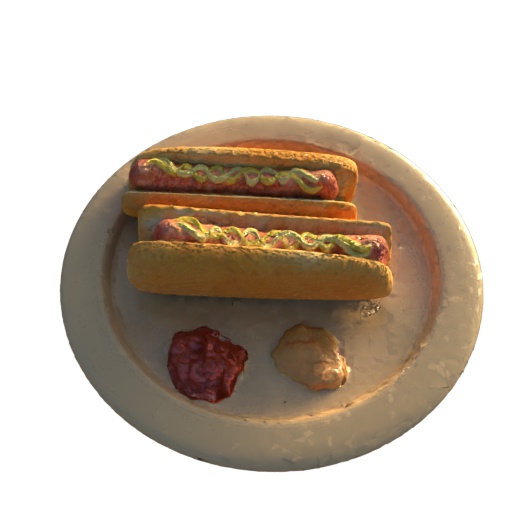}} &
			\raisebox{-0.5\height}{\includegraphics[trim={0cm 1.25cm 0cm 2cm},clip, width=\myfigsize]{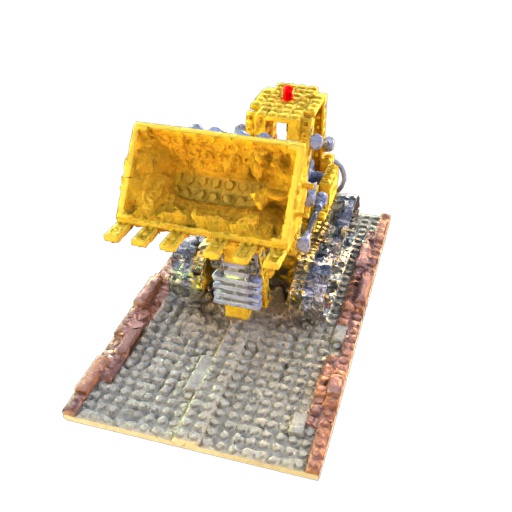}} &
			\raisebox{-0.5\height}{\includegraphics[trim={0cm 1.25cm 0cm 2cm},clip, width=\myfigsize]{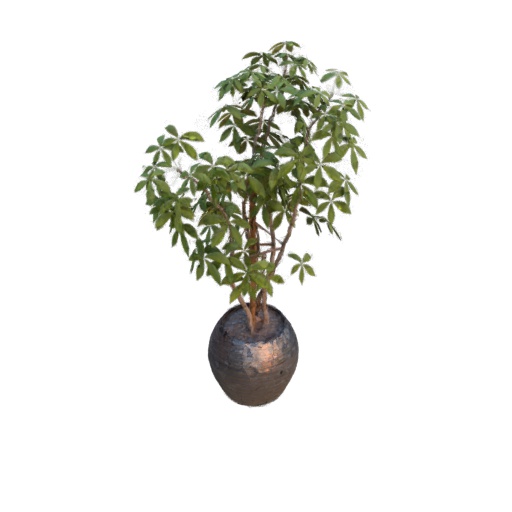}} &
			\raisebox{-0.5\height}{\includegraphics[trim={0cm 1.25cm 0cm 2cm},clip, width=\myfigsize]{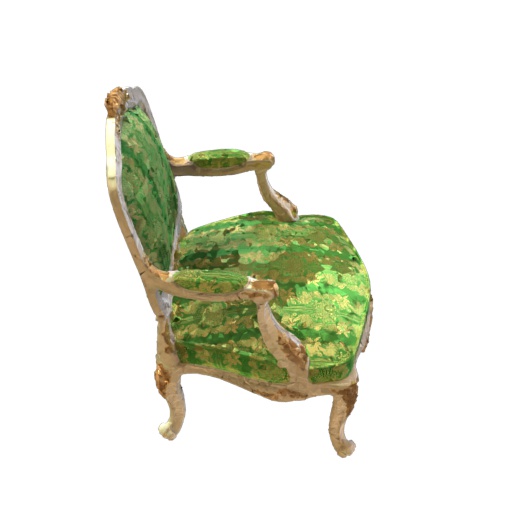}} &
			\raisebox{-0.5\height}{\includegraphics[trim={0cm 1.25cm 0cm 2cm},clip, width=\myfigsize]{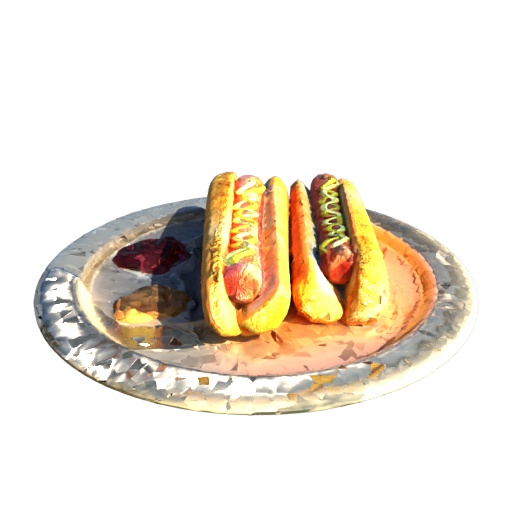}} &
			\raisebox{-0.5\height}{\includegraphics[trim={0cm 1.25cm 0cm 2cm},clip, width=\myfigsize]{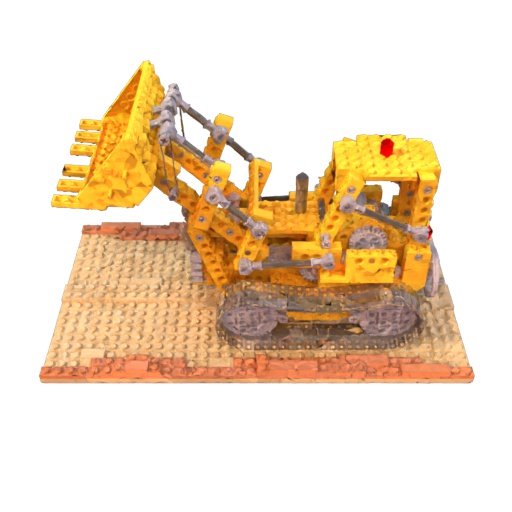}} &
			\raisebox{-0.5\height}{\includegraphics[trim={0cm 1.25cm 0cm 2cm},clip, width=\myfigsize]{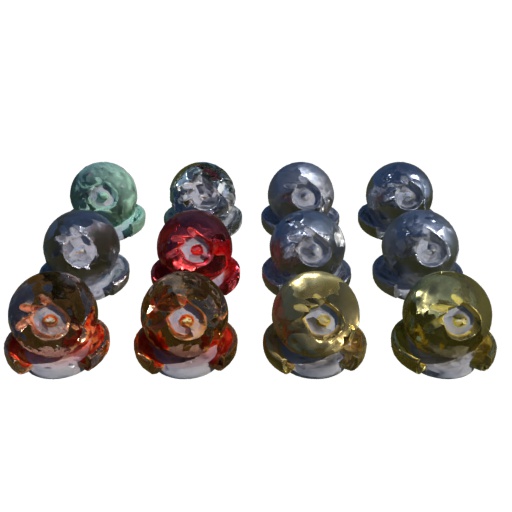}} \\
			
			\rotatebox[origin=c]{90}{Our}  &
			\raisebox{-0.5\height}{\includegraphics[trim={0cm 1.25cm 0cm 2cm},clip, width=\myfigsize]{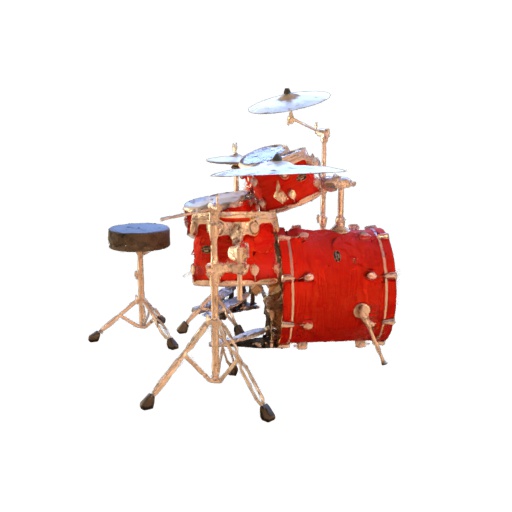}} &
			\raisebox{-0.5\height}{\includegraphics[trim={0cm 1.25cm 0cm 2cm},clip, width=\myfigsize]{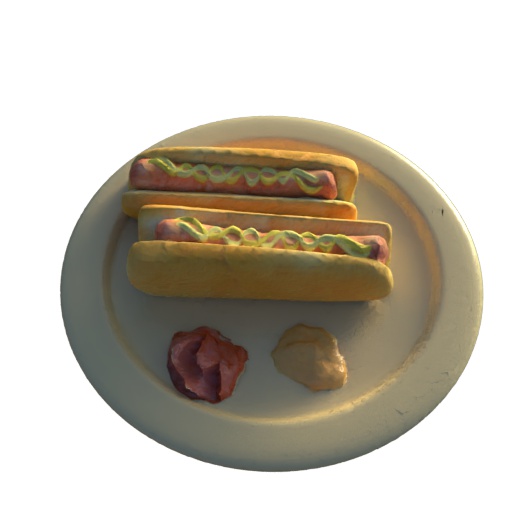}} &
			\raisebox{-0.5\height}{\includegraphics[trim={0cm 1.25cm 0cm 2cm},clip, width=\myfigsize]{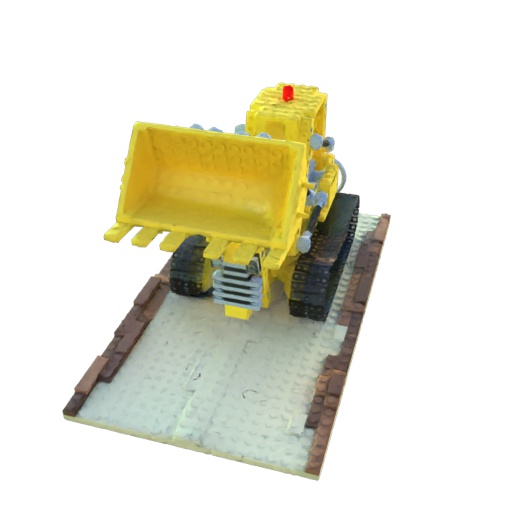}} &
			\raisebox{-0.5\height}{\includegraphics[trim={0cm 1.25cm 0cm 2cm},clip, width=\myfigsize]{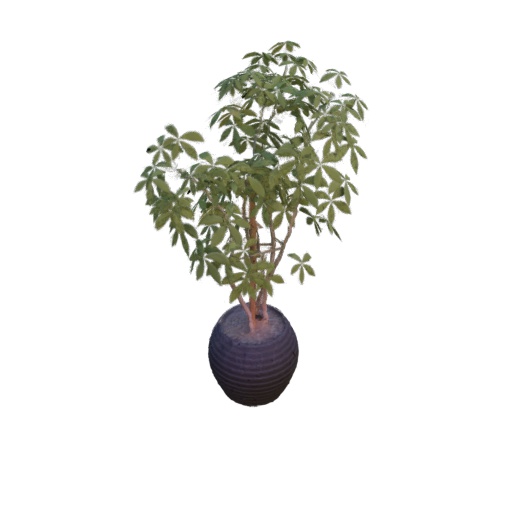}} &
			\raisebox{-0.5\height}{\includegraphics[trim={0cm 1.25cm 0cm 2cm},clip, width=\myfigsize]{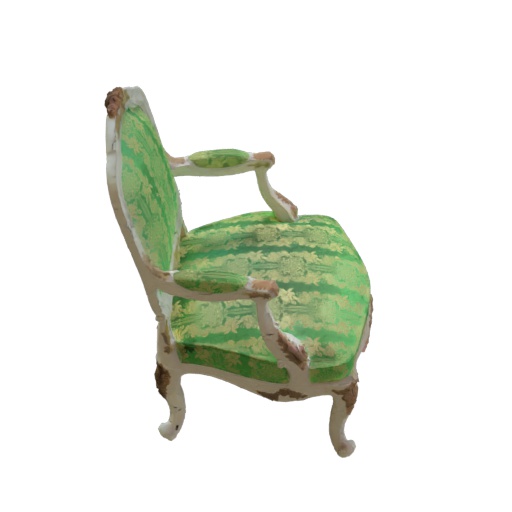}} &
			\raisebox{-0.5\height}{\includegraphics[trim={0cm 1.25cm 0cm 2cm},clip, width=\myfigsize]{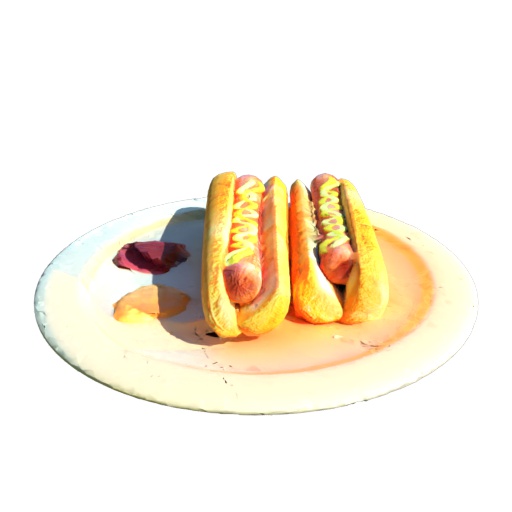}} &
			\raisebox{-0.5\height}{\includegraphics[trim={0cm 1.25cm 0cm 2cm},clip, width=\myfigsize]{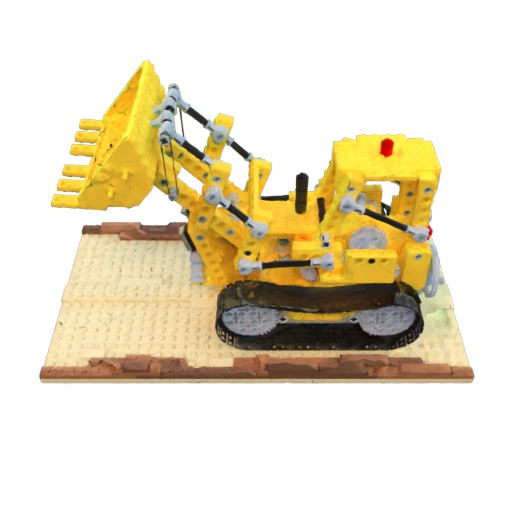}} &
			\raisebox{-0.5\height}{\includegraphics[trim={0cm 1.25cm 0cm 2cm},clip, width=\myfigsize]{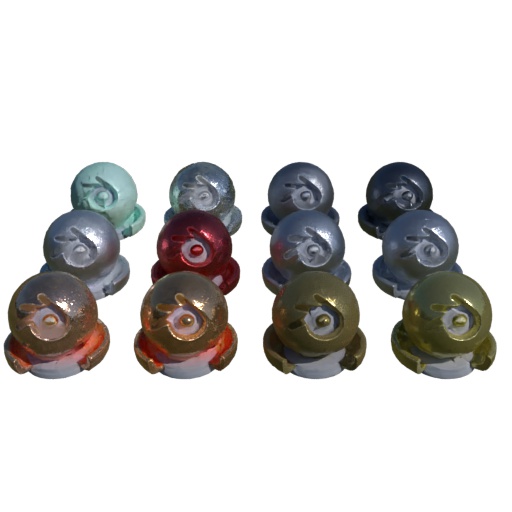}} \\
			
			& \small{Drums} & \small{Hotdog} & \small{Lego} & \small{Ficus} & \small{Chair} & \small{Hotdog} & \small{Lego} & \small{Materials} \\
			& \multicolumn{4}{c@{\hskip 2.0mm}}{\textbf{NeRFactor dataset}} & \multicolumn{4}{c}{\textbf{NeRF synthetic dataset}}		
		\end{tabular}
		\caption{Relighting examples from the NeRFactor and NeRF synthetic datasets. The NeRF dataset contains 
			high frequency lighting and global illumination, and is substantially more challenging than the NeRFactor
			version, which uses downsampled probes.	Our results contain visible artifacts, but outperform the material 
			separation of previous work.}
		\label{fig:relight_nerf_collage}
	\end{figure}
}


\newcommand{\figSceneEdit}{
	\begin{figure}
		\centering
		
		\centering
		\setlength{\myfigsize}{0.32\textwidth}
		
		\setlength{\tabcolsep}{1mm}
		\begin{tabular}{ccc}
			\includegraphics[width=\myfigsize]{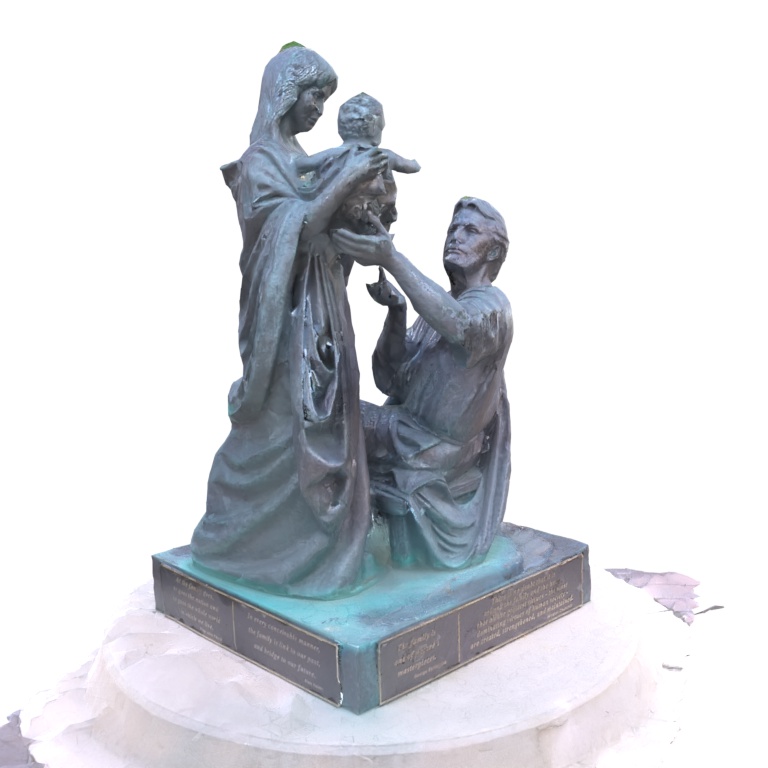} &
			\includegraphics[width=\myfigsize]{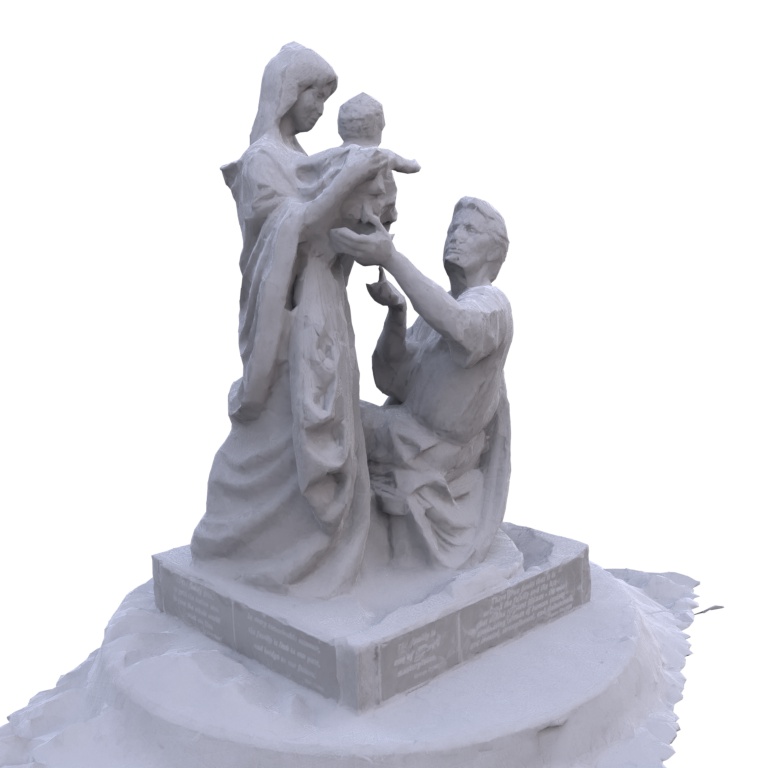} &
			\includegraphics[width=\myfigsize]{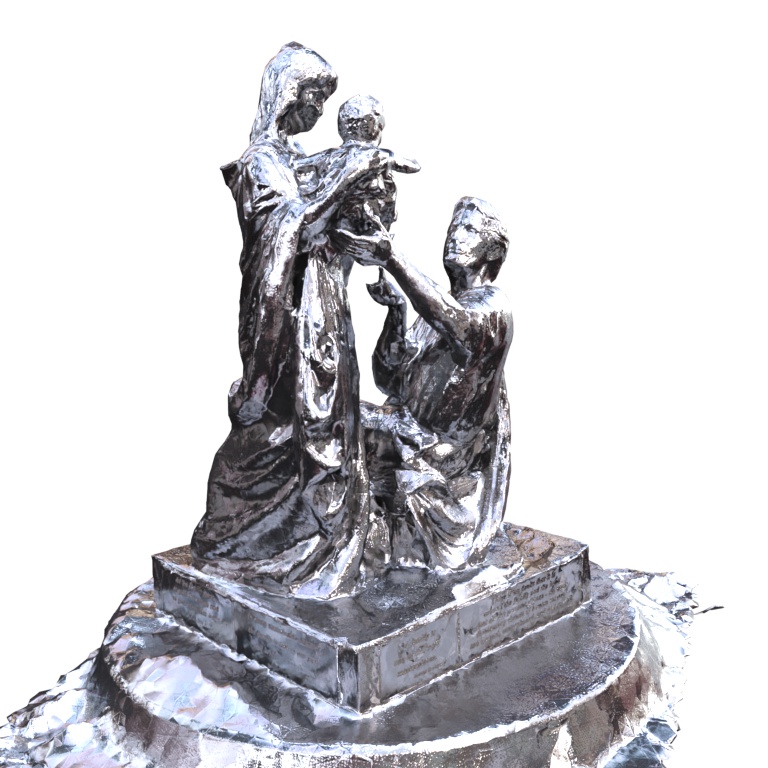} \\
			{Our reconstruction} & \multicolumn{2}{c}{{Material edits}} \\
			\includegraphics[width=\myfigsize]{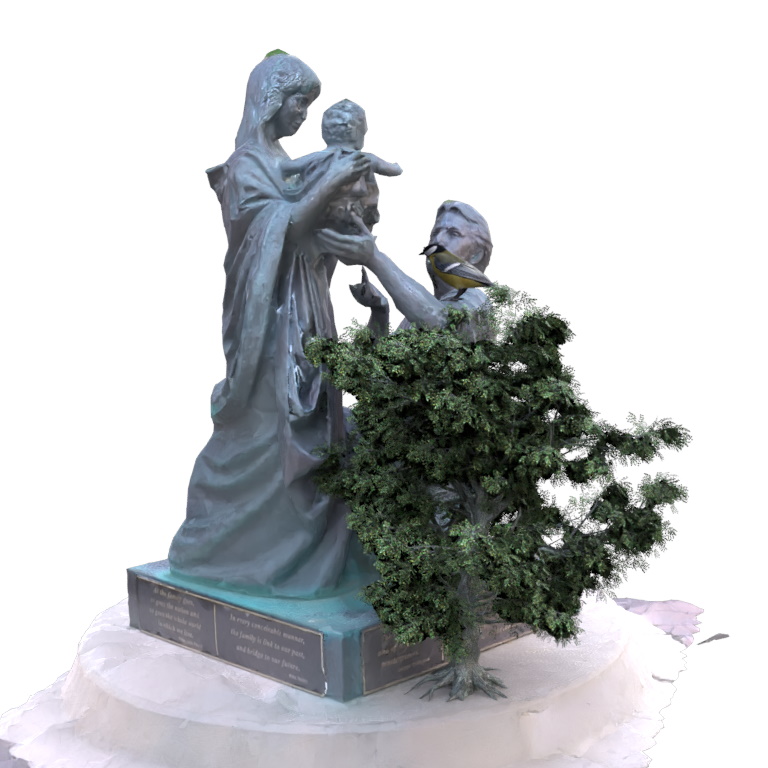} &
			\includegraphics[width=\myfigsize]{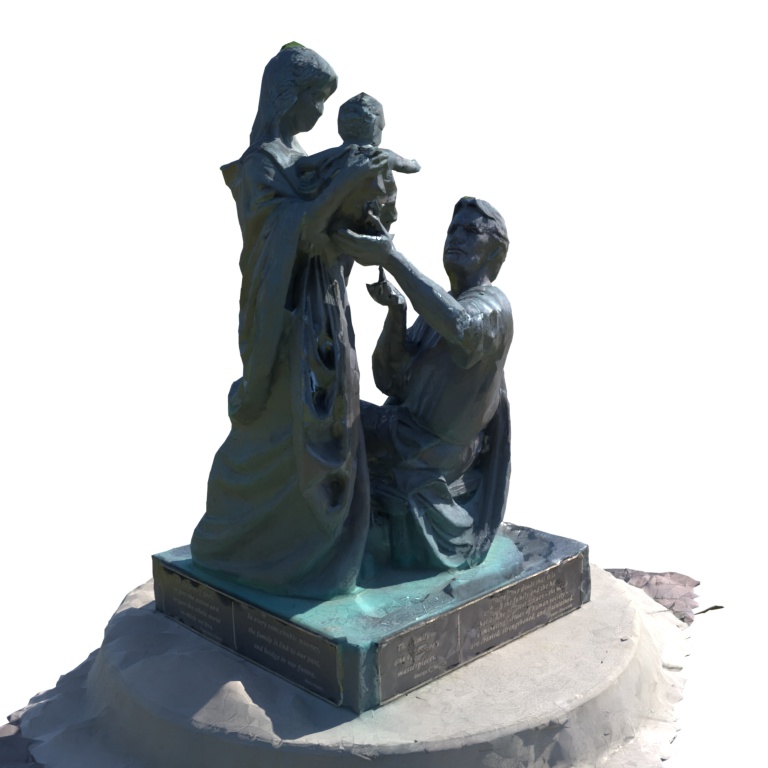} &
			\includegraphics[width=\myfigsize]{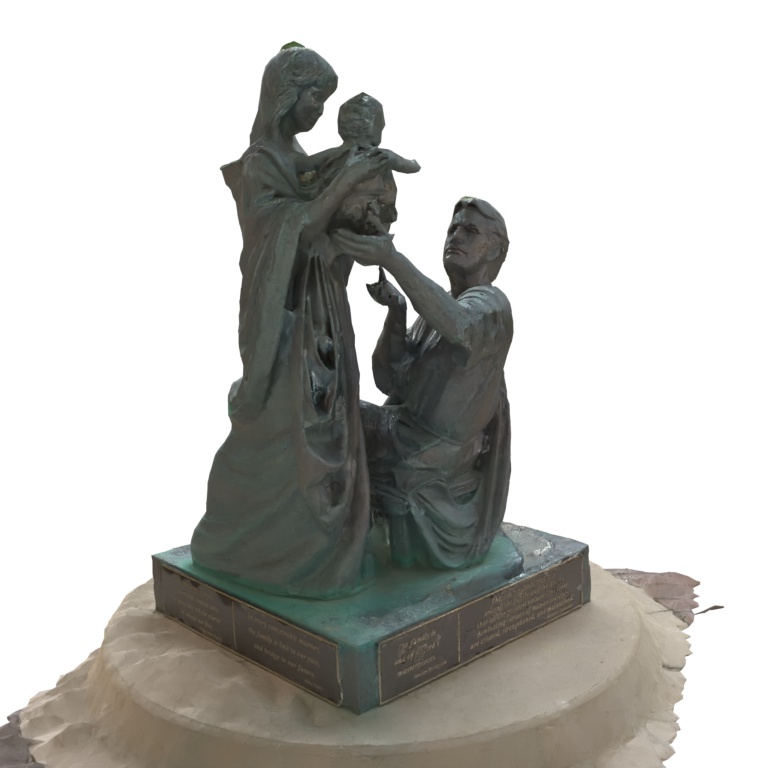} \\
			{Scene edit} & \multicolumn{2}{c}{{Relighting}}
		\end{tabular}
		\caption{Manipulations of our extracted 3D model of the Family dataset in Blender. This scene is part of the Tanks\&Temples~\cite{Knapitsch2017} 
			dataset (CC BY-NC-SA 3.0). Tree and bird models from TurboSquid.}
		\label{fig:scene_edit}
	\end{figure}
}


\newcommand{\figPhotogrammetry}{
	\begin{figure}
		\centering	
		\setlength{\myfigsize}{24mm}		
		\setlength{\tabcolsep}{0mm}
		\def\arraystretch{0}
		\begin{tabular}{lccc@{\hskip1mm}c@{\hskip1mm}ccccc}
			\multirow{2}{*}[0.25\myfigsize]{\rotatebox[origin=c]{90}{Family}}  &
			\multirow{2}{*}[0.5\myfigsize]{\makecropbox{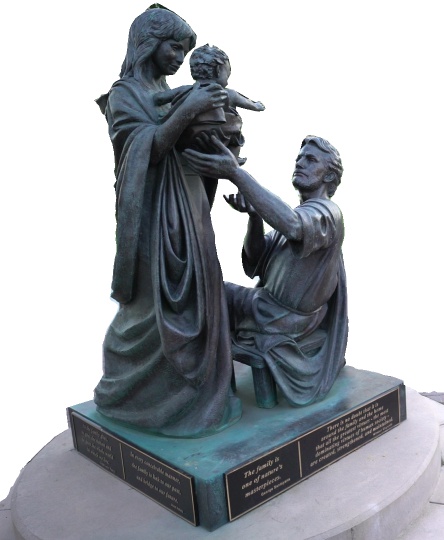}{\myfigsize}{0.4}{0.5}{0.8}{0.75}} &
			\multirow{2}{*}[0.5\myfigsize]{\includegraphics[height=\myfigsize]{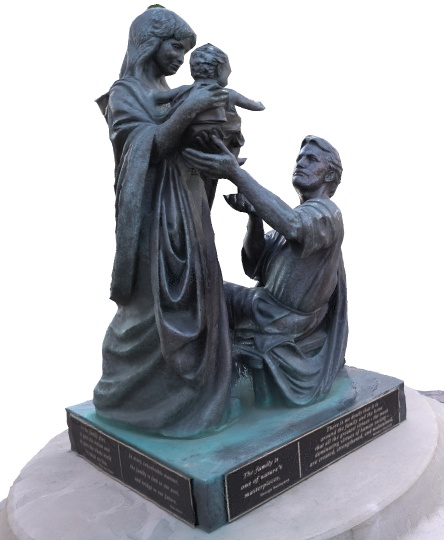}} &
			\multirow{2}{*}[0.5\myfigsize]{\includegraphics[height=\myfigsize]{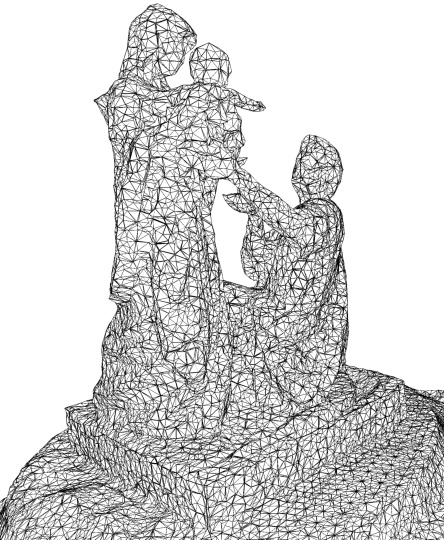}} & 
			
			\multirow{1}{*}[0.32\myfigsize]{\rotatebox[origin=c]{90}{\tiny Our}} &
			\includegraphics[height=0.5\myfigsize]{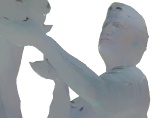} & 
			\includegraphics[height=0.5\myfigsize]{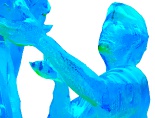} & 
			\includegraphics[height=0.5\myfigsize]{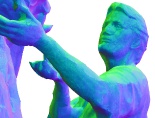} & 
			\includegraphics[height=0.5\myfigsize]{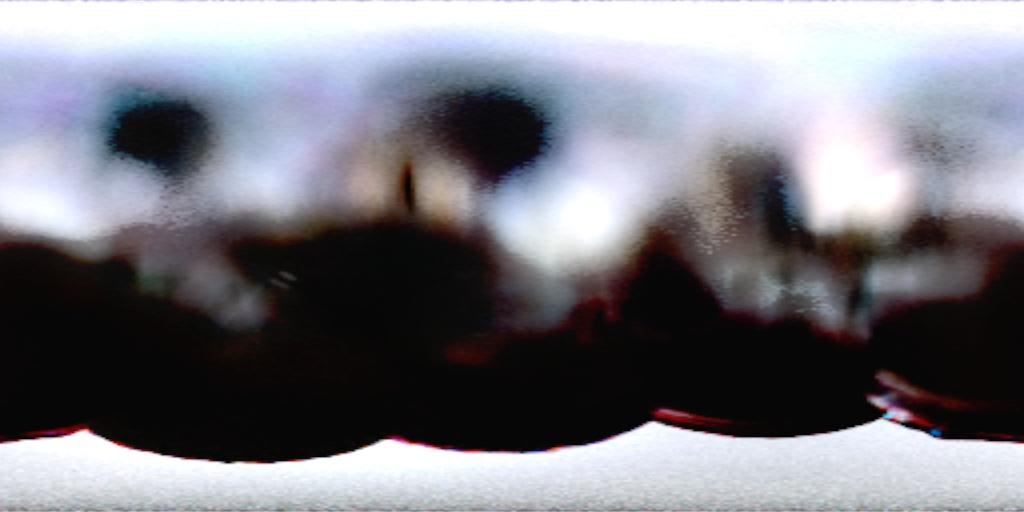} \\
			\vspace{.5mm}
			
			& & & & 
			\multirow{1}{*}[0.45\myfigsize]{\rotatebox[origin=c]{90}{\tiny\textsc{nvdiffrec}}} &
			\includegraphics[height=0.5\myfigsize]{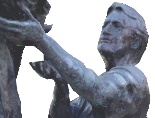} & 
			\includegraphics[height=0.5\myfigsize]{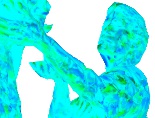} & 
			\includegraphics[height=0.5\myfigsize]{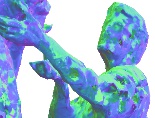} &
			\includegraphics[height=0.5\myfigsize]{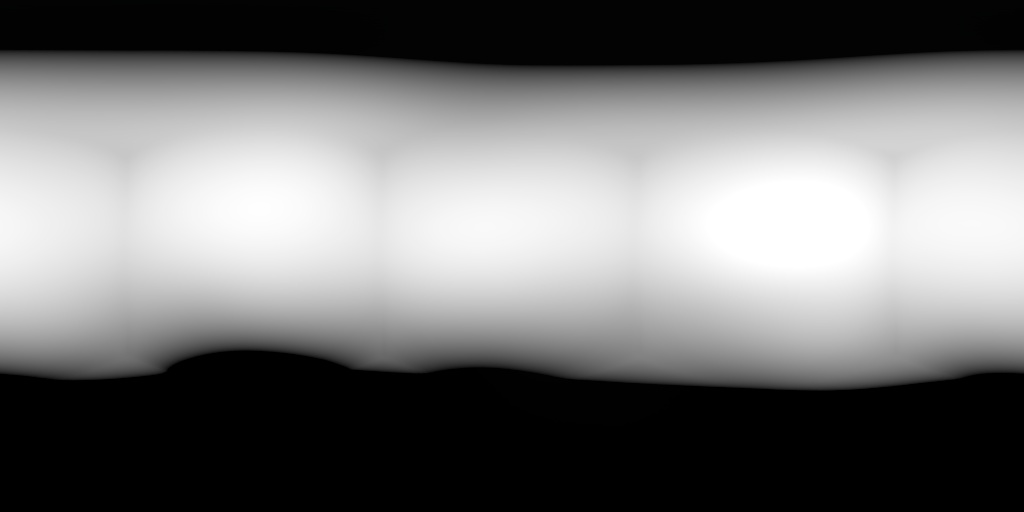} \\
			
			\multirow{2}{*}[0.25\myfigsize]{\rotatebox[origin=c]{90}{Character}}  &
			\multirow{2}{*}[0.5\myfigsize]{\makecropbox{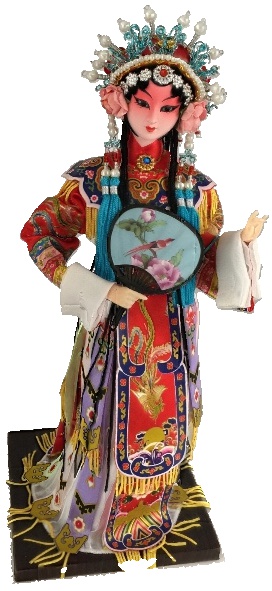}{\myfigsize}{0.17}{0.45}{0.8}{0.66}} &
			\multirow{2}{*}[0.5\myfigsize]{\includegraphics[height=\myfigsize]{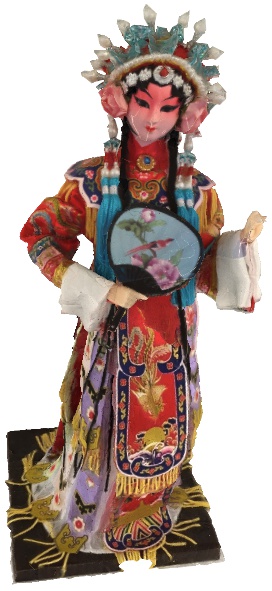}} &
			\multirow{2}{*}[0.5\myfigsize]{\includegraphics[height=\myfigsize]{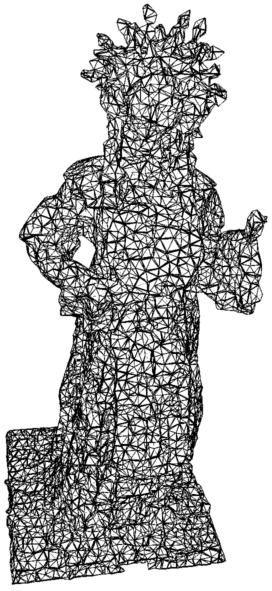}} & 
			
			\multirow{1}{*}[0.32\myfigsize]{\rotatebox[origin=c]{90}{\tiny Our}} &
			\includegraphics[height=0.5\myfigsize]{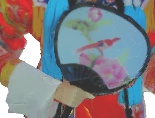} & 
			\includegraphics[height=0.5\myfigsize]{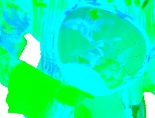} & 
			\includegraphics[height=0.5\myfigsize]{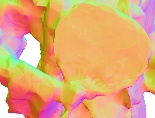} & 
			\includegraphics[height=0.5\myfigsize]{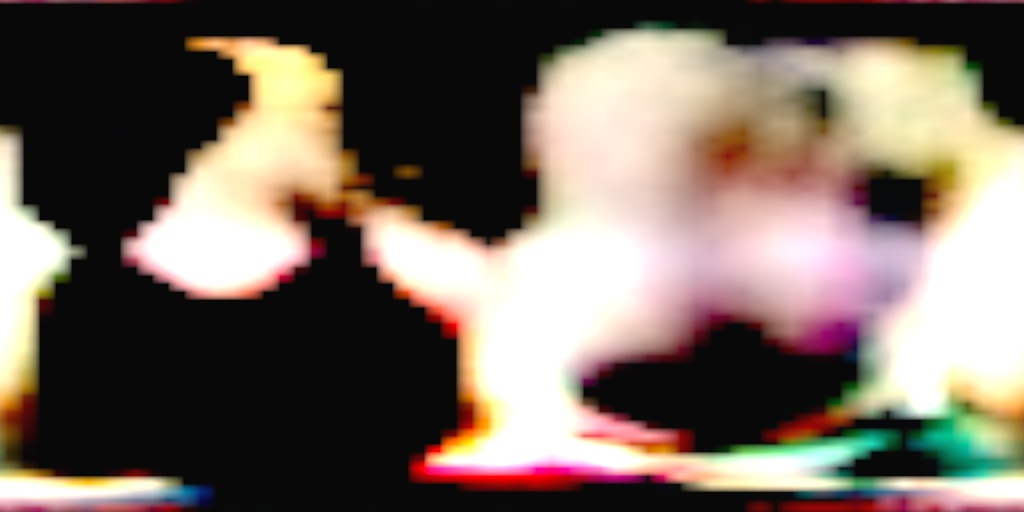} \\
			\vspace{.5mm}
			
			& & & & 
			\multirow{1}{*}[0.45\myfigsize]{\rotatebox[origin=c]{90}{\tiny\textsc{nvdiffrec}}} &
			\includegraphics[height=0.5\myfigsize]{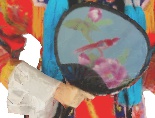} & 
			\includegraphics[height=0.5\myfigsize]{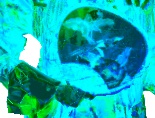} & 
			\includegraphics[height=0.5\myfigsize]{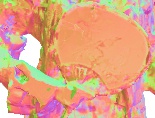} &
			\includegraphics[height=0.5\myfigsize]{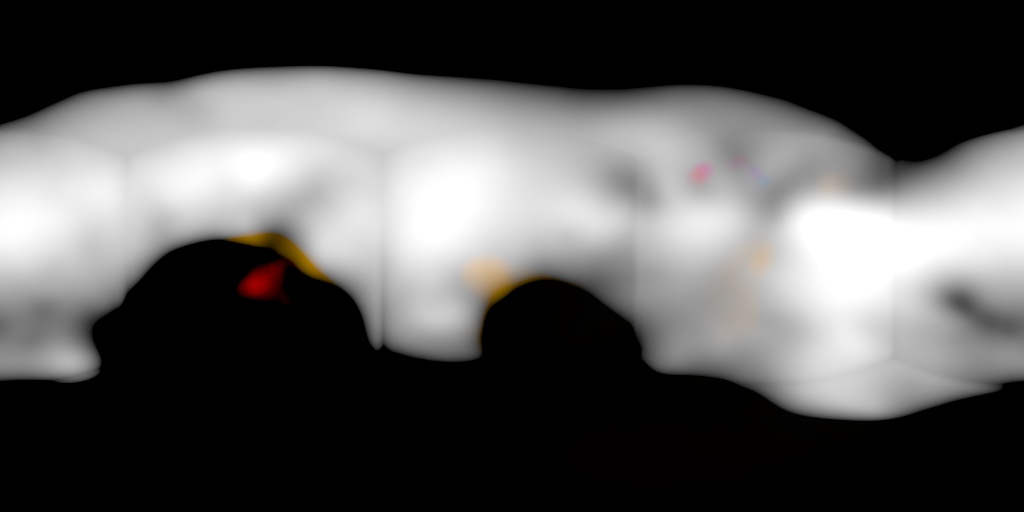} \\
			
			\multirow{2}{*}[0.25\myfigsize]{\rotatebox[origin=c]{90}{Gold Cape}}  &
			\multirow{2}{*}[0.5\myfigsize]{\makecropbox{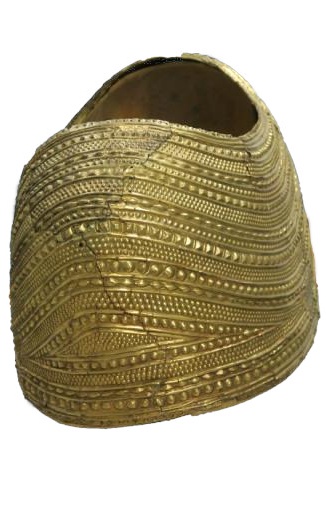}{\myfigsize}{0.1}{0.55}{0.55}{0.78}} &
			\multirow{2}{*}[0.5\myfigsize]{\includegraphics[height=\myfigsize]{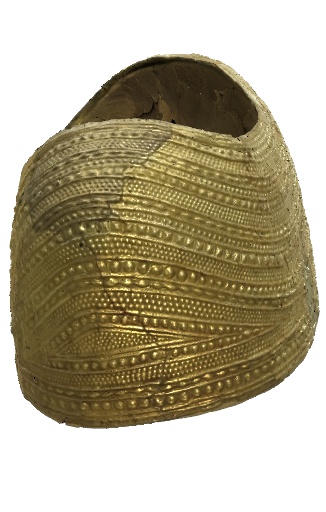}} &
			\multirow{2}{*}[0.5\myfigsize]{\includegraphics[height=\myfigsize]{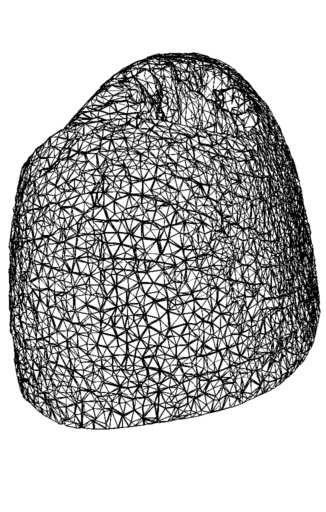}} & 
			
			\multirow{1}{*}[0.32\myfigsize]{\rotatebox[origin=c]{90}{\tiny Our}} &
			\includegraphics[height=0.5\myfigsize]{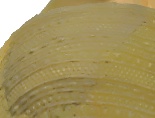} & 
			\includegraphics[height=0.5\myfigsize]{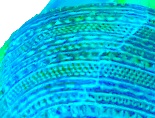} & 
			\includegraphics[height=0.5\myfigsize]{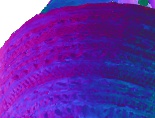} & 
			\includegraphics[height=0.5\myfigsize]{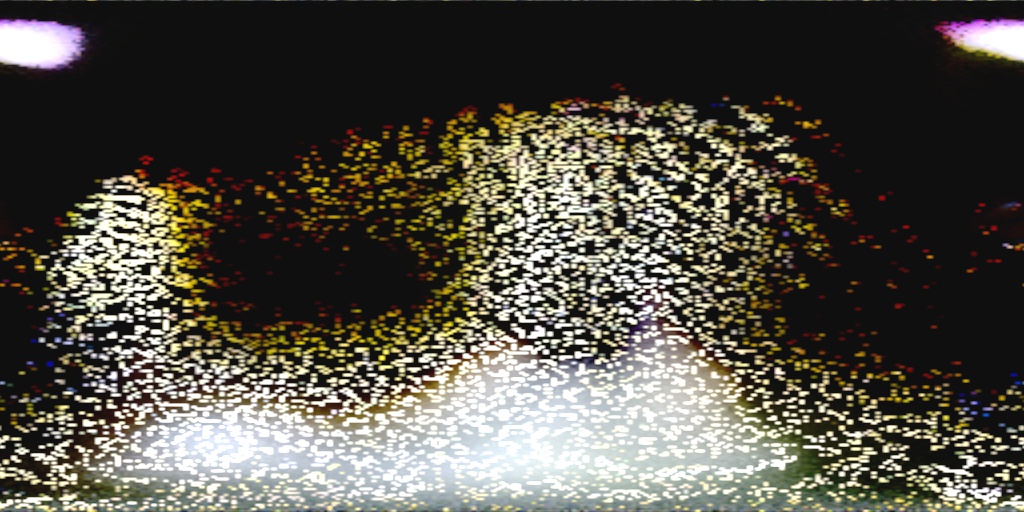} \\
			\vspace{1.5mm}
			
			& & & & 
			\multirow{1}{*}[0.45\myfigsize]{\rotatebox[origin=c]{90}{\tiny\textsc{nvdiffrec}}} &
			\includegraphics[height=0.5\myfigsize]{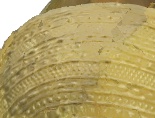} & 
			\includegraphics[height=0.5\myfigsize]{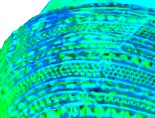} & 
			\includegraphics[height=0.5\myfigsize]{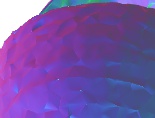} &
			\includegraphics[height=0.5\myfigsize]{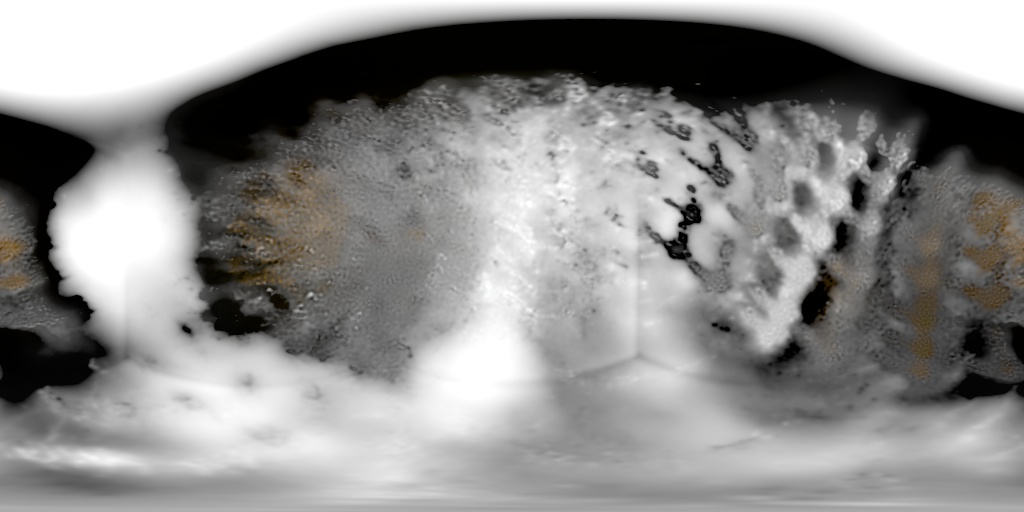} \\
			
			& Ref & Our & Geometry & & $\mathbf{k}_d$ & $\mathbf{k}_{orm}$ & Normal & Probe
		\end{tabular}
		
		\caption{\protect We show explicit decomposition of shape, materials and lighting, directly from photos with known poses. Character is part of the BlendedMVS~\cite{Yao2020blendedmvs} dataset (CC BY-4.0) 
		and Gold Cape is part of the NeRD~\cite{Boss2021} dataset (CC BY-NC-SA 4.0).}
		\label{fig:nsvf}
	\end{figure}
}


\newcommand{\figPorscheDenoising}{
\begin{figure}[tb]
    \centering
    \setlength{\tabcolsep}{1pt}
    \begin{tabular}{cccccc}
		\multirow{2}{*}[1.2cm]{\makeredbox{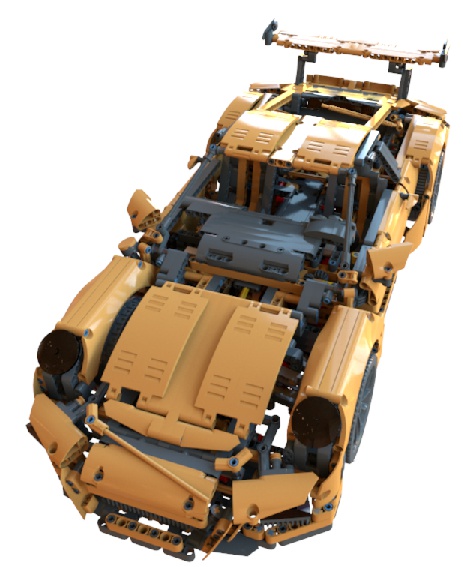}} 
		& \includegraphics[width=0.155\columnwidth]{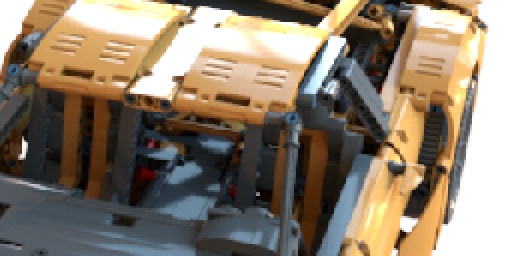} 
		& \includegraphics[width=0.155\columnwidth]{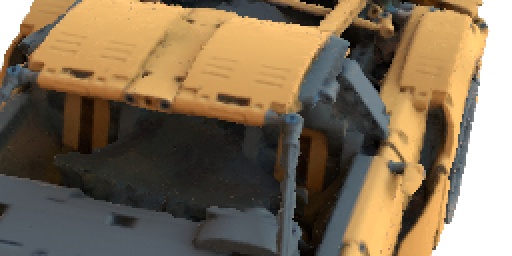}
		& \includegraphics[width=0.155\columnwidth]{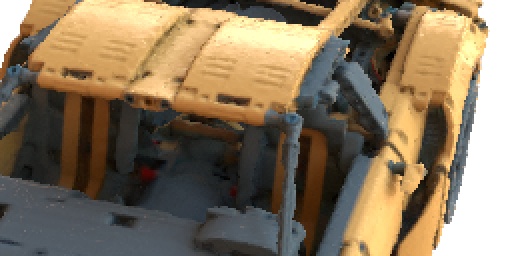}
		& \includegraphics[width=0.155\columnwidth]{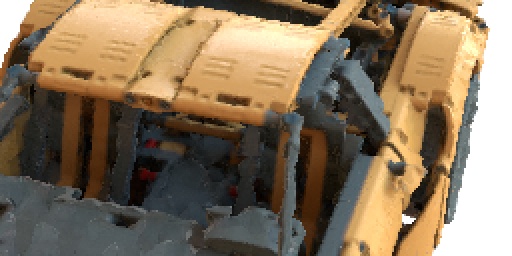}
		& \includegraphics[width=0.155\columnwidth]{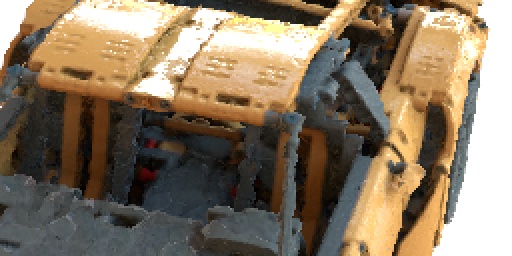} \\

		& \includegraphics[width=0.155\columnwidth]{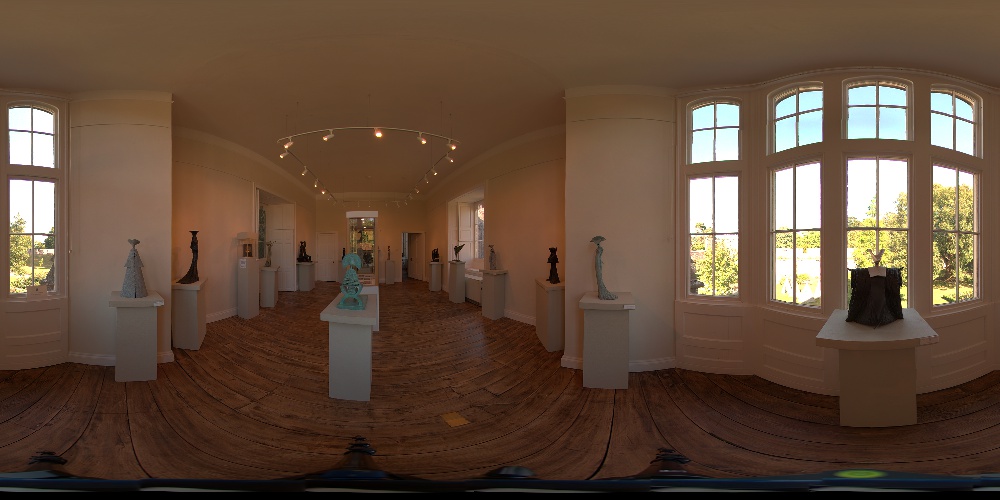} 
		& \includegraphics[width=0.155\columnwidth]{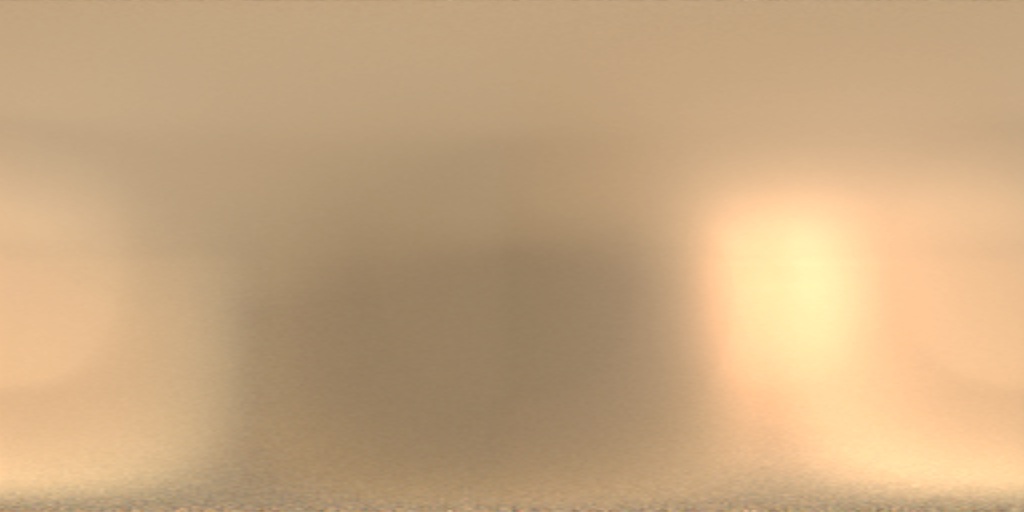}
		& \includegraphics[width=0.155\columnwidth]{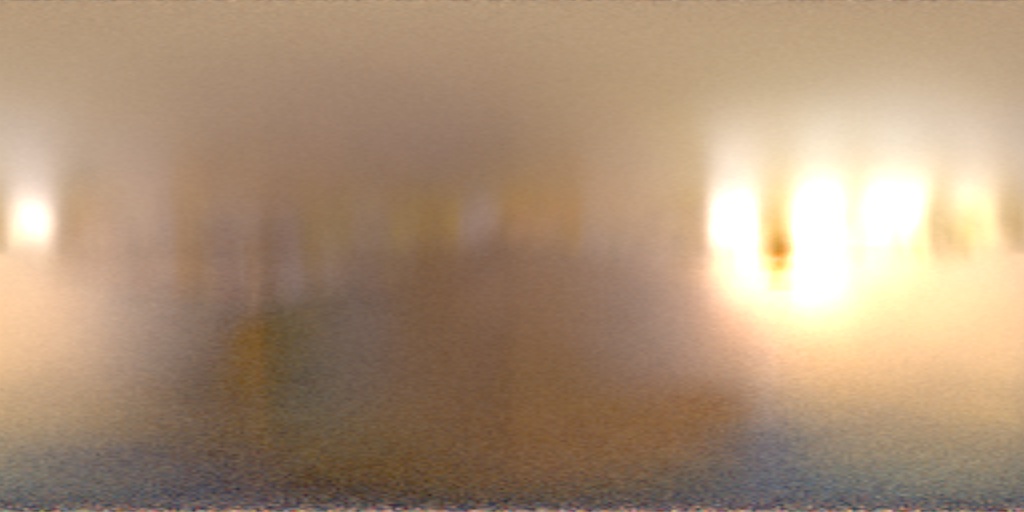}
		& \includegraphics[width=0.155\columnwidth]{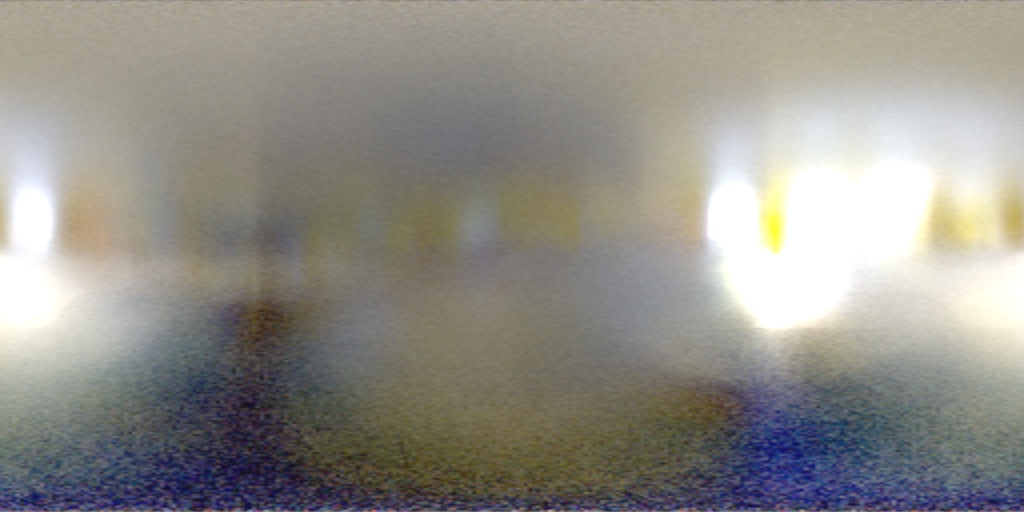} 
		& \includegraphics[width=0.155\columnwidth]{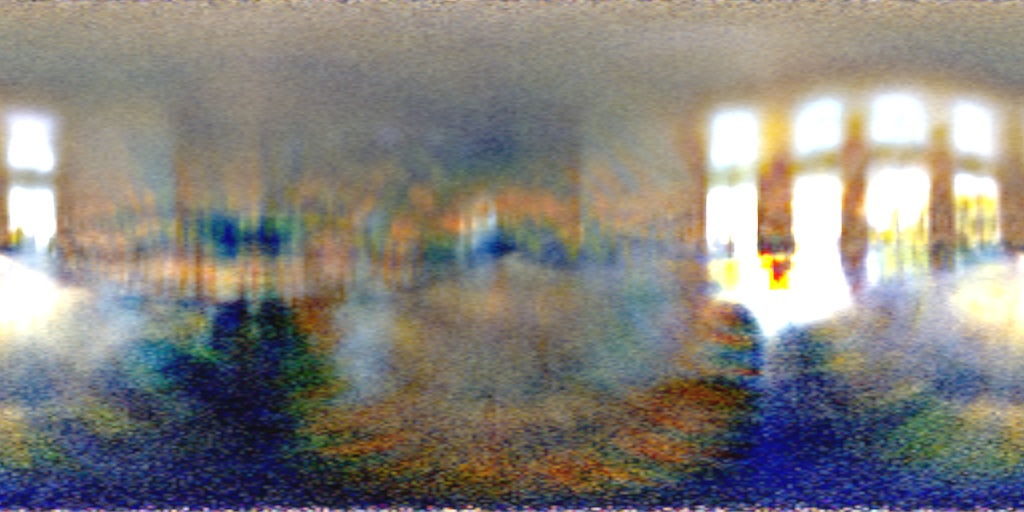} \\
		& PSNR &  \small{24.4~dB} & \small{25.7~dB} & \small{25.6~dB} & \small{26.0~dB} \\
        & \small{Reference} & \small{8~spp} & \small{8~spp denoised} & \small{128~spp} & \small{128~spp denoised} 
    \end{tabular}
    \caption{We show the benefits of denoising on the Porsche scene from LDraw \protect resources~\cite{Lasser2022} (CC BY-2.0). 
    At low sample counts, denoising helps both with geometric reconstruction (in the cockpit) and to capture specular 
    highlights. Even at 128~spp, denoising improves specular highlight and high frequency lighting details.}
    \label{fig:porsche_denoising}
\end{figure}
}



\newcommand{\figLightReg}{
	\begin{figure}
		\centering	
		\setlength{\myfigsize}{52.5mm}		
		\setlength{\tabcolsep}{0mm}
		\def\arraystretch{1}
		\begin{tabular}{lcccc}
			\rotatebox[origin=c]{90}{w/o light reg.} &
			\raisebox{-0.5\height}{\includegraphics[trim={0cm 1.7cm 0cm 3cm}, clip, height=0.5\myfigsize]{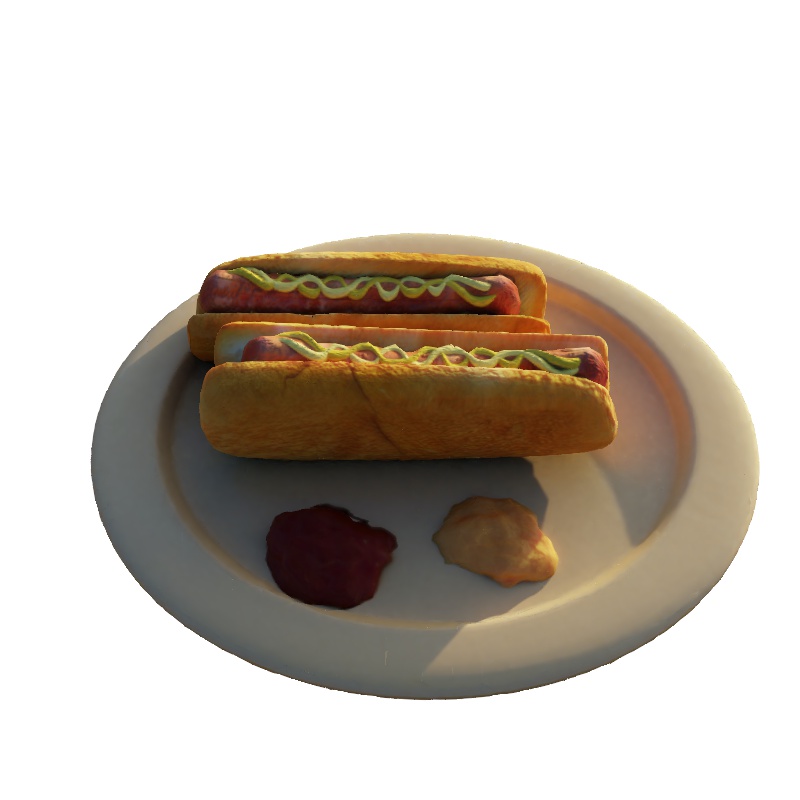}} &
			\raisebox{-0.5\height}{\includegraphics[trim={0cm 1.7cm 0cm 3cm}, clip, height=0.5\myfigsize]{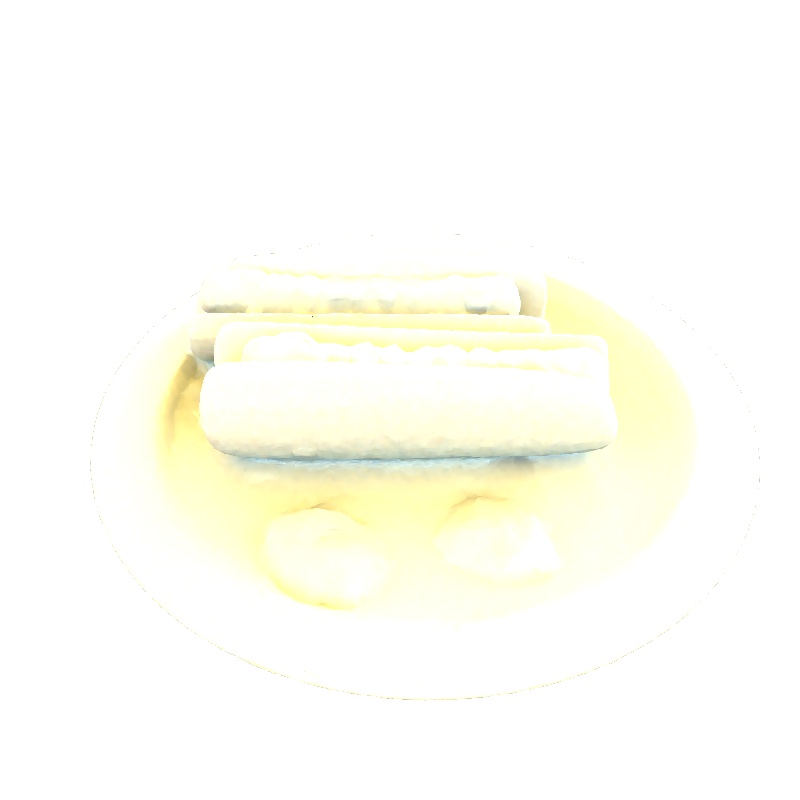}} &
			\raisebox{-0.5\height}{\includegraphics[trim={0cm 1.7cm 0cm 3cm}, clip, height=0.5\myfigsize]{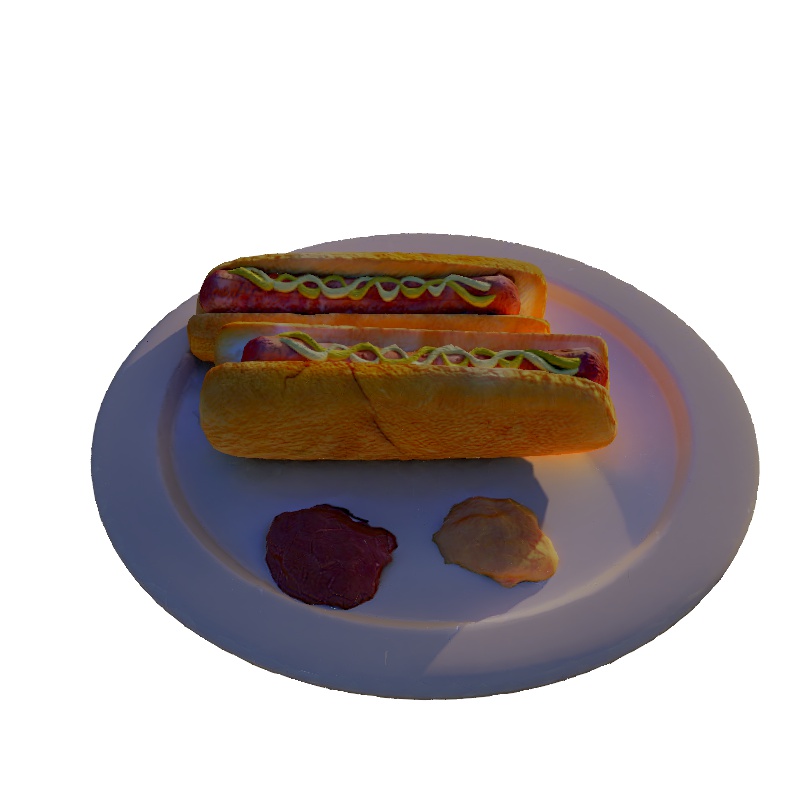}} & 
			\raisebox{-0.5\height}{\includegraphics[trim={0cm 1.7cm 0cm 3cm}, clip, height=0.5\myfigsize]{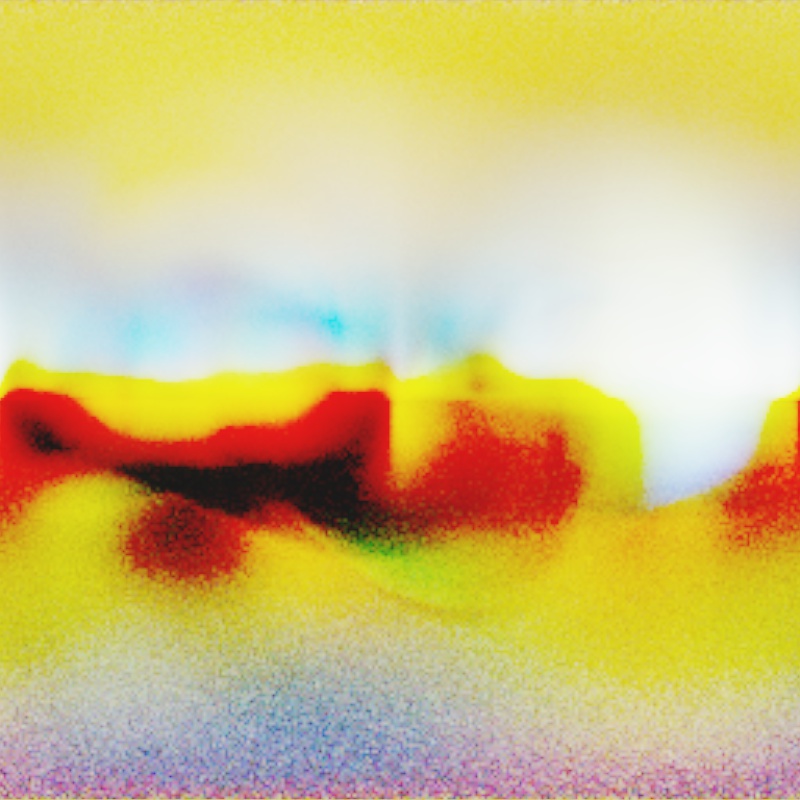}} \\
			
			\rotatebox[origin=c]{90}{w/ light reg.} &
			\raisebox{-0.5\height}{\includegraphics[trim={0cm 1.7cm 0cm 3cm}, clip, height=0.5\myfigsize]{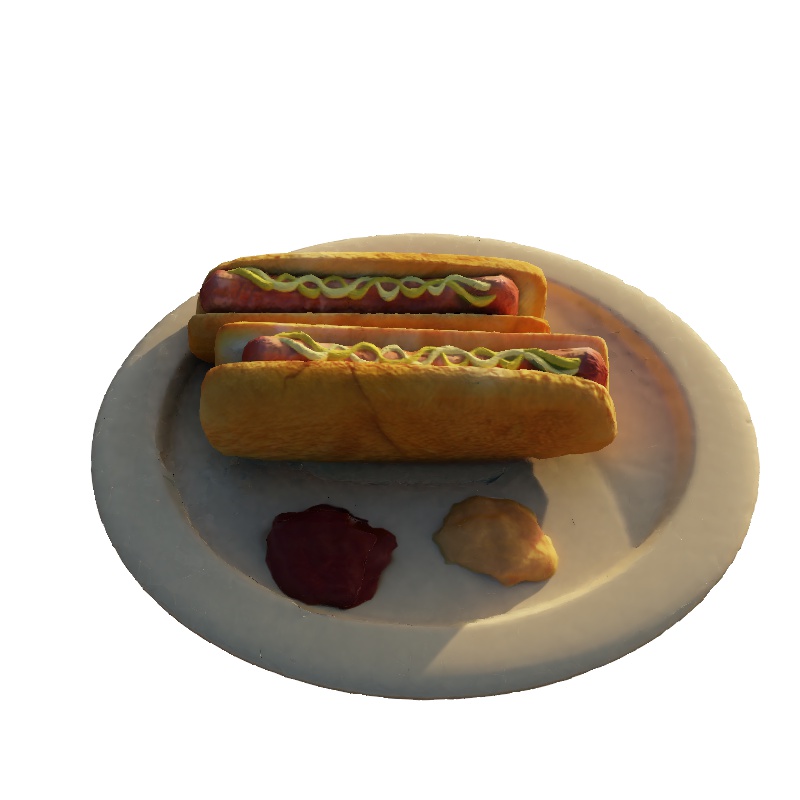}} & 
			\raisebox{-0.5\height}{\includegraphics[trim={0cm 1.7cm 0cm 3cm}, clip, height=0.5\myfigsize]{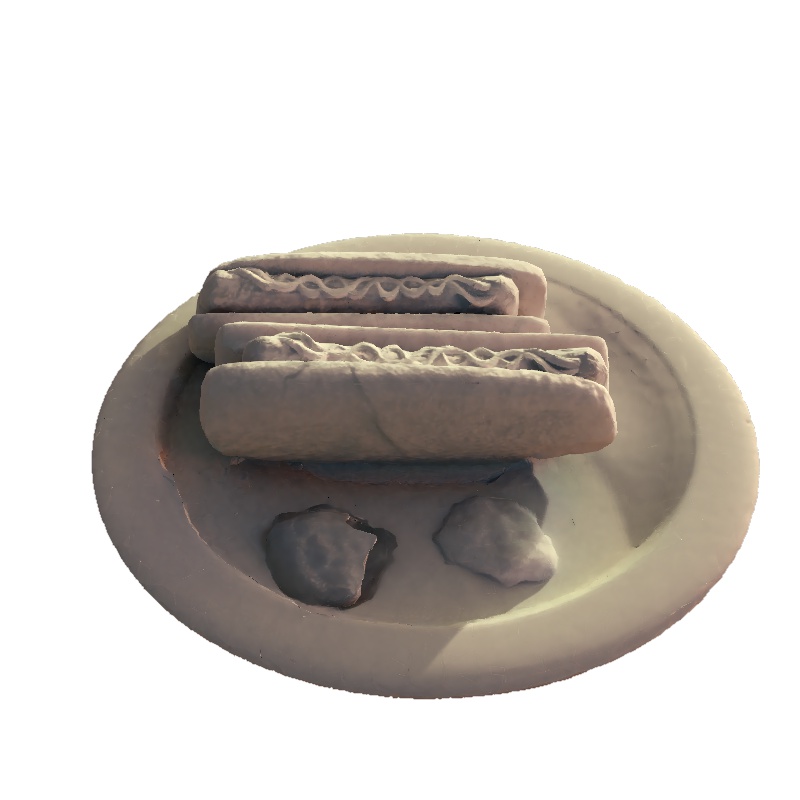}} & 
			\raisebox{-0.5\height}{\includegraphics[trim={0cm 1.7cm 0cm 3cm}, clip, height=0.5\myfigsize]{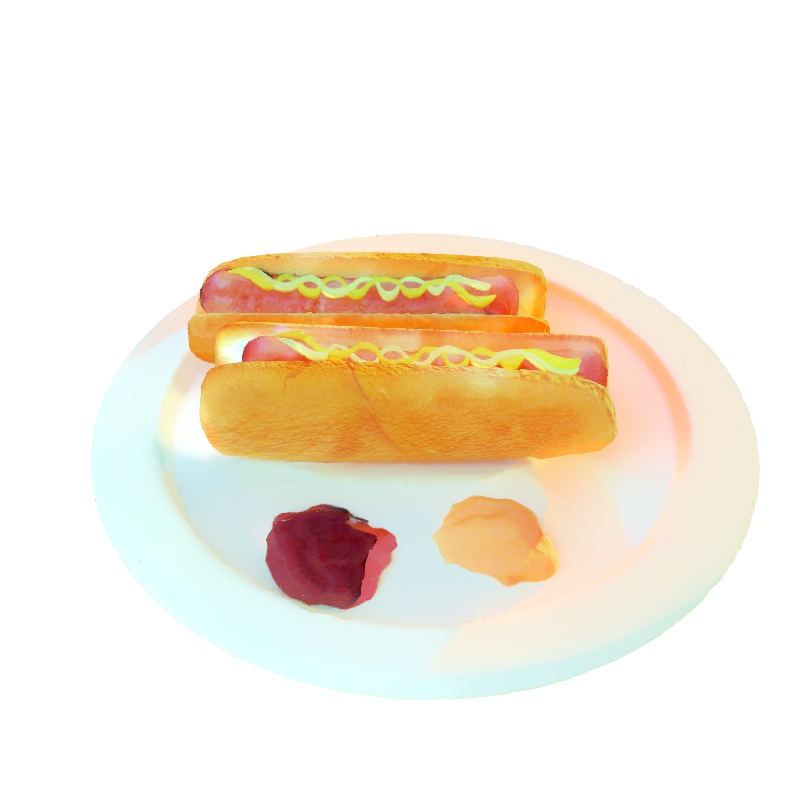}} &
			\raisebox{-0.5\height}{\includegraphics[trim={0cm 1.7cm 0cm 3cm}, clip, height=0.5\myfigsize]{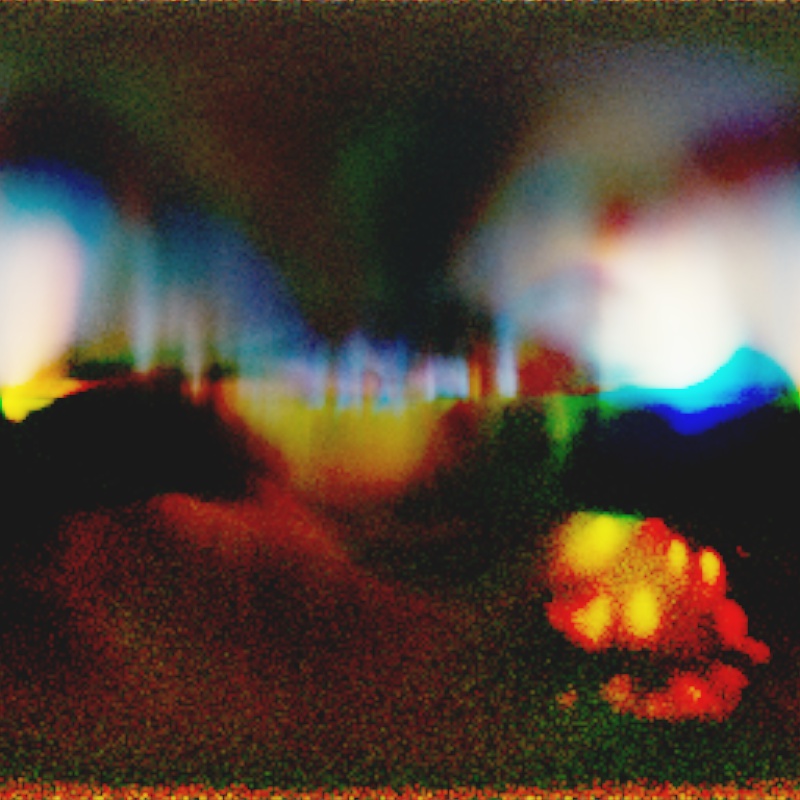}} \\
			
			& Shaded & Diffuse Light & $\kd$ & Probe \\
		\end{tabular}
		\caption{
			The impact of our light regularizer on the Hotdog scene from the synthetic NeRF dataset. The regularizer is particularly 
			effective in this example, as the dataset contains high frequency lighting with sharp shadows. As shown in the top row, without regularization 
			most of the lighting and shadow is baked into the $\kd$ texture because it is the quickest way for the optimizer to minimize loss. While our 
			regularized result is not perfect (some texture detail on the hotdog bun is incorrectly baked into lighting), the lighting is accurately 
			captured and $\kd$ contains mostly chrominance, as expected. The light probe is also significantly more detailed.
		}
		\label{fig:lighting_reg}
	\end{figure}
}

\newcommand{\figProbes}
{
	\begin{figure}
		\centering
		\setlength{\tabcolsep}{0mm}
		\begin{tabular}{cccc}    
			\includegraphics[width=0.2\columnwidth]{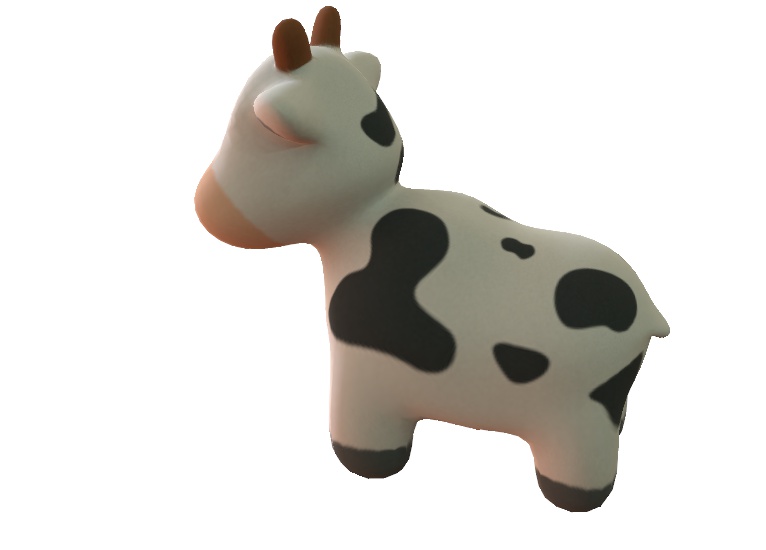} &
			\includegraphics[width=0.3\columnwidth]{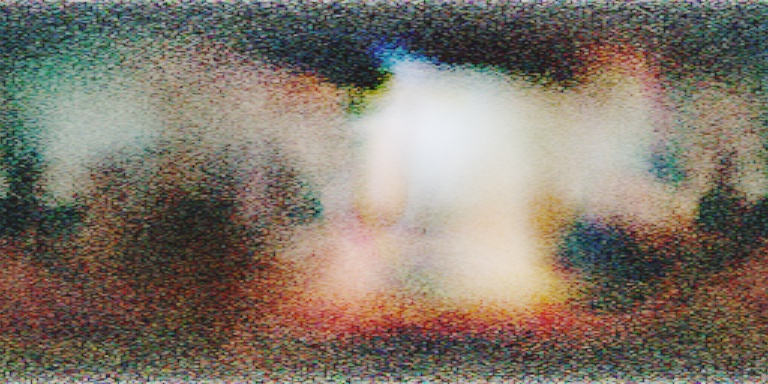} &
			\includegraphics[width=0.2\columnwidth]{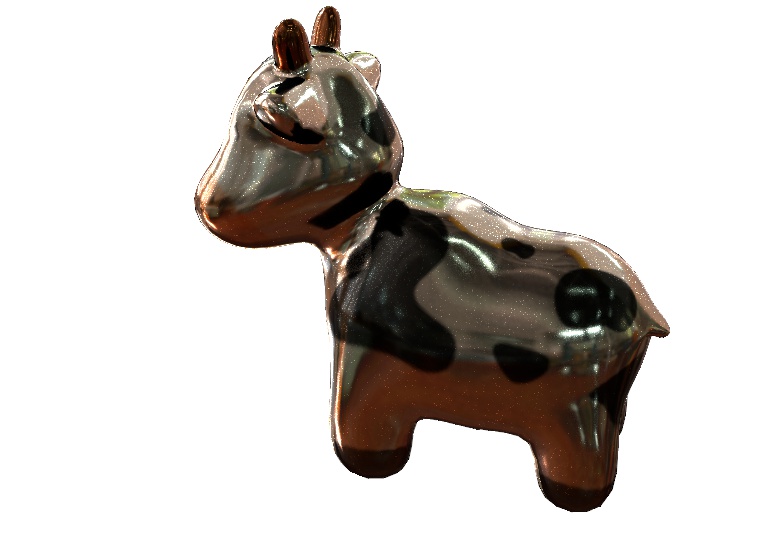} &
			\includegraphics[width=0.3\columnwidth]{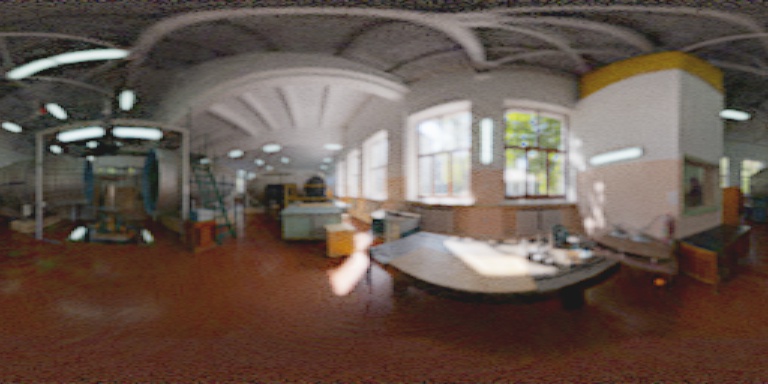} \\
			\multicolumn{2}{c}{\small{Diffuse}} & \multicolumn{2}{c}{\small{Specular}}
		\end{tabular}
		\caption{We compute direct lighting by integrating the BSDF over the hemisphere using Monte Carlo. 
			For diffuse scenes (left), the BSDF has a wide lobe, and the extracted 
			probe often contains significant noise. For specular scenes, the BSDF lobe is sharper, 
			which allow us to recover high frequency lighting.} \label{fig:probes}
	\end{figure}
}


\newcommand{\figShadowRamp}{
	\begin{figure}
		\centering	
		\setlength{\myfigsize}{45mm}		
		\setlength{\tabcolsep}{0mm}
		\def\arraystretch{1}
		\begin{tabular}{lcccc}
			\rotatebox[origin=c]{90}{w/o blend} &
			\raisebox{-0.5\height}{\includegraphics[trim={0cm 1.7cm 0cm 3cm}, clip, height=0.5\myfigsize]{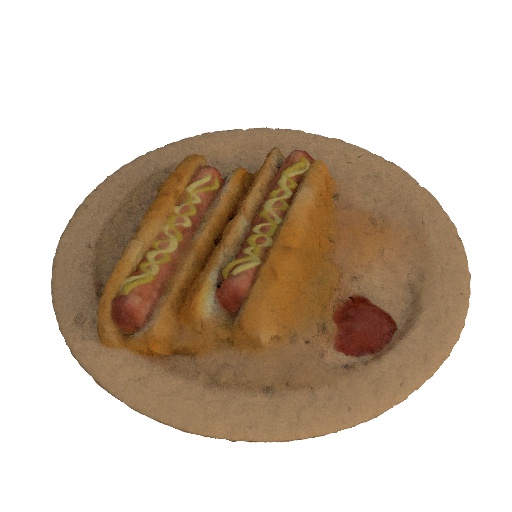}} &
			\raisebox{-0.5\height}{\includegraphics[trim={0cm 1.7cm 0cm 3cm}, clip, height=0.5\myfigsize]{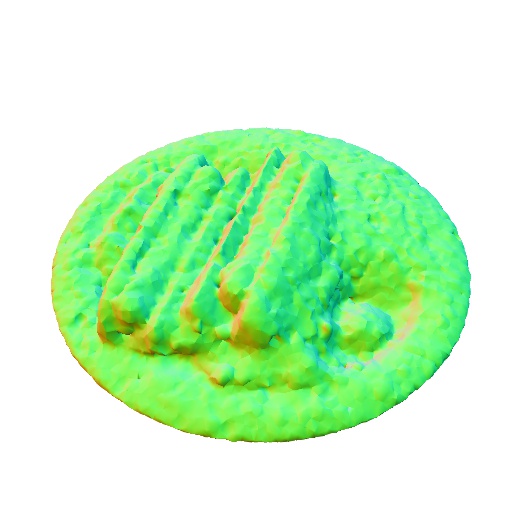}} &
			\raisebox{-0.5\height}{\includegraphics[trim={0cm 1.7cm 0cm 3cm}, clip, height=0.5\myfigsize]{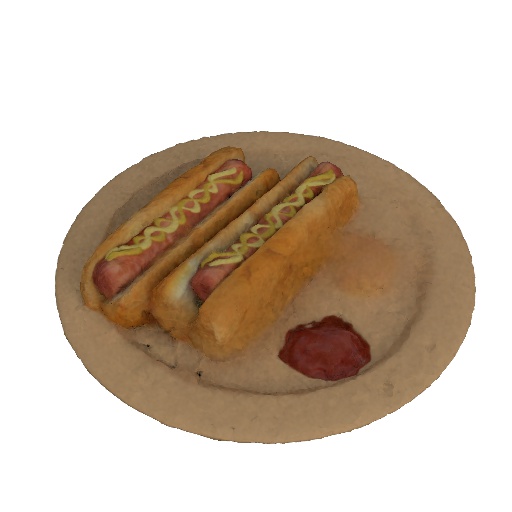}} & 
			\raisebox{-0.5\height}{\includegraphics[trim={0cm 1.7cm 0cm 3cm}, clip, height=0.5\myfigsize]{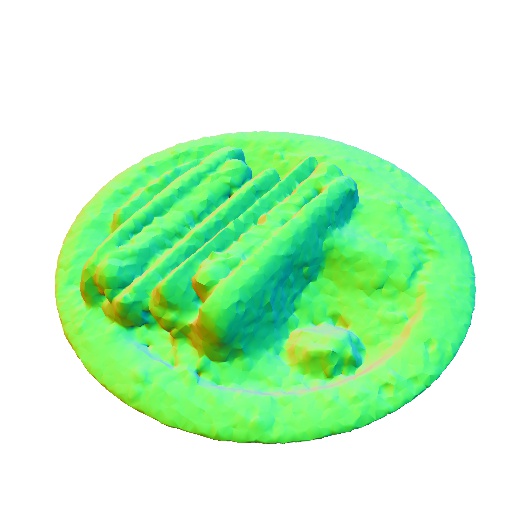}} \\
			
			\rotatebox[origin=c]{90}{w/ blend} &
			\raisebox{-0.5\height}{\includegraphics[trim={0cm 1.7cm 0cm 3cm}, clip, height=0.5\myfigsize]{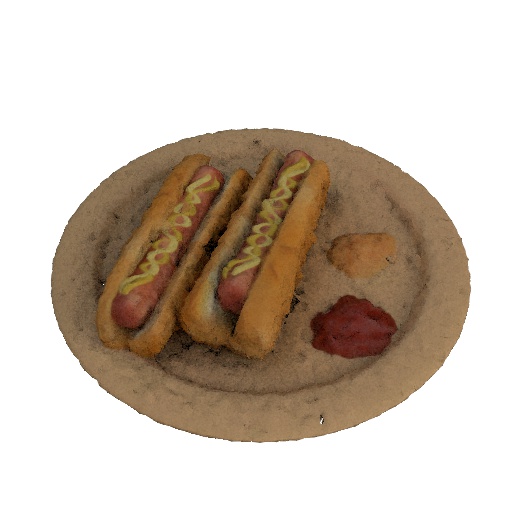}} & 
			\raisebox{-0.5\height}{\includegraphics[trim={0cm 1.7cm 0cm 3cm}, clip, height=0.5\myfigsize]{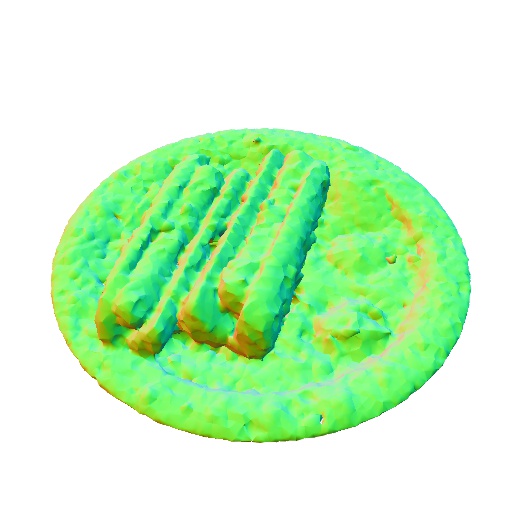}} & 
			\raisebox{-0.5\height}{\includegraphics[trim={0cm 1.7cm 0cm 3cm}, clip, height=0.5\myfigsize]{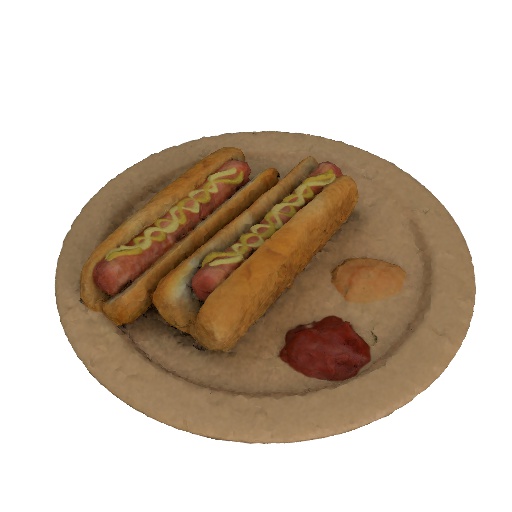}} &
			\raisebox{-0.5\height}{\includegraphics[trim={0cm 1.7cm 0cm 3cm}, clip, height=0.5\myfigsize]{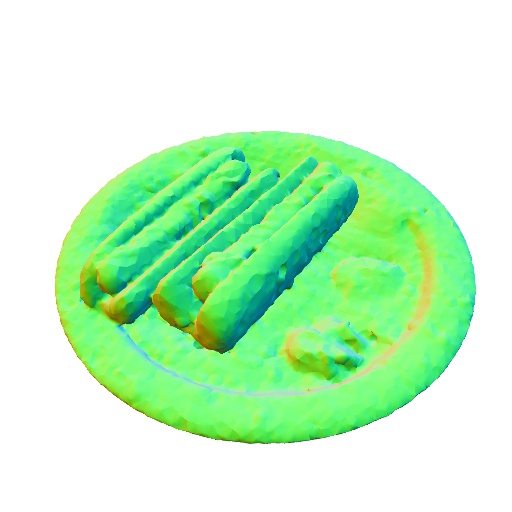}} \\
			
			& \small{Shaded} & \small{Normals} & \small{Shaded} & \small{Normals} \\
			& \multicolumn{2}{c}{\small{500 iterations}} & \multicolumn{2}{c}{\small{1000 iterations}}
		\end{tabular}
		\caption{
			Illustration of the impact of gradually blending in the shadow term in the early stages of optimization. 
			In early stages, there are large changes in topology, and local changes in geometry can have a large (global) impact 
			on color and shading. As shown in the top row, early convergence suffers, and the optimization fails to carve out some geometry.
			By incrementally ramping up the shadow term, as shown in the bottom row, we improve geometry reconstruction.	
		}
		\label{fig:shadow_ramp}
	\end{figure}
}


\newcommand{\figAblationDenoisePlotsLiveTrain}{
	\begin{figure}
		\centering
		\includegraphics[width=\columnwidth]{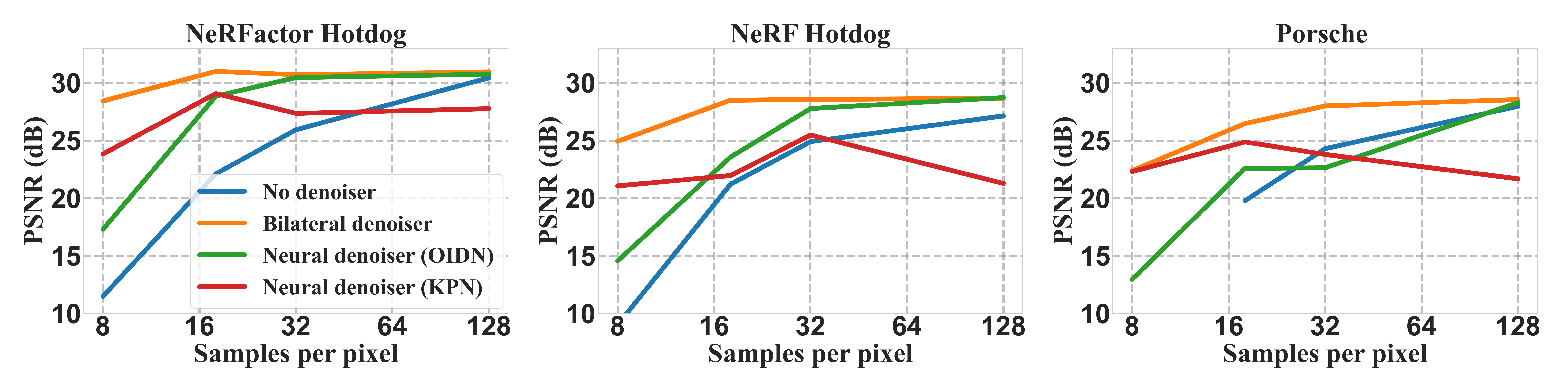}
		\caption{
			The effect of using different denoising algorithms during optimization at low sample counts on three different scenes of
			increasing complexity (from left to right). We plot averaged PSNR scores over 200 novel views,
			rendered without denoising, using high sample counts. 
			The most complex scene failed to converge at 8~spp without denoising.
			This experiment is identical to the denoising ablation in the paper,
			except including results from a jointly optimized single-frame version of the kernel prediction network (KPN) 
			architecture from Hasselgren et~al.~\cite{Hasselgren2020}.
			Inherent problems of live-training the denoiser, such as baking features into the denoiser weights 
			instead of the desired parameters, become apparent from deteriorating results, even at higher sample counts.}
		\label{fig:ablation-denoise-plots-live}
	\end{figure}
}


\newcommand{\figAblationDenoiseLight}{
	\begin{figure}
		\centering
		\setlength{\tabcolsep}{1pt}
		\begin{tabular}{ccccc}
			\toprule
			\includegraphics[width=0.19\columnwidth]{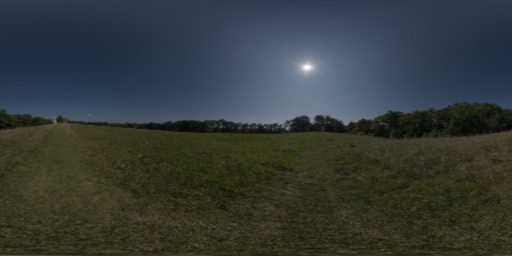} &
			\includegraphics[width=0.19\columnwidth]{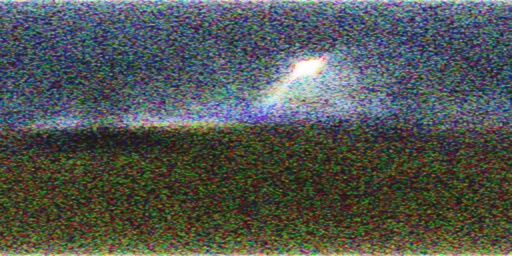} &
			\includegraphics[width=0.19\columnwidth]{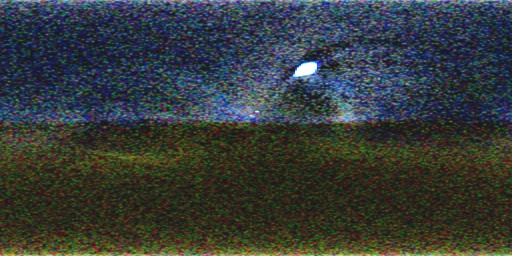} &
			\includegraphics[width=0.19\columnwidth]{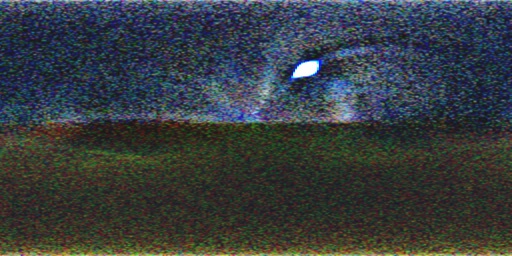} &
			\includegraphics[width=0.19\columnwidth]{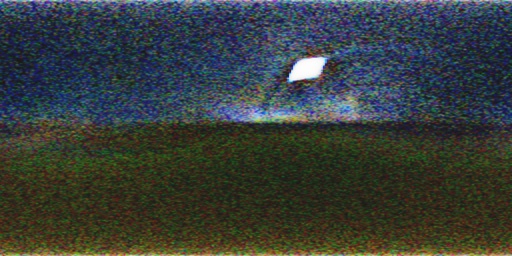} \\
			\small{Reference} & \small{No denoising} & \small{Bilateral} & \small{OIDN (pre-trained)} & \small{KPN (live-trained)} \\
			\midrule
			\textbf{8~spp}   & 26.24 & 36.15 & 35.94 & 35.21 \\
			18~spp  & 33.98 & 35.86 & 35.74 & 35.36 \\
			32~spp  & 35.87 & 35.72 & 35.65 & 35.19 \\
			128~spp & 36.08 & 35.43 & 35.51 & 35.21 \\
			\bottomrule
		\end{tabular}
		\caption{	
			The effect of denoising when optimizing HDR environment lighting, 
			using a diffuse version of the Hotdog scene. In this test, geometry and material are fixed to study the effect of
			optimizing the lighting in isolation. 
			\textbf{Top}: Resulting light probes at 8~spp for each method tabulated above.
			\textbf{Bottom}: Tabulated average PSNR over 200 novel views using high sample count without denoising.
		}
		\label{fig:ablation-denoise-light}
	\end{figure}
}


\newcommand{\figAblationDenoiseGeom}{
	\begin{figure}
		\centering
		\setlength{\tabcolsep}{1pt}
		\begin{tabular}{ccccc}
			\toprule
			\includegraphics[trim={0cm 2cm 0cm 3cm}, clip, width=0.19\columnwidth]{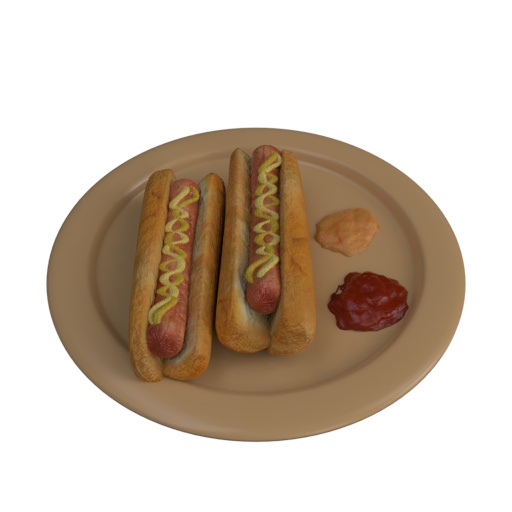} &
			\includegraphics[trim={0cm 2cm 0cm 3cm}, clip, width=0.19\columnwidth]{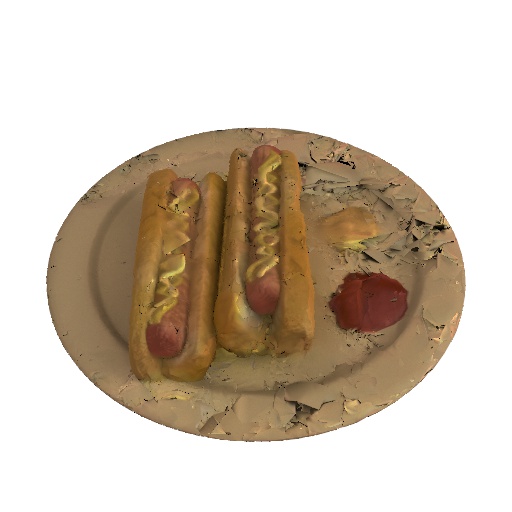} &
			\includegraphics[trim={0cm 2cm 0cm 3cm}, clip, width=0.19\columnwidth]{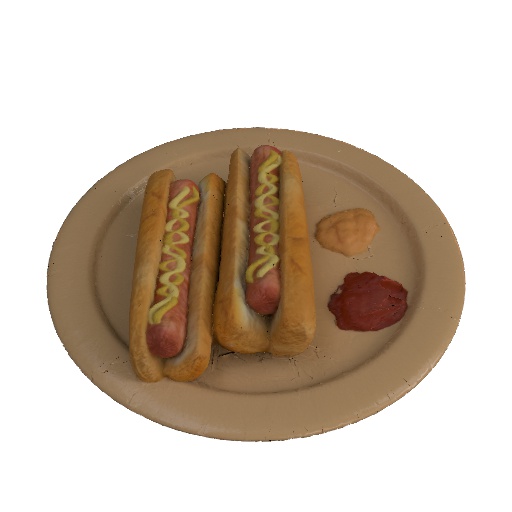} &
			\includegraphics[trim={0cm 2cm 0cm 3cm}, clip, width=0.19\columnwidth]{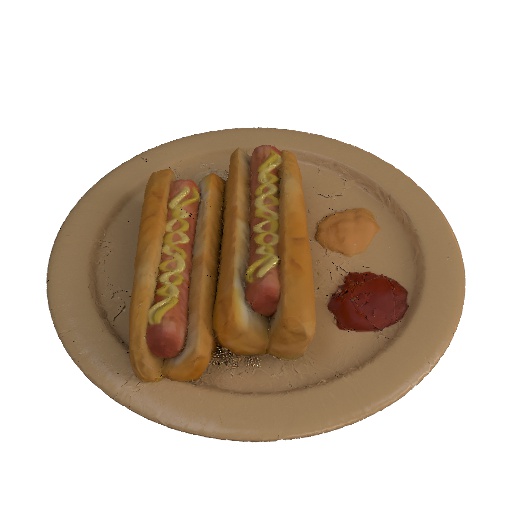} &
			\includegraphics[trim={0cm 2cm 0cm 3cm}, clip, width=0.19\columnwidth]{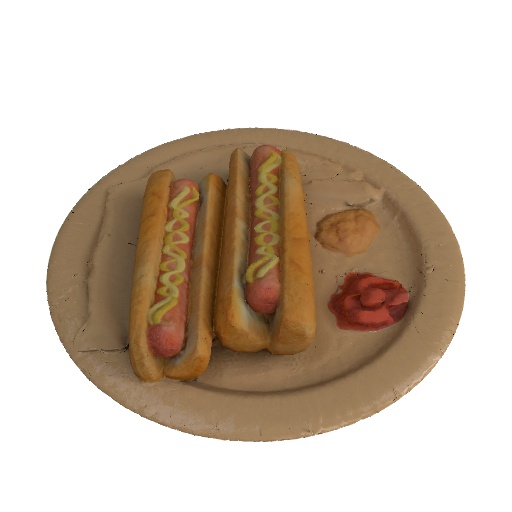} \\
			\small{Reference} & \small{No denoising} & \small{Bilateral} & \small{OIDN (pre-trained)} & \small{KPN (live-trained)} \\
			\midrule
			8~spp & 11.50 & 28.43 & 17.31 & 23.84 \\ 
			18~spp & 22.12 & 30.99 & 28.85 & 29.07 \\ 
			\textbf{32~spp} & 25.93 & 30.71 & 30.46 & 27.36 \\
			128~spp & 30.44 & 30.95 & 30.76 & 27.76 \\
			\bottomrule
		\end{tabular}
		\caption{
			Impact of denoising when jointly optimizing geometry, materials and environment lighting, 
			using the NeRFactor Hotdog scene.
			\textbf{Top}: Reconstruction results at 32~spp for each of the methods tabulated above.
			\textbf{Bottom}: Tabulated average PSNR (dB) over 200 novel views using high sample count without denoising.
		}
		\label{fig:ablation-denoise-geom}
	\end{figure}
}


\newcommand{\figShadowGradients}
{
	\begin{figure}[tb]
		\centering
		\small
		\setlength{\myfigsize}{0.24\textwidth}	
		\setlength{\tabcolsep}{0.0mm}
		\begin{tabular}{ccccc}		
			\rotatebox[origin=c]{90}{Ref} &
			\raisebox{-0.5\height}{\includegraphics[trim={0cm 0cm 0cm 1cm}, clip, width=\myfigsize]{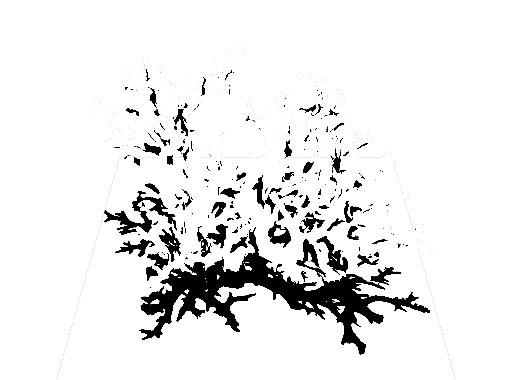}} &
			\raisebox{-0.5\height}{\includegraphics[trim={0cm 0cm 0cm 1cm}, clip, width=\myfigsize]{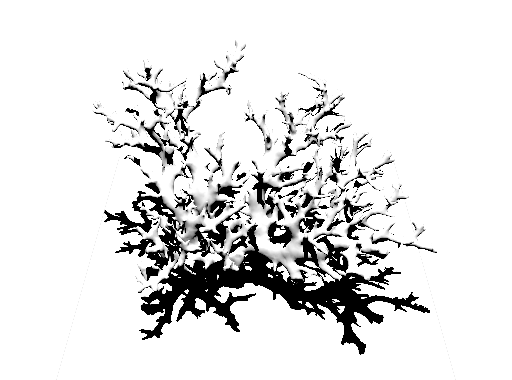}} &
			\raisebox{-0.5\height}{\includegraphics[trim={0cm 0cm 0cm 1cm}, clip, width=\myfigsize]{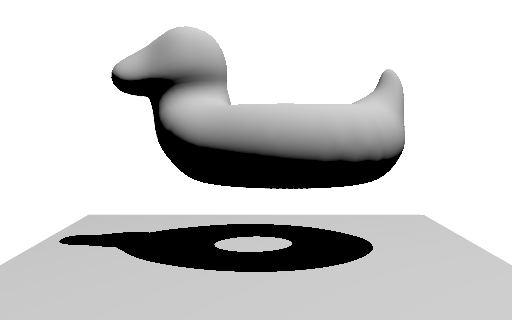}} &
			\raisebox{-0.5\height}{\includegraphics[trim={0cm 0cm 0cm 1cm}, clip, width=\myfigsize]{figures/supplemental/shadow_grad/single_ref.png}} \\
			
			\rotatebox[origin=c]{90}{w/o gradients} &
			\raisebox{-0.5\height}{\includegraphics[trim={0cm 0cm 0cm 1cm}, clip, width=\myfigsize]{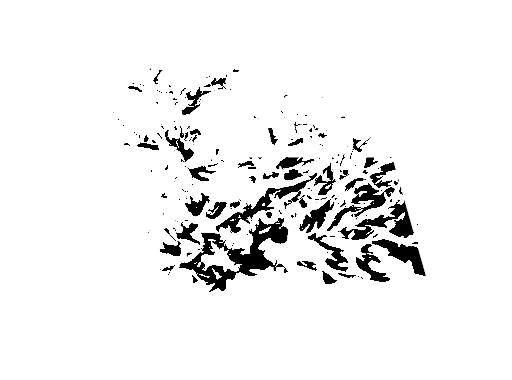}} &
			\raisebox{-0.5\height}{\includegraphics[trim={0cm 0cm 0cm 1cm}, clip, width=\myfigsize]{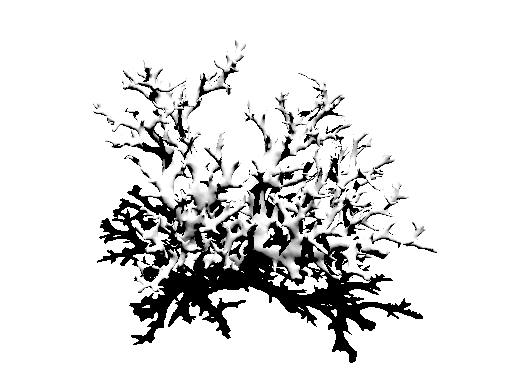}} &
			\raisebox{-0.5\height}{\includegraphics[trim={0cm 0cm 0cm 1cm}, clip, width=\myfigsize]{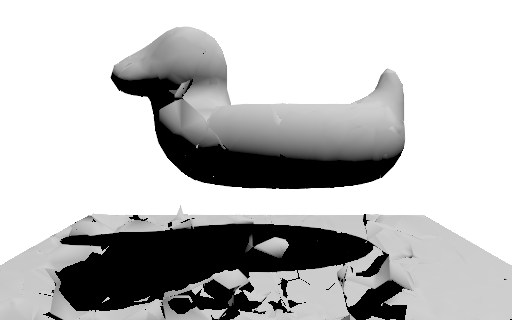}} &
			\raisebox{-0.5\height}{\includegraphics[trim={0cm 0cm 0cm 1cm}, clip, width=\myfigsize]{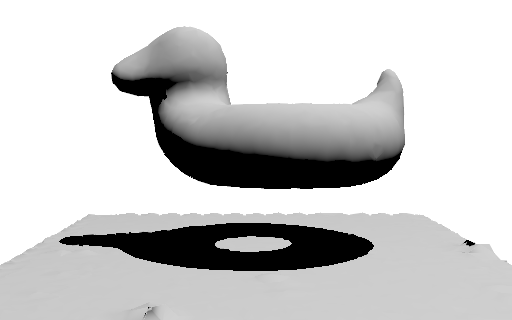}} \\
			
			\rotatebox[origin=c]{90}{w/ gradients} &
			\raisebox{-0.5\height}{\includegraphics[trim={0cm 0cm 0cm 1cm}, clip, width=\myfigsize]{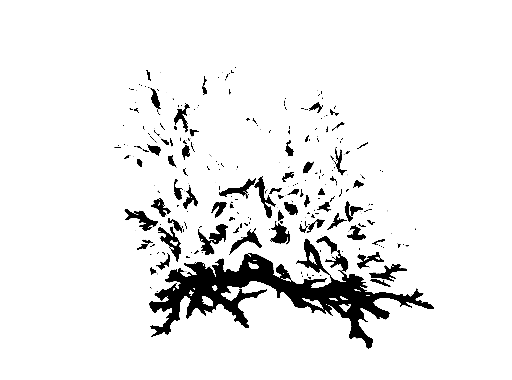}} &
			\raisebox{-0.5\height}{\includegraphics[trim={0cm 0cm 0cm 1cm}, clip, width=\myfigsize]{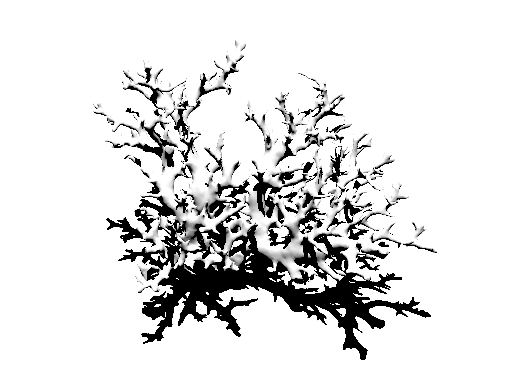}} &
			\raisebox{-0.5\height}{\includegraphics[trim={0cm 0cm 0cm 1cm}, clip, width=\myfigsize]{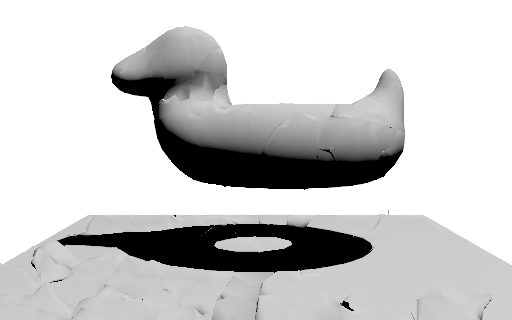}} &
			\raisebox{-0.5\height}{\includegraphics[trim={0cm 0cm 0cm 1cm}, clip, width=\myfigsize]{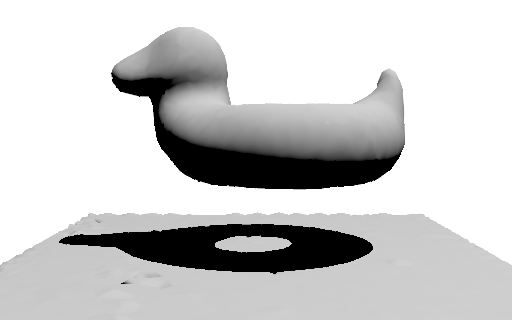}} \\
			
			& \small{White} & \small{Lambert} & \small{Single-view} & \small{Multi-view}  \\
			& \multicolumn{2}{c@{\hskip 2.0mm}}{\textbf{Light position optimization}} & \multicolumn{2}{c}{\textbf{Shape optimization}} \\
		\end{tabular}
		\caption{
			We show a few examples illustrating the benefits of shadow visibility gradients for inverse rendering setups. 
			\emph{White}: we optimize the position of a point light source, using a pure white material. 
			Here, shadow gradients are necessary for the optimization to succeed. 
			\emph{Lambert}: we redo the same experiment, but with a Lambertian material including an $\mathbf{n} \cdot \mathbf{l}$ 
			shading term. Now, both the shadow and shading terms provide gradients to the light position, and the optimization succeeds even 
			without shadow gradients. \emph{Single-view}: we optimize shape from a single view (purposely chosen to hide 
			the hole in the model), using a known point light source. Here, the shadow gradients provide useful information, and improves the 
			reconstruction to capture the hole in the model.
			\emph{Multi-view}: the same setup, but optimized using multiple views. Here, there is no clear benefit of using shadow
			gradients (the primary visibility gradients from each view dominate the geometry reconstruction).
		}
		\label{fig:shadow_gradients}
	\end{figure}
}


\newcommand{\figShadowGradientsPorsche}
{
	\begin{figure}[tb]
		\centering
		\small
		\setlength{\myfigsize}{0.49\textwidth}	
		\setlength{\tabcolsep}{1.0mm}
		\begin{tabular}{cc}		
			\includegraphics[width=\myfigsize]{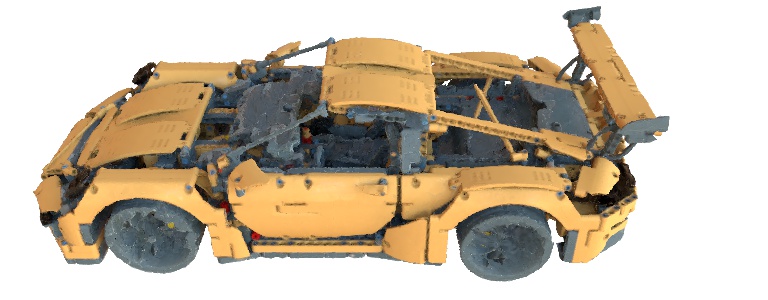} &
			\includegraphics[width=\myfigsize]{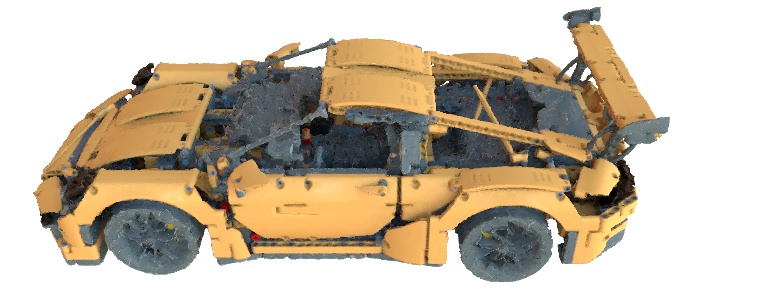} \\
			\includegraphics[width=\myfigsize]{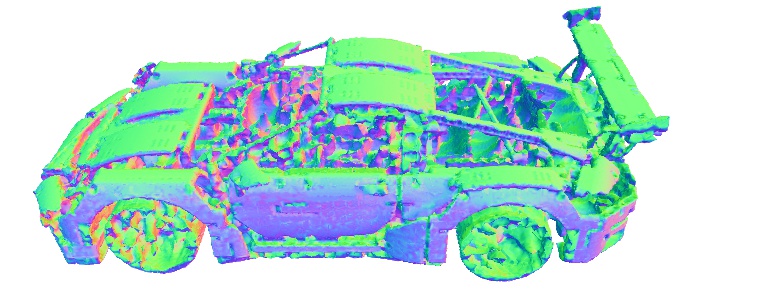} &
			\includegraphics[width=\myfigsize]{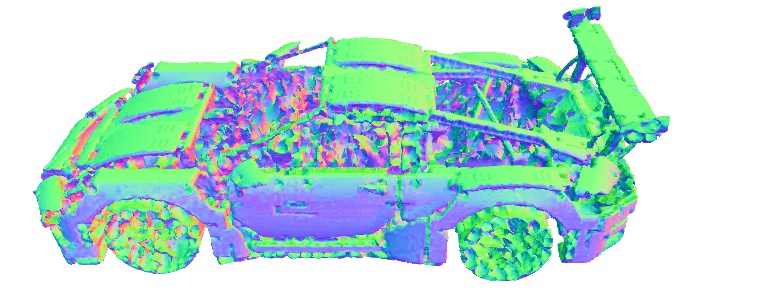} \\
			PSNR | SSIM: 28.21 | 0.945 & PSNR | SSIM: 27.82 | 0.941 \\
			\textbf{w/o shadow gradients} & \textbf{w/ shadow gradients} \\
		\end{tabular}
		\caption{Comparing the influence of shadow visibility gradients in our full pipeline with joint optimization of shape, 
			materials, and lighting. We observe slightly higher noise levels on the geometry reconstructions with shadow visibility
			gradients enabled, as can be seen in the visualization of surface normals. 
			The error metrics are average view interpolation scores over 200 novel views.}
		\label{fig:shadow_gradients_porsche}
	\end{figure}
}


\newcommand{\figAblationMIS}{
	\begin{figure}
		\centering
		\begin{tabular}{lccc}
			\toprule
			& \includegraphics[width=0.2\columnwidth]{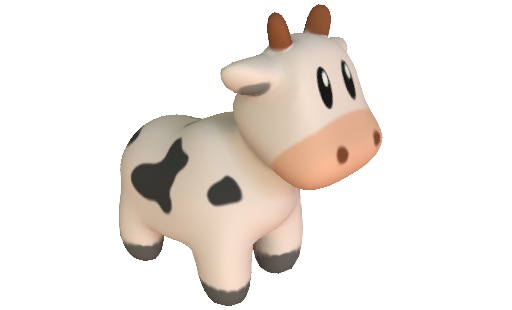}
			& \includegraphics[width=0.2\columnwidth]{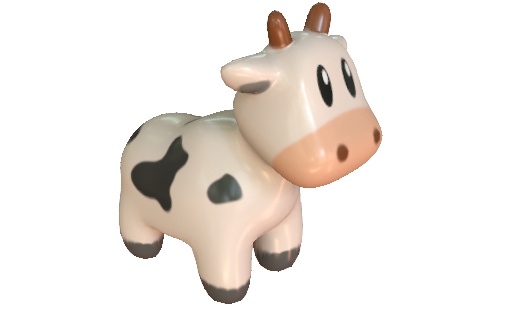}
			& \includegraphics[width=0.2\columnwidth]{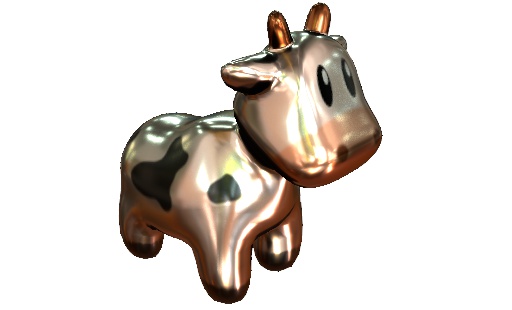} \\
			& \small{Diffuse} & \small{Plastic} & \small{Metallic} \\ \hline
			\small{Cosine sampling} & 46.81 & 34.88 & 25.61 \\
			\small{BSDF sampling}   & 46.81 & 34.78 & 31.66 \\
			\small{Light sampling}  & 42.70 & 34.23 & 24.38 \\
			\small{MIS}	            & 46.50 & 35.40 & 35.82 \\
			\bottomrule		
		\end{tabular}
		\caption{Comparison of different imaportance sampling strategies, using a diffuse, plastic and metal version of the Spot model. 
			Lighting and materials are optimized, and geometry is kept fixed. The hemisphere is sampled using 32~spp, and denoising is disabled for this test.
			Tabulated average validation PSNR of the forward renderings using high sample count and no denoising over 200 frames.}
		\label{tab:ablation-MIS}
	\end{figure}
}


\newcommand{\figAblationCorrelationPlot}{
	\begin{figure}
		\centering
		\includegraphics[width=.8\columnwidth]{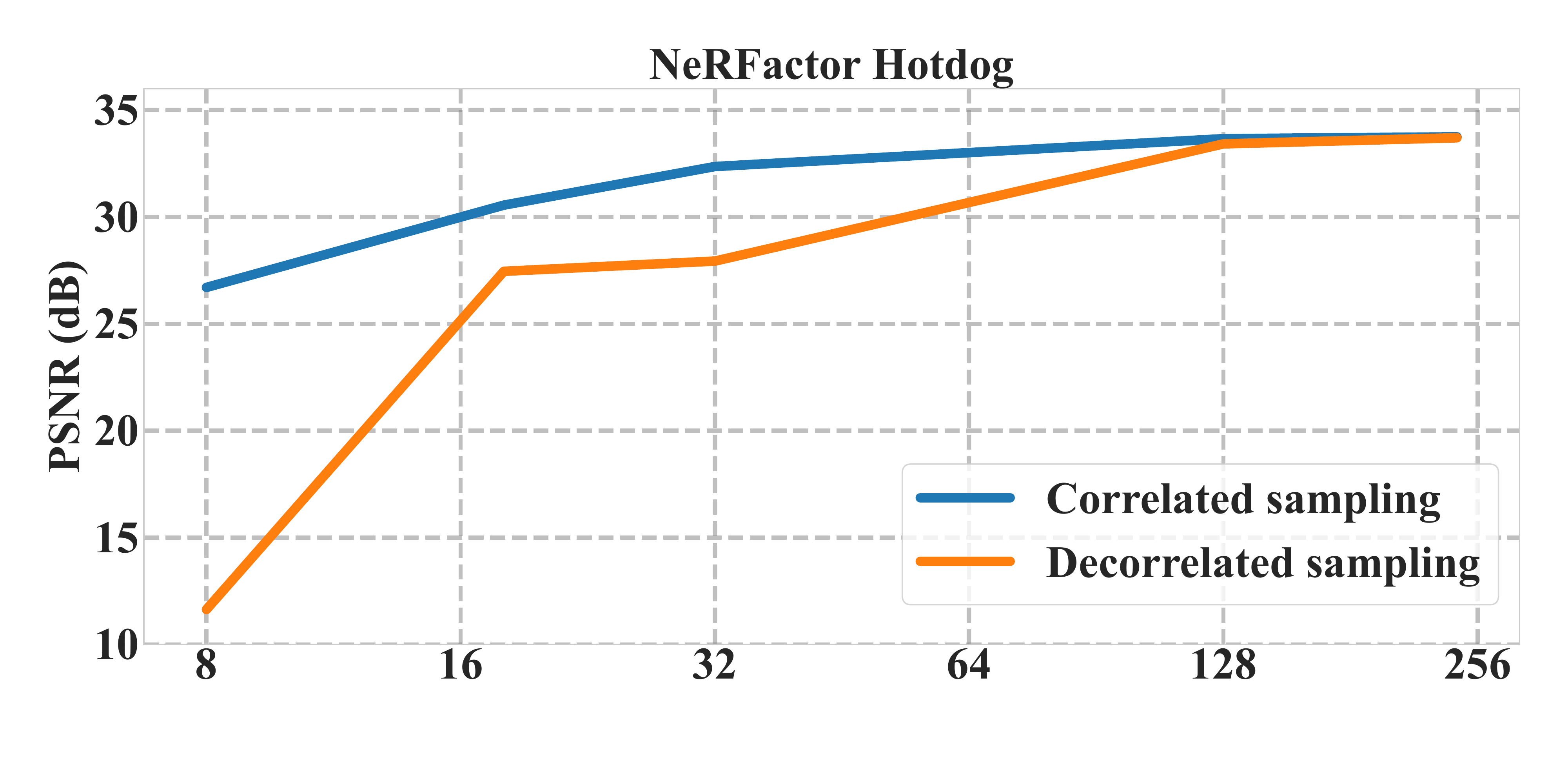}
		\caption{
			Correlated vs. decorrelated samples during gradient backpropagation.
			Although clearly \emph{biased}, convergence at lower sample counts improves drastically when explicitly re-using samples in the backward pass (correlated), 
			as gradients are propagated to the exact same set of parameters which contributed to the forward rendering. 
	}
		\label{fig:ablation-correlation-plot}
	\end{figure}
}


\newcommand{\figNerfactorRelight}[6]{
	\begin{figure}[tb]
		\centering
		\small
		\setlength{\myfigsize}{0.17\textwidth}	
		\setlength{\tabcolsep}{0.0mm}
		\begin{tabular}{ccccccc}		
			\rotatebox[origin=c]{90}{Ref} &
			\raisebox{-0.5\height}{\includegraphics[trim={0cm #3 0cm #4}, clip, width=\myfigsize]{#1/city_#2_ref.jpg}} &
			\raisebox{-0.5\height}{\includegraphics[trim={0cm #3 0cm #4}, clip, width=\myfigsize]{#1/courtyard_#2_ref.jpg}} &
			\raisebox{-0.5\height}{\includegraphics[trim={0cm #3 0cm #4}, clip, width=\myfigsize]{#1/forest_#2_ref.jpg}} &
			\raisebox{-0.5\height}{\includegraphics[trim={0cm #3 0cm #4}, clip, width=\myfigsize]{#1/studio_#2_ref.jpg}} &
			\raisebox{-0.5\height}{\includegraphics[trim={0cm #3 0cm #4}, clip, width=\myfigsize]{#1/sunrise_#2_ref.jpg}} &
			\raisebox{-0.5\height}{\includegraphics[trim={0cm #3 0cm #4}, clip, width=\myfigsize]{#1/sunset_#2_ref.jpg}} \\

			\rotatebox[origin=c]{90}{Our} &
			\raisebox{-0.5\height}{\includegraphics[trim={0cm #3 0cm #4}, clip, width=\myfigsize]{#1/city_#2_our.jpg}} &
			\raisebox{-0.5\height}{\includegraphics[trim={0cm #3 0cm #4}, clip, width=\myfigsize]{#1/courtyard_#2_our.jpg}} &
			\raisebox{-0.5\height}{\includegraphics[trim={0cm #3 0cm #4}, clip, width=\myfigsize]{#1/forest_#2_our.jpg}} &
			\raisebox{-0.5\height}{\includegraphics[trim={0cm #3 0cm #4}, clip, width=\myfigsize]{#1/studio_#2_our.jpg}} &
			\raisebox{-0.5\height}{\includegraphics[trim={0cm #3 0cm #4}, clip, width=\myfigsize]{#1/sunrise_#2_our.jpg}} &
			\raisebox{-0.5\height}{\includegraphics[trim={0cm #3 0cm #4}, clip, width=\myfigsize]{#1/sunset_#2_our.jpg}} \\

			\rotatebox[origin=c]{90}{\textsc{nvdiffrec}} &
			\raisebox{-0.5\height}{\includegraphics[trim={0cm #3 0cm #4}, clip, width=\myfigsize]{#1/city_#2_nvdiffrec.jpg}} &
			\raisebox{-0.5\height}{\includegraphics[trim={0cm #3 0cm #4}, clip, width=\myfigsize]{#1/courtyard_#2_nvdiffrec.jpg}} &
			\raisebox{-0.5\height}{\includegraphics[trim={0cm #3 0cm #4}, clip, width=\myfigsize]{#1/forest_#2_nvdiffrec.jpg}} &
			\raisebox{-0.5\height}{\includegraphics[trim={0cm #3 0cm #4}, clip, width=\myfigsize]{#1/studio_#2_nvdiffrec.jpg}} &
			\raisebox{-0.5\height}{\includegraphics[trim={0cm #3 0cm #4}, clip, width=\myfigsize]{#1/sunrise_#2_nvdiffrec.jpg}} &
			\raisebox{-0.5\height}{\includegraphics[trim={0cm #3 0cm #4}, clip, width=\myfigsize]{#1/sunset_#2_nvdiffrec.jpg}} \\		

			\rotatebox[origin=c]{90}{\textsc{NeRFactor}} &
			\raisebox{-0.5\height}{\includegraphics[trim={0cm #3 0cm #4}, clip, width=\myfigsize]{#1/city_#2_nerfactor.jpg}} &
			\raisebox{-0.5\height}{\includegraphics[trim={0cm #3 0cm #4}, clip, width=\myfigsize]{#1/courtyard_#2_nerfactor.jpg}} &
			\raisebox{-0.5\height}{\includegraphics[trim={0cm #3 0cm #4}, clip, width=\myfigsize]{#1/forest_#2_nerfactor.jpg}} &
			\raisebox{-0.5\height}{\includegraphics[trim={0cm #3 0cm #4}, clip, width=\myfigsize]{#1/studio_#2_nerfactor.jpg}} &
			\raisebox{-0.5\height}{\includegraphics[trim={0cm #3 0cm #4}, clip, width=\myfigsize]{#1/sunrise_#2_nerfactor.jpg}} &
			\raisebox{-0.5\height}{\includegraphics[trim={0cm #3 0cm #4}, clip, width=\myfigsize]{#1/sunset_#2_nerfactor.jpg}} \\		
		
			& City & Courtyard & Forest & Studio & Sunrise & Sunset \\
		\end{tabular}
		\caption{Relighting examples from \texttt{#5} scene of the NeRFactor synthetic dataset. We show a single view relit by 
			six of the eight different low frequency light probes of the test set.}
		\label{#6}
	\end{figure}
}

\newcommand{\figNerfactorRelightProbe}[6]{
	\begin{figure}[tb]
		\centering
		\small
		\setlength{\myfigsize}{0.17\textwidth}	
		\setlength{\tabcolsep}{0.0mm}
		\begin{tabular}{ccccccc}		
			\rotatebox[origin=c]{90}{Ref} &
			\raisebox{-0.5\height}{\includegraphics[trim={0cm #3 0cm #4}, clip, width=\myfigsize]{#1/city_#2_ref.jpg}} &
			\raisebox{-0.5\height}{\includegraphics[trim={0cm #3 0cm #4}, clip, width=\myfigsize]{#1/courtyard_#2_ref.jpg}} &
			\raisebox{-0.5\height}{\includegraphics[trim={0cm #3 0cm #4}, clip, width=\myfigsize]{#1/forest_#2_ref.jpg}} &
			\raisebox{-0.5\height}{\includegraphics[trim={0cm #3 0cm #4}, clip, width=\myfigsize]{#1/studio_#2_ref.jpg}} &
			\raisebox{-0.5\height}{\includegraphics[trim={0cm #3 0cm #4}, clip, width=\myfigsize]{#1/sunrise_#2_ref.jpg}} &
			\raisebox{-0.5\height}{\includegraphics[trim={0cm #3 0cm #4}, clip, width=\myfigsize]{#1/sunset_#2_ref.jpg}} \\

			\rotatebox[origin=c]{90}{Our} &
			\raisebox{-0.5\height}{\includegraphics[trim={0cm #3 0cm #4}, clip, width=\myfigsize]{#1/city_#2_our.jpg}} &
			\raisebox{-0.5\height}{\includegraphics[trim={0cm #3 0cm #4}, clip, width=\myfigsize]{#1/courtyard_#2_our.jpg}} &
			\raisebox{-0.5\height}{\includegraphics[trim={0cm #3 0cm #4}, clip, width=\myfigsize]{#1/forest_#2_our.jpg}} &
			\raisebox{-0.5\height}{\includegraphics[trim={0cm #3 0cm #4}, clip, width=\myfigsize]{#1/studio_#2_our.jpg}} &
			\raisebox{-0.5\height}{\includegraphics[trim={0cm #3 0cm #4}, clip, width=\myfigsize]{#1/sunrise_#2_our.jpg}} &
			\raisebox{-0.5\height}{\includegraphics[trim={0cm #3 0cm #4}, clip, width=\myfigsize]{#1/sunset_#2_our.jpg}} \\

			\rotatebox[origin=c]{90}{\textsc{nvdiffrec}} &
			\raisebox{-0.5\height}{\includegraphics[trim={0cm #3 0cm #4}, clip, width=\myfigsize]{#1/city_#2_nvdiffrec.jpg}} &
			\raisebox{-0.5\height}{\includegraphics[trim={0cm #3 0cm #4}, clip, width=\myfigsize]{#1/courtyard_#2_nvdiffrec.jpg}} &
			\raisebox{-0.5\height}{\includegraphics[trim={0cm #3 0cm #4}, clip, width=\myfigsize]{#1/forest_#2_nvdiffrec.jpg}} &
			\raisebox{-0.5\height}{\includegraphics[trim={0cm #3 0cm #4}, clip, width=\myfigsize]{#1/studio_#2_nvdiffrec.jpg}} &
			\raisebox{-0.5\height}{\includegraphics[trim={0cm #3 0cm #4}, clip, width=\myfigsize]{#1/sunrise_#2_nvdiffrec.jpg}} &
			\raisebox{-0.5\height}{\includegraphics[trim={0cm #3 0cm #4}, clip, width=\myfigsize]{#1/sunset_#2_nvdiffrec.jpg}} \\		

			\rotatebox[origin=c]{90}{\textsc{NeRFactor}} &
			\raisebox{-0.5\height}{\includegraphics[trim={0cm #3 0cm #4}, clip, width=\myfigsize]{#1/city_#2_nerfactor.jpg}} &
			\raisebox{-0.5\height}{\includegraphics[trim={0cm #3 0cm #4}, clip, width=\myfigsize]{#1/courtyard_#2_nerfactor.jpg}} &
			\raisebox{-0.5\height}{\includegraphics[trim={0cm #3 0cm #4}, clip, width=\myfigsize]{#1/forest_#2_nerfactor.jpg}} &
			\raisebox{-0.5\height}{\includegraphics[trim={0cm #3 0cm #4}, clip, width=\myfigsize]{#1/studio_#2_nerfactor.jpg}} &
			\raisebox{-0.5\height}{\includegraphics[trim={0cm #3 0cm #4}, clip, width=\myfigsize]{#1/sunrise_#2_nerfactor.jpg}} &
			\raisebox{-0.5\height}{\includegraphics[trim={0cm #3 0cm #4}, clip, width=\myfigsize]{#1/sunset_#2_nerfactor.jpg}} \\		
		
			\rotatebox[origin=c]{90}{\textsc{Probes}} &
			\raisebox{-0.5\height}{\includegraphics[width=0.8\myfigsize]{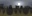}} &
			\raisebox{-0.5\height}{\includegraphics[width=0.8\myfigsize]{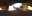}} &
			\raisebox{-0.5\height}{\includegraphics[width=0.8\myfigsize]{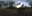}} &
			\raisebox{-0.5\height}{\includegraphics[width=0.8\myfigsize]{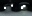}} &
			\raisebox{-0.5\height}{\includegraphics[width=0.8\myfigsize]{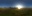}} &
			\raisebox{-0.5\height}{\includegraphics[width=0.8\myfigsize]{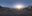}} \\
			& City & Courtyard & Forest & Studio & Sunrise & Sunset \\
		\end{tabular}
		\caption{Relighting examples from \texttt{#5} scene of the NeRFactor synthetic dataset. We show a single view relit by 
			six of the eight different low frequency light probes from the test set.}
		\label{#6}
	\end{figure}
}


\newcommand{\figNerfRelight}[6]{
	\begin{figure}[tb]
		\centering
		\small
		\setlength{\myfigsize}{0.24\textwidth}	
		\setlength{\tabcolsep}{0.0mm}
		\begin{tabular}{ccccc}		
			\rotatebox[origin=c]{90}{Ref} &
			\raisebox{-0.5\height}{\includegraphics[trim={0cm #3 0cm #4}, clip, width=\myfigsize]{#1/aerodynamics_workshop_2k_#2_ref.jpg}} &
			\raisebox{-0.5\height}{\includegraphics[trim={0cm #3 0cm #4}, clip, width=\myfigsize]{#1/boiler_room_2k_#2_ref.jpg}} &
			\raisebox{-0.5\height}{\includegraphics[trim={0cm #3 0cm #4}, clip, width=\myfigsize]{#1/dreifaltigkeitsberg_2k_#2_ref.jpg}} &
			\raisebox{-0.5\height}{\includegraphics[trim={0cm #3 0cm #4}, clip, width=\myfigsize]{#1/music_hall_01_2k_#2_ref.jpg}} \\
			
			\rotatebox[origin=c]{90}{Our} &
			\raisebox{-0.5\height}{\includegraphics[trim={0cm #3 0cm #4}, clip, width=\myfigsize]{#1/aerodynamics_workshop_2k_#2_our.jpg}} &
			\raisebox{-0.5\height}{\includegraphics[trim={0cm #3 0cm #4}, clip, width=\myfigsize]{#1/boiler_room_2k_#2_our.jpg}} &
			\raisebox{-0.5\height}{\includegraphics[trim={0cm #3 0cm #4}, clip, width=\myfigsize]{#1/dreifaltigkeitsberg_2k_#2_our.jpg}} &
			\raisebox{-0.5\height}{\includegraphics[trim={0cm #3 0cm #4}, clip, width=\myfigsize]{#1/music_hall_01_2k_#2_our.jpg}} \\
			
			\rotatebox[origin=c]{90}{\textsc{nvdiffrec}} &
			\raisebox{-0.5\height}{\includegraphics[trim={0cm #3 0cm #4}, clip, width=\myfigsize]{#1/aerodynamics_workshop_2k_#2_nvdiffrec.jpg}} &
			\raisebox{-0.5\height}{\includegraphics[trim={0cm #3 0cm #4}, clip, width=\myfigsize]{#1/boiler_room_2k_#2_nvdiffrec.jpg}} &
			\raisebox{-0.5\height}{\includegraphics[trim={0cm #3 0cm #4}, clip, width=\myfigsize]{#1/dreifaltigkeitsberg_2k_#2_nvdiffrec.jpg}} &
			\raisebox{-0.5\height}{\includegraphics[trim={0cm #3 0cm #4}, clip, width=\myfigsize]{#1/music_hall_01_2k_#2_nvdiffrec.jpg}} \\		
			
			& Aerodynamics Workshop & Boiler room & Dreifaltigkeitsberg & Music hall \\
		\end{tabular}
		\caption{Relighting examples from \texttt{#5} scene of the NeRF synthetic dataset. We show a single view relit by the four 
			different high frequency light probes.}
		\label{#6}
	\end{figure}
}

\newcommand{\figNerfRelightProbe}[6]{
	\begin{figure}[tb]
		\centering
		\small
		\setlength{\myfigsize}{0.24\textwidth}	
		\setlength{\tabcolsep}{0.0mm}
		\begin{tabular}{ccccc}		
			\rotatebox[origin=c]{90}{Ref} &
			\raisebox{-0.5\height}{\includegraphics[trim={0cm #3 0cm #4}, clip, width=\myfigsize]{#1/aerodynamics_workshop_2k_#2_ref.jpg}} &
			\raisebox{-0.5\height}{\includegraphics[trim={0cm #3 0cm #4}, clip, width=\myfigsize]{#1/boiler_room_2k_#2_ref.jpg}} &
			\raisebox{-0.5\height}{\includegraphics[trim={0cm #3 0cm #4}, clip, width=\myfigsize]{#1/dreifaltigkeitsberg_2k_#2_ref.jpg}} &
			\raisebox{-0.5\height}{\includegraphics[trim={0cm #3 0cm #4}, clip, width=\myfigsize]{#1/music_hall_01_2k_#2_ref.jpg}} \\
			
			\rotatebox[origin=c]{90}{Our} &
			\raisebox{-0.5\height}{\includegraphics[trim={0cm #3 0cm #4}, clip, width=\myfigsize]{#1/aerodynamics_workshop_2k_#2_our.jpg}} &
			\raisebox{-0.5\height}{\includegraphics[trim={0cm #3 0cm #4}, clip, width=\myfigsize]{#1/boiler_room_2k_#2_our.jpg}} &
			\raisebox{-0.5\height}{\includegraphics[trim={0cm #3 0cm #4}, clip, width=\myfigsize]{#1/dreifaltigkeitsberg_2k_#2_our.jpg}} &
			\raisebox{-0.5\height}{\includegraphics[trim={0cm #3 0cm #4}, clip, width=\myfigsize]{#1/music_hall_01_2k_#2_our.jpg}} \\
			
			\rotatebox[origin=c]{90}{\textsc{nvdiffrec}} &
			\raisebox{-0.5\height}{\includegraphics[trim={0cm #3 0cm #4}, clip, width=\myfigsize]{#1/aerodynamics_workshop_2k_#2_nvdiffrec.jpg}} &
			\raisebox{-0.5\height}{\includegraphics[trim={0cm #3 0cm #4}, clip, width=\myfigsize]{#1/boiler_room_2k_#2_nvdiffrec.jpg}} &
			\raisebox{-0.5\height}{\includegraphics[trim={0cm #3 0cm #4}, clip, width=\myfigsize]{#1/dreifaltigkeitsberg_2k_#2_nvdiffrec.jpg}} &
			\raisebox{-0.5\height}{\includegraphics[trim={0cm #3 0cm #4}, clip, width=\myfigsize]{#1/music_hall_01_2k_#2_nvdiffrec.jpg}} \\		
			\\
			\rotatebox[origin=c]{90}{\textsc{Probes}} &
			\raisebox{-0.5\height}{\includegraphics[width=0.8\myfigsize]{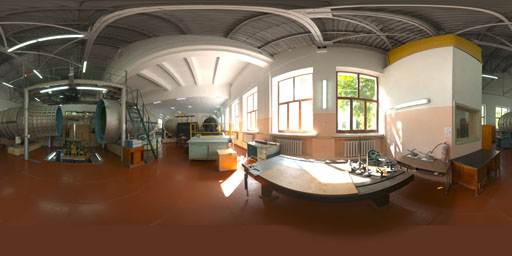}} &
			\raisebox{-0.5\height}{\includegraphics[width=0.8\myfigsize]{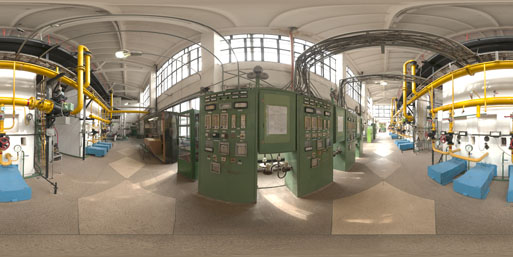}} &
			\raisebox{-0.5\height}{\includegraphics[width=0.8\myfigsize]{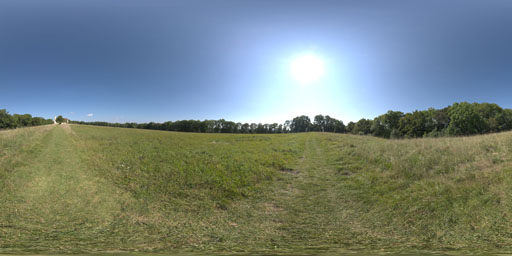}} &
			\raisebox{-0.5\height}{\includegraphics[width=0.8\myfigsize]{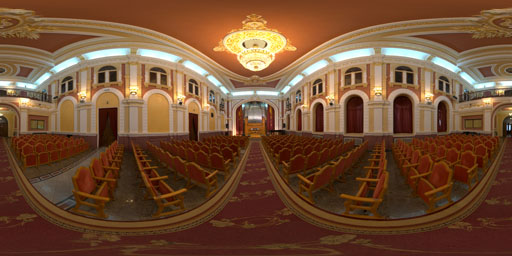}} \\
			& Aerodynamics Workshop & Boiler room & Dreifaltigkeitsberg & Music hall \\
		\end{tabular}
		\caption{Relighting examples from \texttt{#5} scene of the NeRF synthetic dataset. We show a single view relit by the four 
			different high frequency light probes from Poly Haven (CC0).}
		\label{#6}
	\end{figure}
}


\newcommand{\tabRelightingBreakdown}
{
	\begin{table}
		\caption{
			Relighting results for all synthetic datasets with per-scene quality metrics. The synthetic NeRFactor and NeRF train sets use 
			100 images from random viewpoints, and our dataset use 200 images for \texttt{Apollo} and 256 images for \texttt{Porsche} and \texttt{Roller}. 
			The test sets used to compute relighting quality use eight novel views. 
			Each view is relit by eight low resolution ($32\times16$ pixel) light probes for NeRFactor, and four high resolution 
			($2048 \times 1024$ pixels) light probes for NeRF and our dataset.
		}
		\centering
		{
			\footnotesize
			\setlength{\tabcolsep}{1pt}
			\begin{tabularx}{\textwidth}{lYYYY@{\hskip 4mm}YYYY@{\hskip 4mm}YYYY}
				\toprule
				& \multicolumn{12}{c}{\textbf{NeRFactor synthetic dataset}} \\
				& \multicolumn{4}{c}{PSNR$\uparrow$} & \multicolumn{4}{c}{SSIM$\uparrow$} & \multicolumn{4}{c}{LPIPS$\downarrow$} \\
				& 
				{Drums} & {Ficus} & {Hotdog} & {Lego} & 
				{Drums} & {Ficus} & {Hotdog} & {Lego} & 
				{Drums} & {Ficus} & {Hotdog} & {Lego} \\
				\midrule
				Our                 & 23.6 & 29.3 & 30.0 & 21.2 & 0.926 & 0.968 & 0.952 & 0.853 & 0.062 & 0.025 & 0.040 & 0.113 \\
				\textsc{nvdiffrec}  & 22.4 & 27.8 & 29.5 & 19.4 & 0.915 & 0.962 & 0.936 & 0.825 & 0.066 & 0.028 & 0.048 & 0.109 \\
				\textsc{NeRFactor}  & 21.6 & 21.6 & 25.5 & 20.3 & 0.911 & 0.921 & 0.916 & 0.839 & 0.065 & 0.075 & 0.087 & 0.121 \\
				\bottomrule
			\end{tabularx}
		}	
		{
			\footnotesize
			\setlength{\tabcolsep}{0.75pt}
			\begin{tabularx}{\textwidth}{lYYYYY@{\hskip 4mm}YYYYY@{\hskip 4mm}YYYYY}
				\toprule
				& \multicolumn{15}{c}{\textbf{NeRF synthetic dataset}} \\
				& \multicolumn{5}{c}{PSNR$\uparrow$} & \multicolumn{5}{c}{SSIM$\uparrow$} & \multicolumn{5}{c}{LPIPS$\downarrow$} \\
				& 
				{Chair} & {Hotd.} & {Lego} & {Mats.} & {Mic} &
				{Chair} & {Hotd.} & {Lego} & {Mats.} & {Mic} &
				{Chair} & {Hotd.} & {Lego} & {Mats.} & {Mic} \\
				\midrule
				Our                 & 27.0 & 25.9 & 23.4 & 25.1 & 30.9 & 0.946 & 0.929 & 0.890 & 0.923 & 0.972 & 0.044 & 0.073 & 0.088 & 0.047 & 0.025 \\
				\textsc{nvdiffrec}  & 25.1 & 20.3 & 21.4 & 19.9 & 29.7 & 0.921 & 0.853 & 0.861 & 0.847 & 0.967 & 0.060 & 0.126 & 0.089 & 0.079 & 0.026 \\
				\bottomrule
			\end{tabularx}
		}
		{
			\footnotesize
			\setlength{\tabcolsep}{1pt}
			\begin{tabularx}{\textwidth}{lYYY@{\hskip 4mm}YYY@{\hskip 4mm}YYY}
				\toprule
				& \multicolumn{9}{c}{\textbf{Our synthetic dataset}} \\
				& \multicolumn{3}{c}{PSNR$\uparrow$} & \multicolumn{3}{c}{SSIM$\uparrow$} & \multicolumn{3}{c}{LPIPS$\downarrow$}\\
				& 
				{Apollo} & {Porsche} & {Roller} &
				{Apollo} & {Porsche} & {Roller} &
				{Apollo} & {Porsche} & {Roller} \\
				\midrule
				Our                 & 25.6 & 28.2 & 27.5 & 0.931 & 0.965 & 0.952 & 0.045 & 0.030 & 0.027 \\
				\textsc{nvdiffrec}  & 24.7 & 26.1 & 20.4 & 0.926 & 0.952 & 0.898 & 0.049 & 0.029 & 0.067 \\
				\bottomrule
			\end{tabularx}
		}
		\label{tab:all_relighting}
	\end{table}
}


\newcommand{\tabViewBreakdown}
{
	\begin{table}
		\caption
		{
			View interpolation results for all synthetic datasets with per-scene quality metrics. The synthetic NeRFactor and NeRF 
			training sets use 100 images from random viewpoints, and our dataset use 200 images for \texttt{Apollo} and 256 images for \texttt{Porsche} and 
			\texttt{Roller}. The quality metric for a scene is the arithmetic average over all 200 test images.
		}
		\centering
		{
			\setlength{\tabcolsep}{1pt}
			\begin{tabularx}{\textwidth}{lYYYY@{\hskip 8mm}YYYY}
				\toprule
				& \multicolumn{8}{c}{\textbf{NeRFactor synthetic dataset}} \\			
				& \multicolumn{4}{c}{PSNR$\uparrow$} & \multicolumn{4}{c}{SSIM$\uparrow$} \\
				& 
				{Drums} & {Ficus} & {Hotdog} & {Lego} & 
				{Drums} & {Ficus} & {Hotdog} & {Lego} \\
				\midrule
				Our                 & 26.6 & 30.8 & 33.9 & 29.1 & 0.940 & 0.975 & 0.967 & 0.930 \\
				\textsc{nvdiffrec}  & 28.5 & 31.2 & 36.3 & 30.7 & 0.959 & 0.978 & 0.981 & 0.951 \\
				\textsc{NeRFactor}  & 24.6 & 23.1 & 31.6 & 28.1 & 0.933 & 0.937 & 0.948 & 0.900 \\
				\bottomrule
			\end{tabularx}
		}
		{
			\setlength{\tabcolsep}{1pt}
			\begin{tabularx}{\textwidth}{lYYYYY@{\hskip 8mm}YYYYY}
				\toprule
				& \multicolumn{10}{c}{\textbf{NeRF synthetic dataset}} \\			
				& \multicolumn{5}{c}{PSNR$\uparrow$} & \multicolumn{5}{c}{SSIM$\uparrow$} \\
				& 
				{Chair} & {Hotdog} & {Lego} & {Mats.} & {Mic} &
				{Chair} & {Hotdog} & {Lego} & {Mats.} & {Mic} \\
				\midrule
				Our                 & 29.0 & 30.4 & 26.8 & 25.7 & 29.0 & 0.947 & 0.942 & 0.917 & 0.913 & 0.963 \\
				\textsc{nvdiffrec}  & 31.6 & 33.0 & 29.1 & 26.7 & 30.8 & 0.969 & 0.973 & 0.949 & 0.923 & 0.977 \\
				\bottomrule
			\end{tabularx}
		}
		{
			\setlength{\tabcolsep}{1pt}
			\begin{tabularx}{\textwidth}{lYYY@{\hskip 8mm}YYY}
				\toprule
				& \multicolumn{6}{c}{\textbf{Our synthetic dataset}} \\
				& \multicolumn{3}{c}{PSNR$\uparrow$} & \multicolumn{3}{c}{SSIM$\uparrow$} \\
				& 
				{Apollo} & {Porsche} & {Roller} &
				{Apollo} & {Porsche} & {Roller} \\
				\midrule
				Our                 & 23.4 & 28.4 & 25.0 & 0.908 & 0.953 & 0.941 \\
				\textsc{nvdiffrec}  & 23.9 & 28.5 & 25.1 & 0.923 & 0.958 & 0.951 \\
				\bottomrule
			\end{tabularx}
		}
		\label{tab:all_view_interp}
	\end{table}
}


\newcommand{\figMeshesAO}{
	\begin{figure}
		\centering
		\setlength{\tabcolsep}{1pt}
		\begin{tabular}{ccccc}
			\rotatebox[origin=r]{90}{Hotdog} &
			\raisebox{-0.5\height}{\includegraphics[width=0.24\columnwidth]{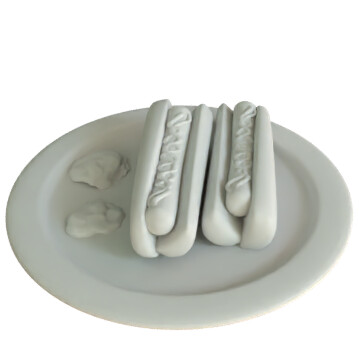}} &
			\raisebox{-0.5\height}{\includegraphics[width=0.24\columnwidth]{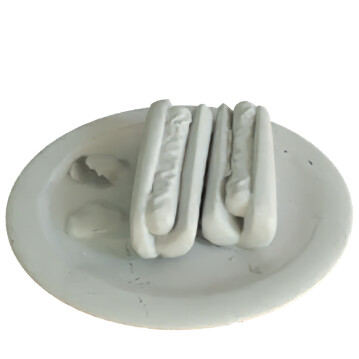}} &
			\raisebox{-0.5\height}{\includegraphics[width=0.24\columnwidth]{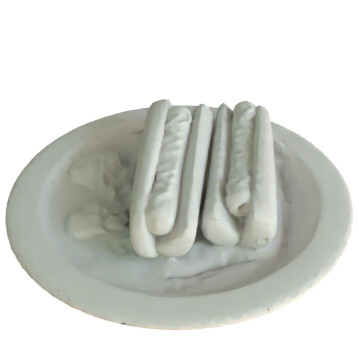}} &
			\raisebox{-0.5\height}{\includegraphics[width=0.24\columnwidth]{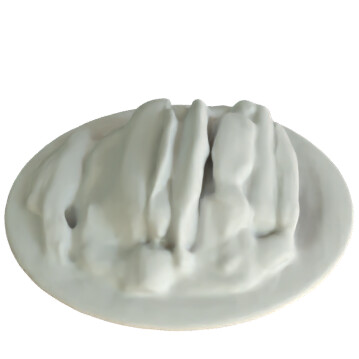}} \\
			& \small{66k tris} & \small{57k tris} & \small{57k tris} & \small{725k tris} \\
			
			\rotatebox[origin=c]{90}{Chair} &
			\raisebox{-0.5\height}{\includegraphics[width=0.24\columnwidth]{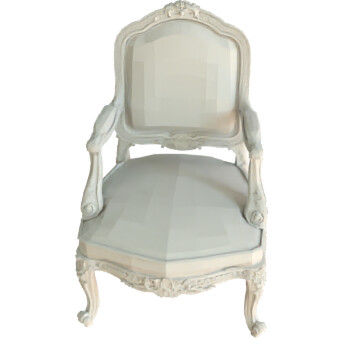}} &
			\raisebox{-0.5\height}{\includegraphics[width=0.24\columnwidth]{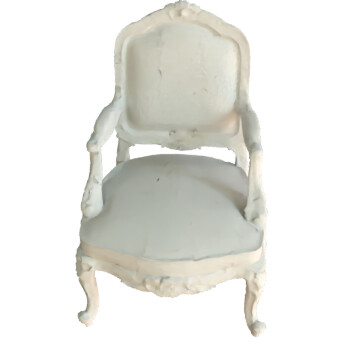}} &
			\raisebox{-0.5\height}{\includegraphics[width=0.24\columnwidth]{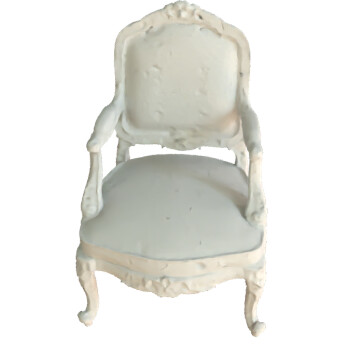}} &
			\raisebox{-0.5\height}{\includegraphics[width=0.24\columnwidth]{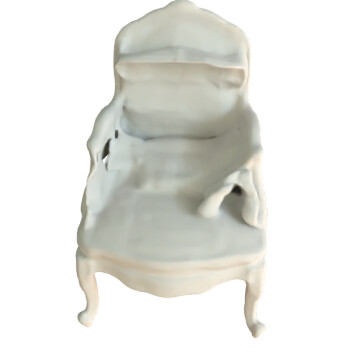}} \\
			& \small{259k tris} & \small{101k tris} & \small{102k tris} & \small{353k tris} \\
			
			\rotatebox[origin=c]{90}{Lego} &
			\raisebox{-0.5\height}{\includegraphics[width=0.24\columnwidth]{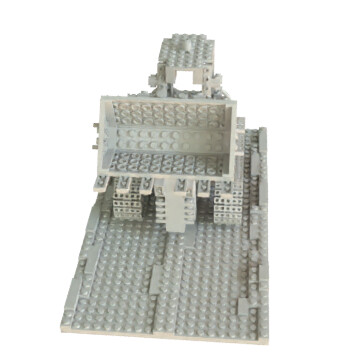}} &
			\raisebox{-0.5\height}{\includegraphics[width=0.24\columnwidth]{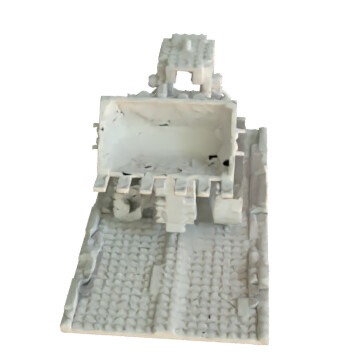}} &
			\raisebox{-0.5\height}{\includegraphics[width=0.24\columnwidth]{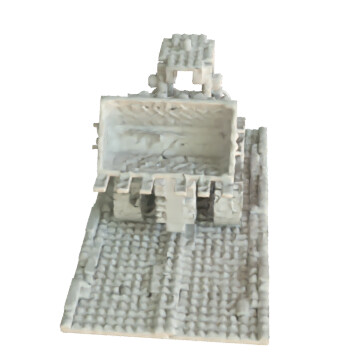}} &
			\raisebox{-0.5\height}{\includegraphics[width=0.24\columnwidth]{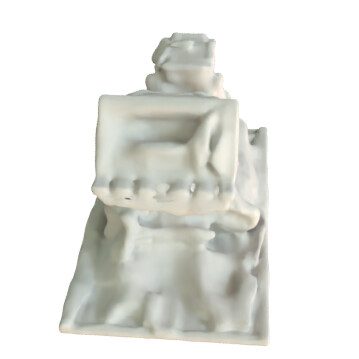}} \\
			& \small{2035k tris} & \small{104k tris} & \small{111k tris} & \small{498k tris} \\
			
			& Reference & Our & \textsc{nvdiffrec} & PhySG \\
		\end{tabular}
		\caption{
			Extracted mesh quality visualization examples using the NeRF synthetic \texttt{Chair}, \texttt{Hotdog}, 
			and \texttt{Lego} datasets. For these visualizations, the normal mapping shading term which simulates small geometric detail 
			(used by Our and \textsc{nvdiffrec}) is disabled.
		}
		\label{fig:ao-meshes}
	\end{figure}
}

\newcommand{\figNewreg}{
	\setlength{\myfigsize}{0.152\columnwidth}
	\begin{figure}
		\centering
		\setlength{\tabcolsep}{1pt}
		\begin{tabular}{ccccccc}
			Shaded & Diffuse light & Specular light & $\kd$ & $\korm$ & Probe \\
			
			\rotatebox[origin=c]{90}{w/ reg} 
			\raisebox{-0.5\height}{\includegraphics[width=\myfigsize]{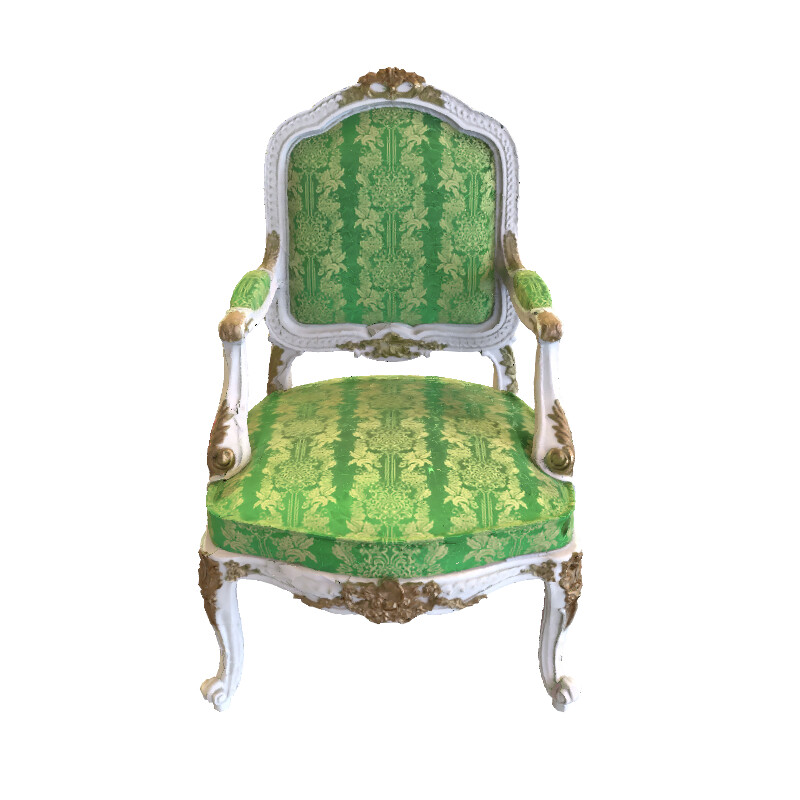}} &
			\raisebox{-0.5\height}{\includegraphics[width=\myfigsize]{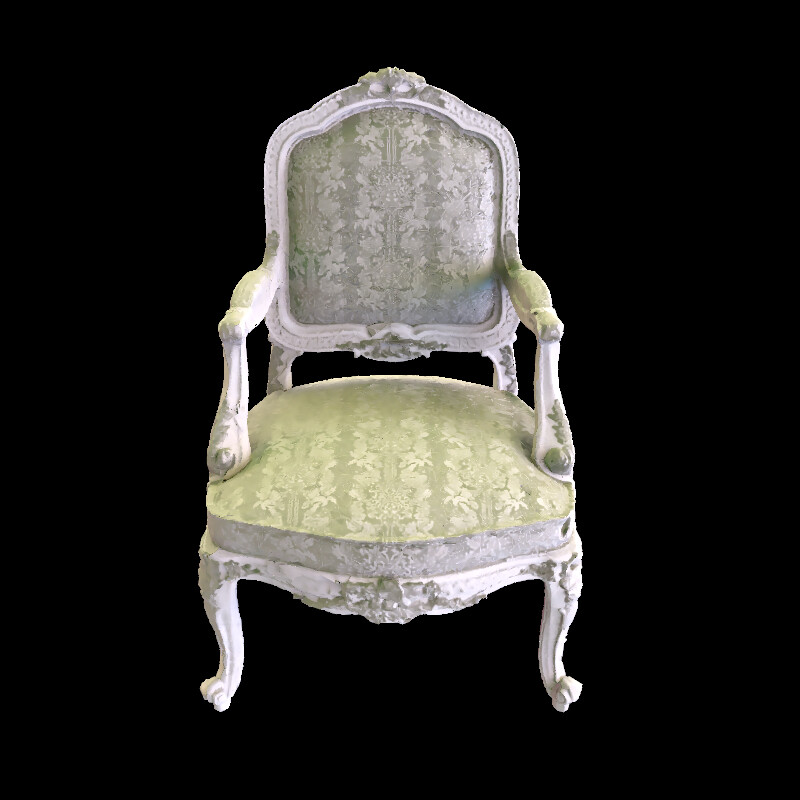}} &
			\raisebox{-0.5\height}{\includegraphics[width=\myfigsize]{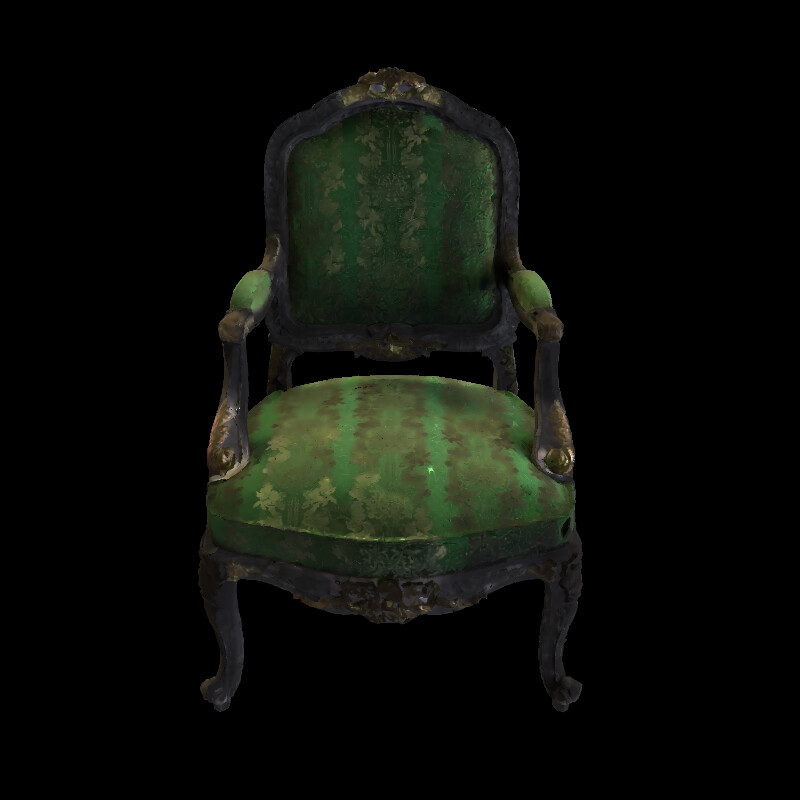}} &
			\raisebox{-0.5\height}{\includegraphics[width=\myfigsize]{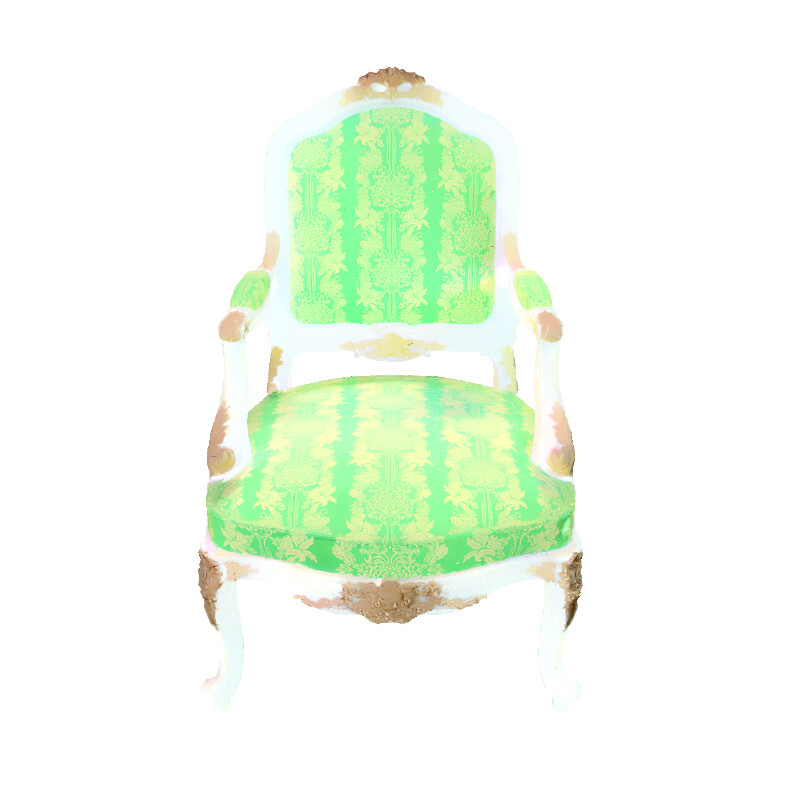}} &
			\raisebox{-0.5\height}{\includegraphics[width=\myfigsize]{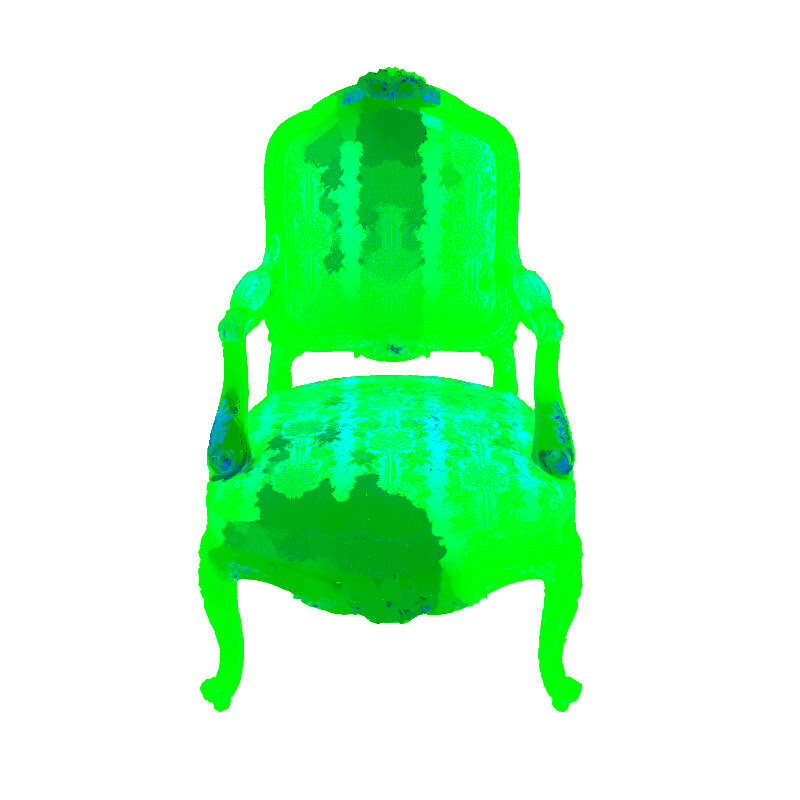}} &
			\raisebox{-0.5\height}{\includegraphics[width=\myfigsize]{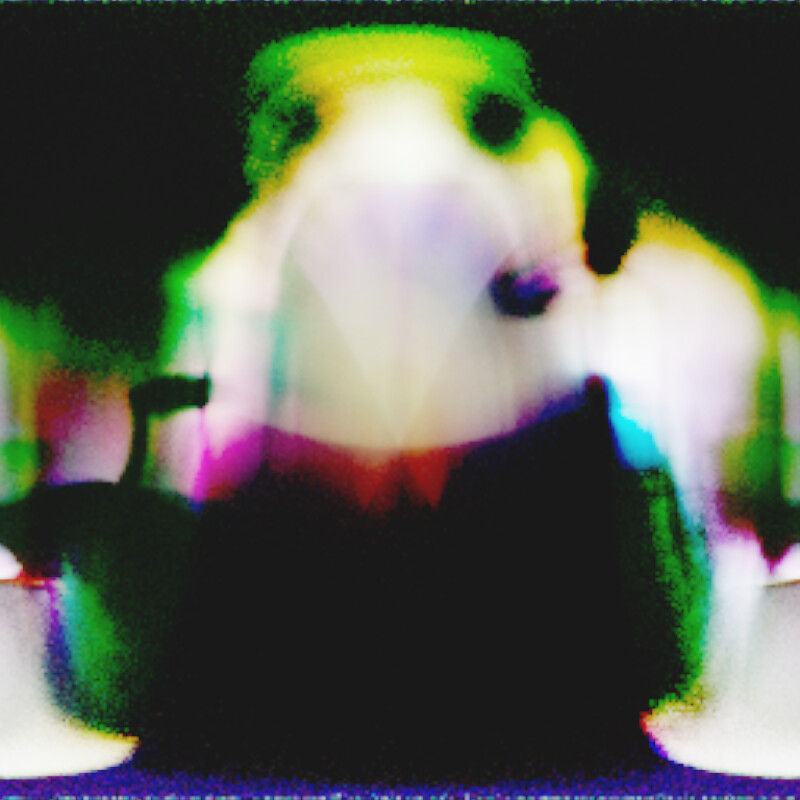}} \\

			\rotatebox[origin=c]{90}{w/o reg} 
			\raisebox{-0.5\height}{\includegraphics[width=\myfigsize]{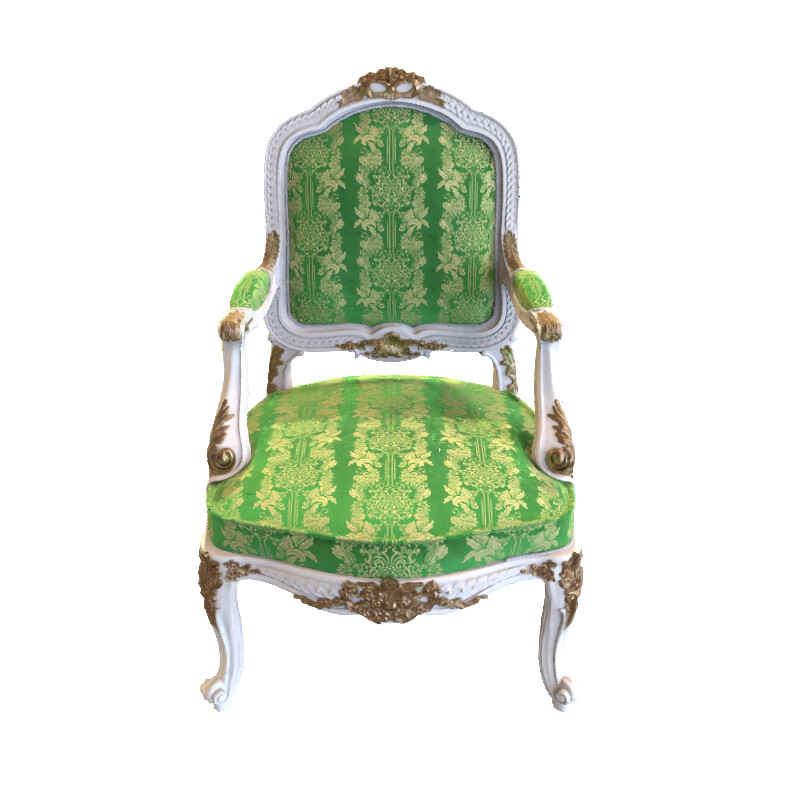}} &
			\raisebox{-0.5\height}{\includegraphics[width=\myfigsize]{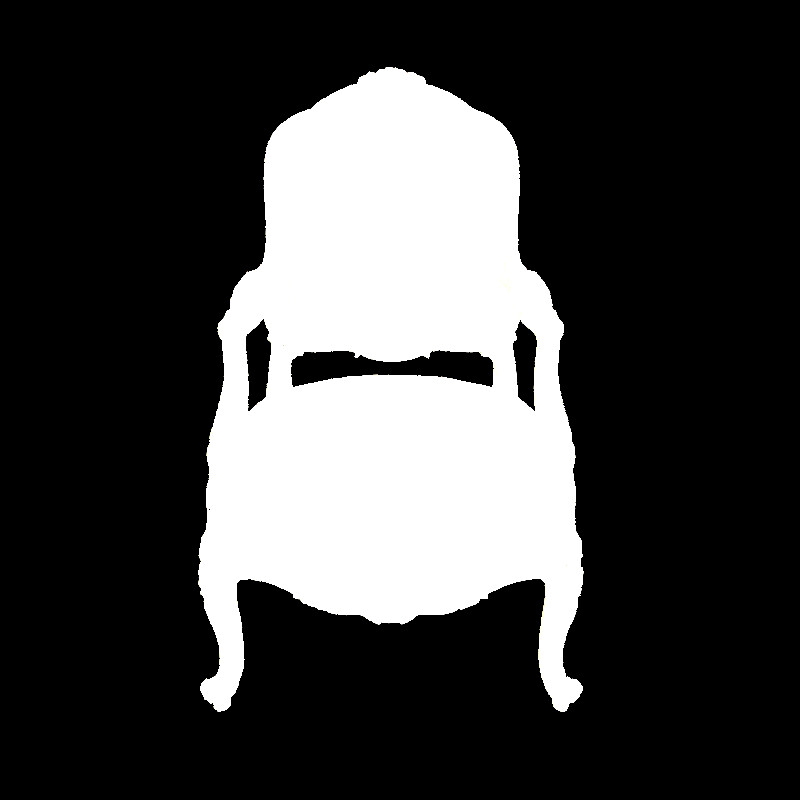}} &
			\raisebox{-0.5\height}{\includegraphics[width=\myfigsize]{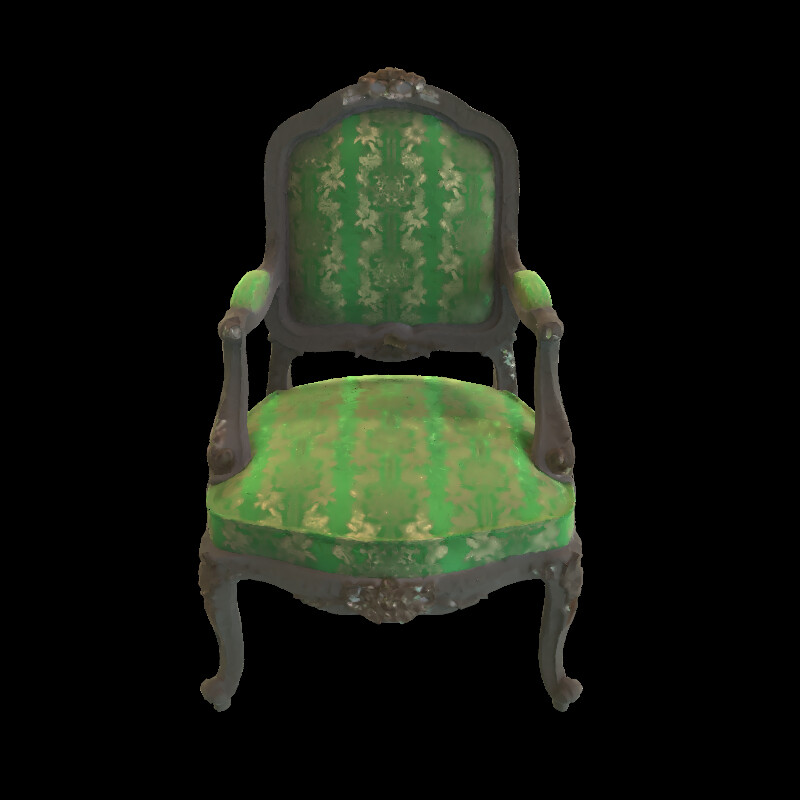}} &
			\raisebox{-0.5\height}{\includegraphics[width=\myfigsize]{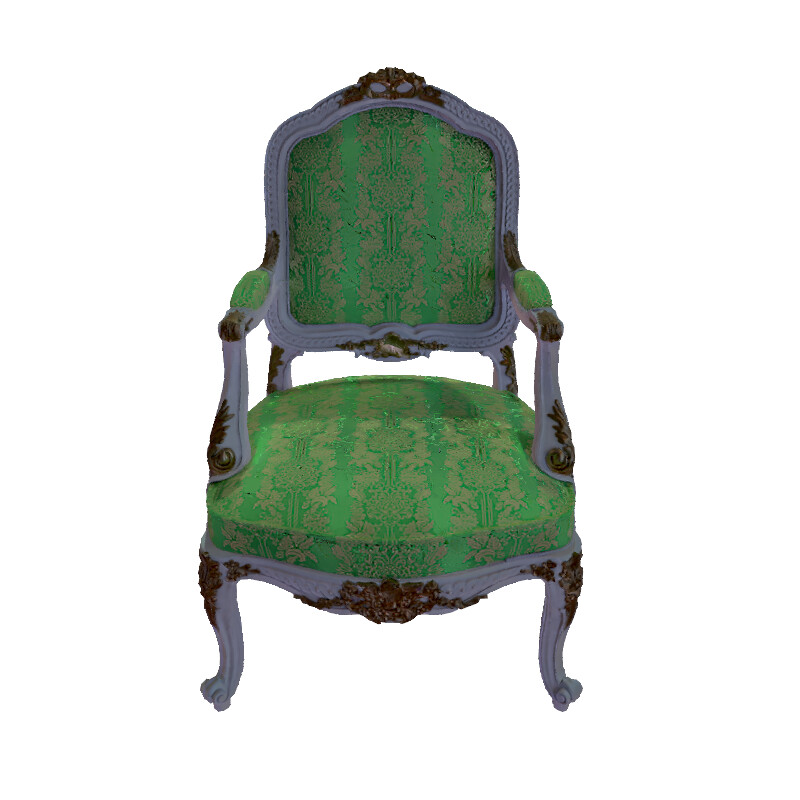}} &
			\raisebox{-0.5\height}{\includegraphics[width=\myfigsize]{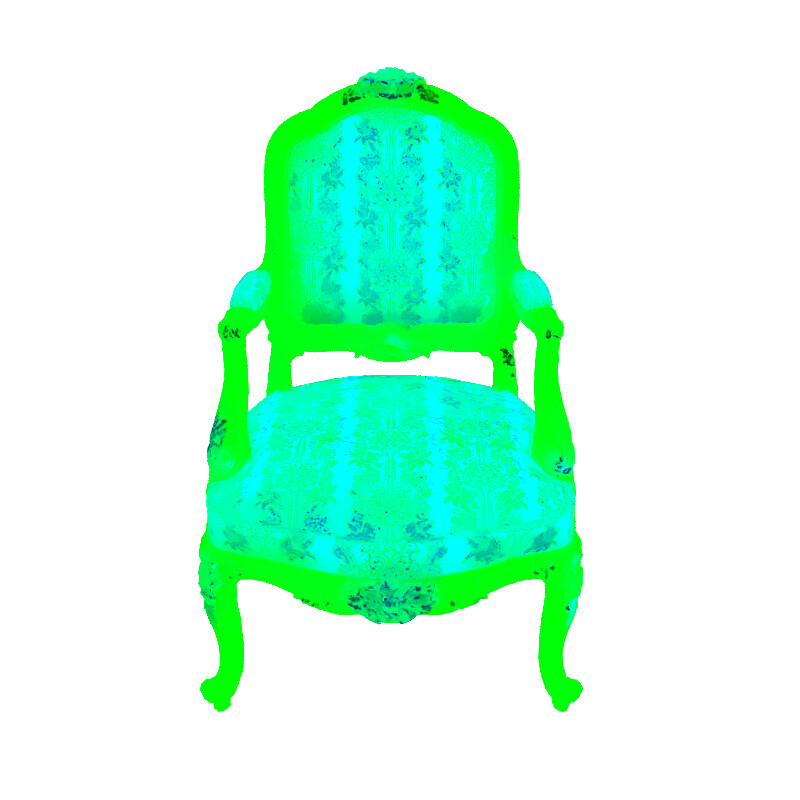}} &
			\raisebox{-0.5\height}{\includegraphics[width=\myfigsize]{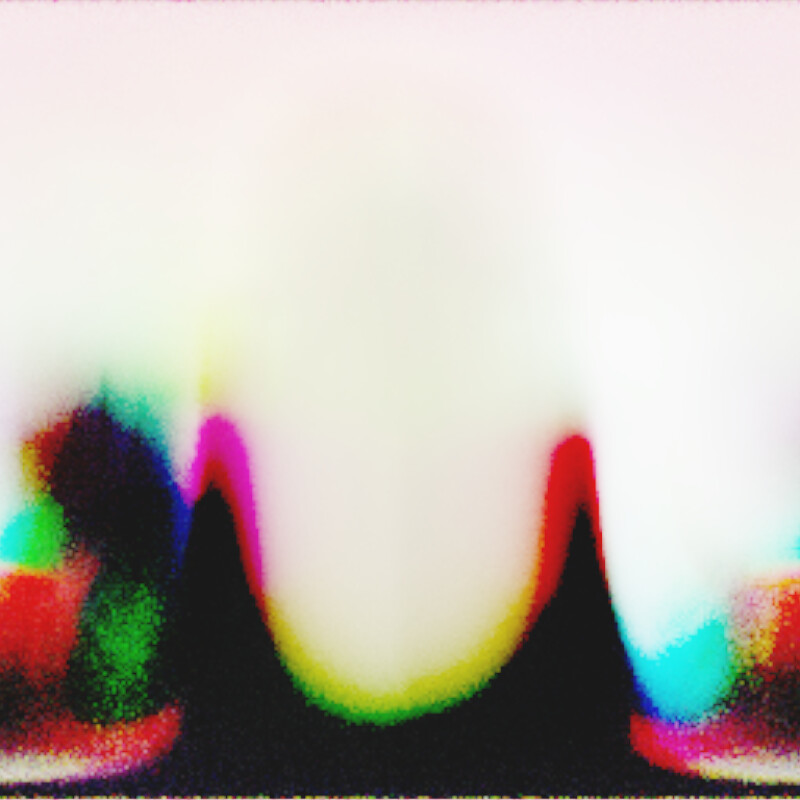}} \\
			
			\rotatebox[origin=c]{90}{w/ reg} 
			\raisebox{-0.5\height}{\includegraphics[width=\myfigsize]{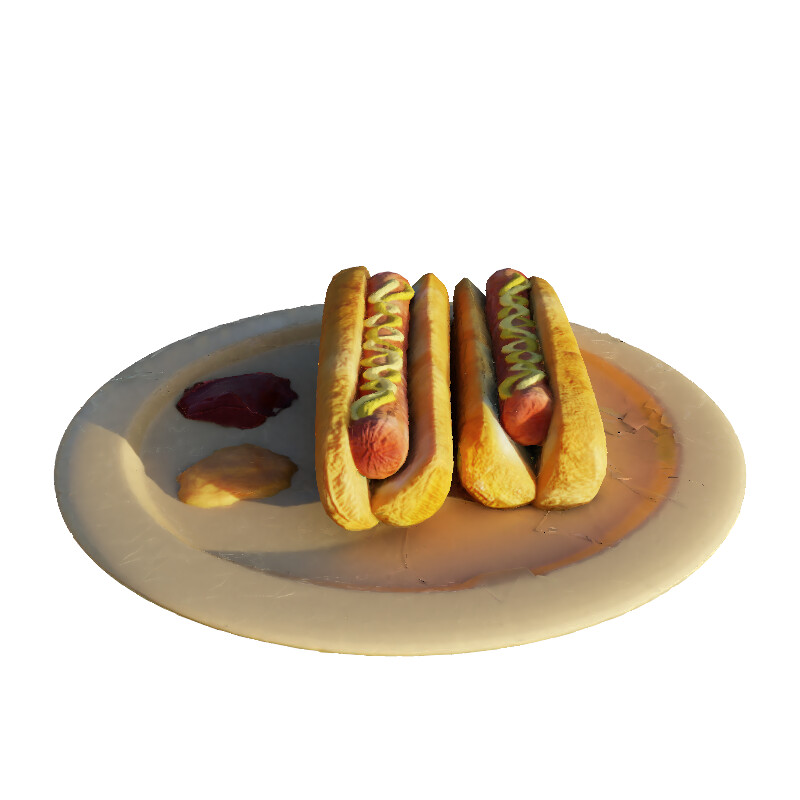}} &
			\raisebox{-0.5\height}{\includegraphics[width=\myfigsize]{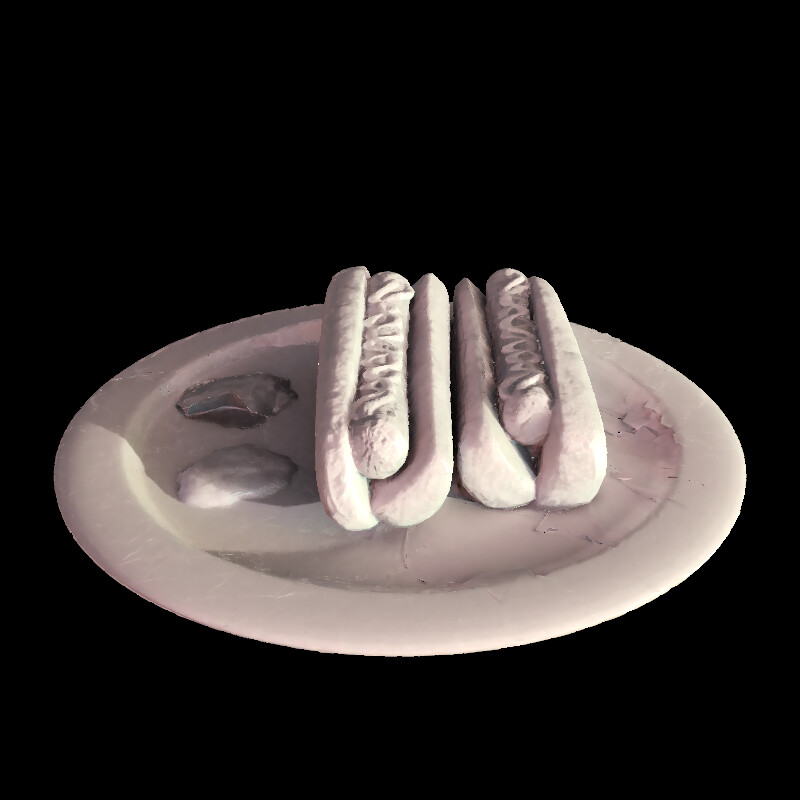}} &
			\raisebox{-0.5\height}{\includegraphics[width=\myfigsize]{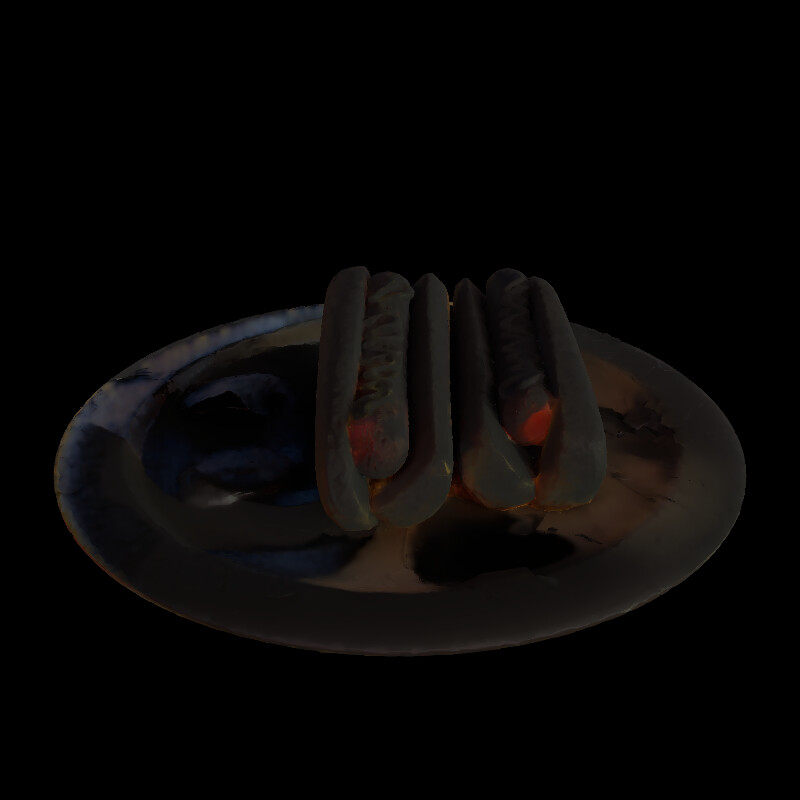}} &
			\raisebox{-0.5\height}{\includegraphics[width=\myfigsize]{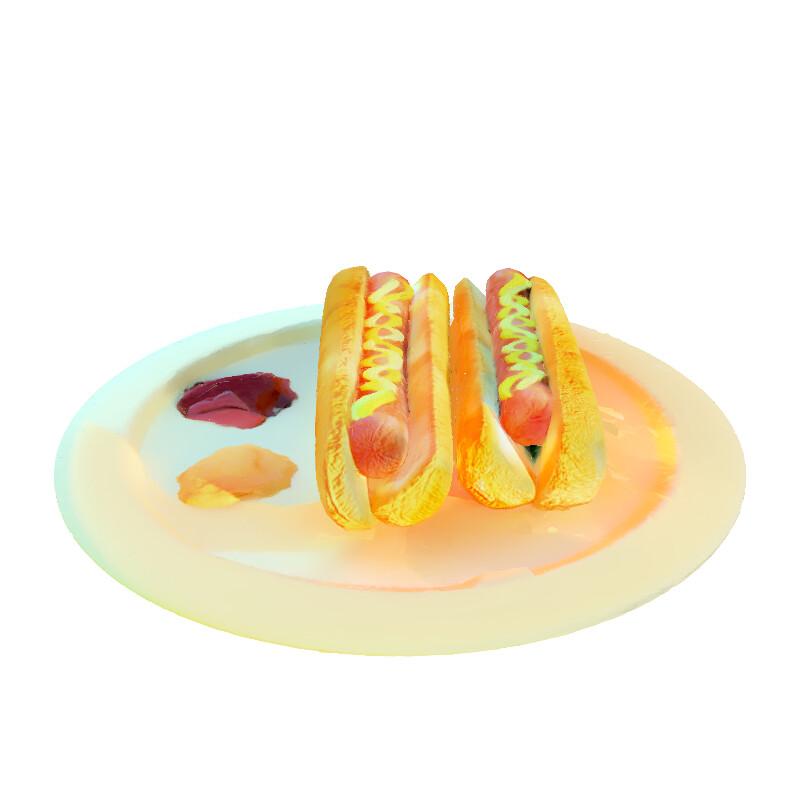}} &
			\raisebox{-0.5\height}{\includegraphics[width=\myfigsize]{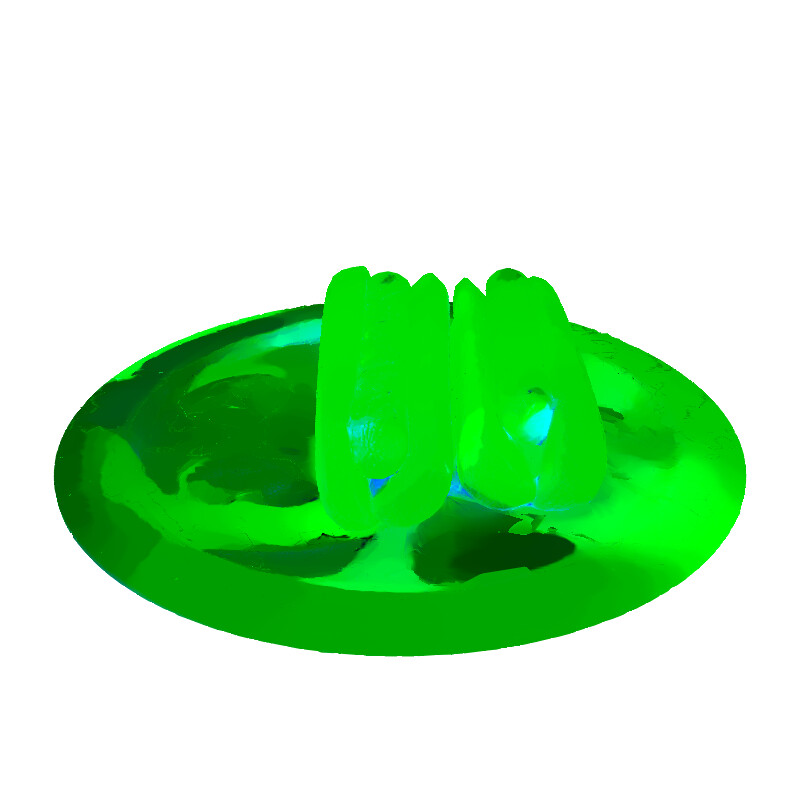}} &
			\raisebox{-0.5\height}{\includegraphics[width=\myfigsize]{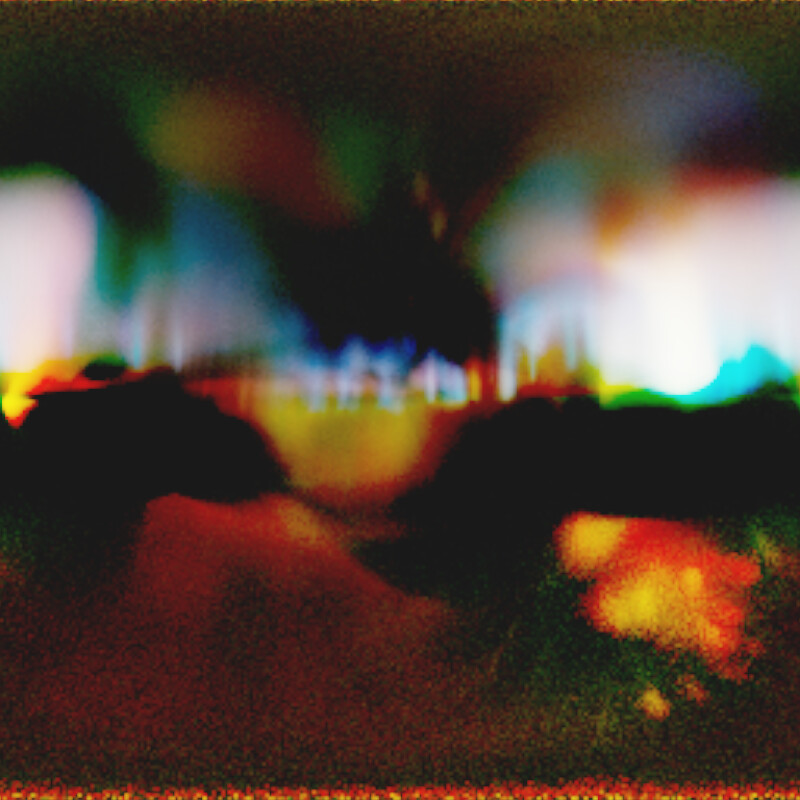}} \\

			\rotatebox[origin=c]{90}{w/o reg} 
			\raisebox{-0.5\height}{\includegraphics[width=\myfigsize]{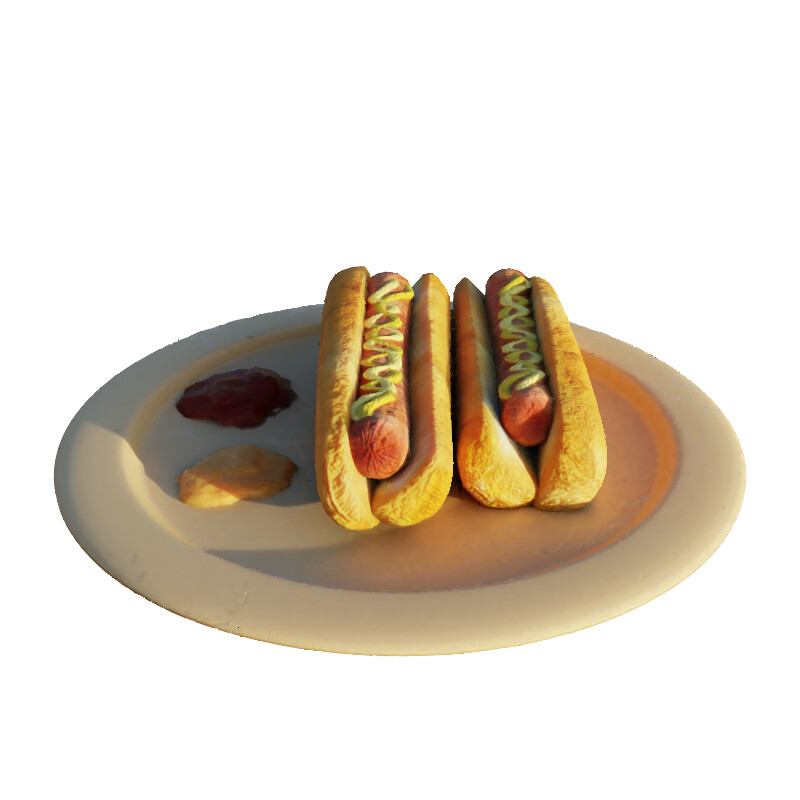}} &
			\raisebox{-0.5\height}{\includegraphics[width=\myfigsize]{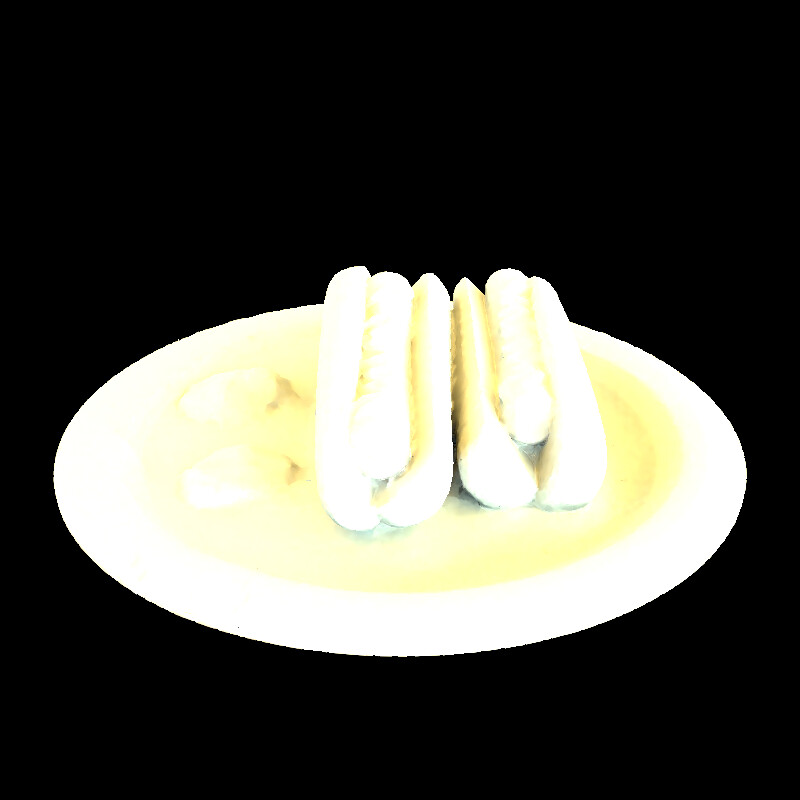}} &
			\raisebox{-0.5\height}{\includegraphics[width=\myfigsize]{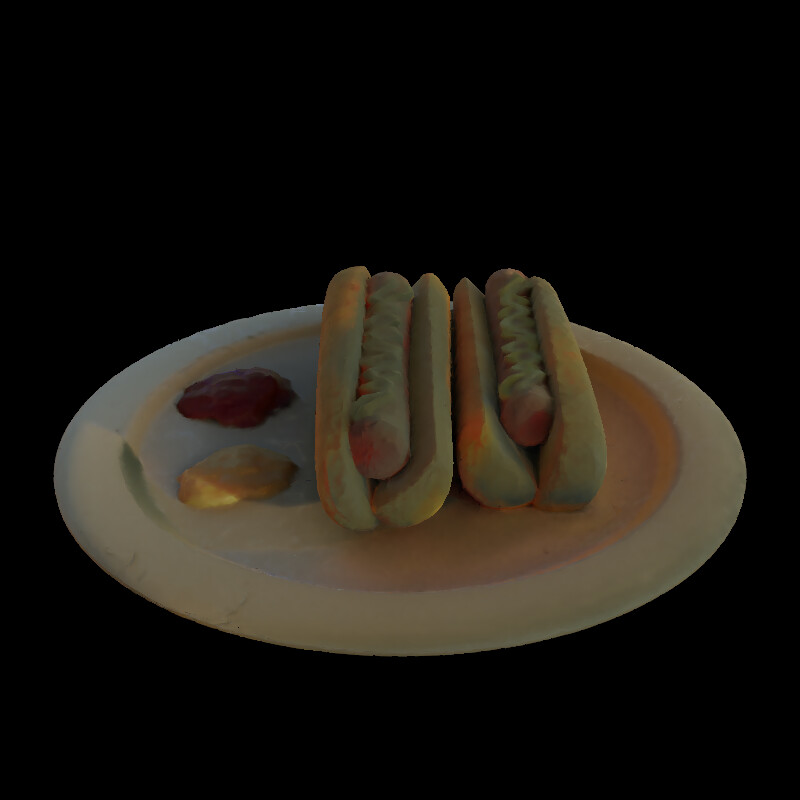}} &
			\raisebox{-0.5\height}{\includegraphics[width=\myfigsize]{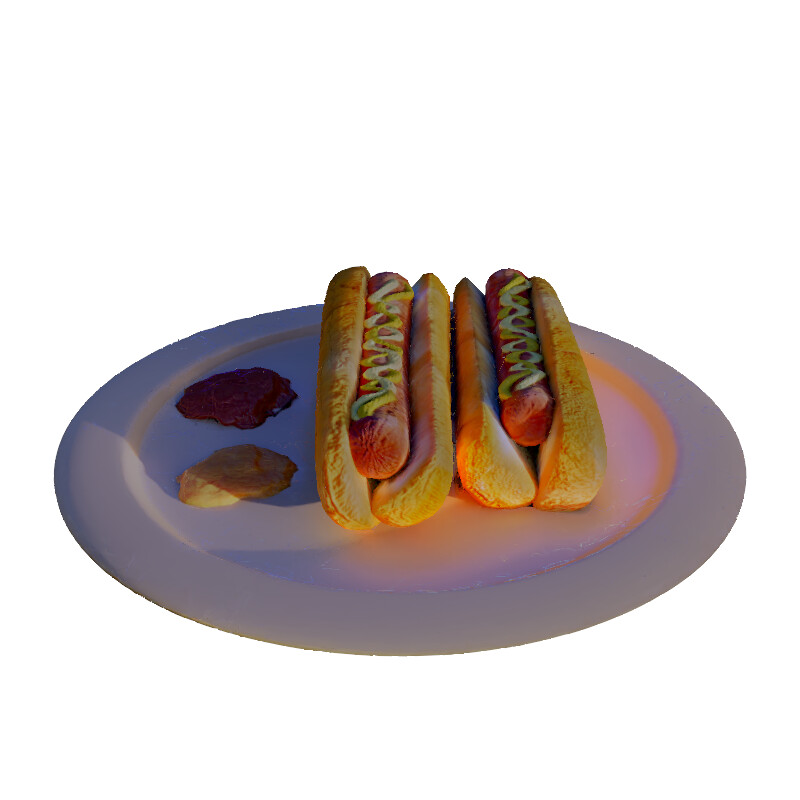}} &
			\raisebox{-0.5\height}{\includegraphics[width=\myfigsize]{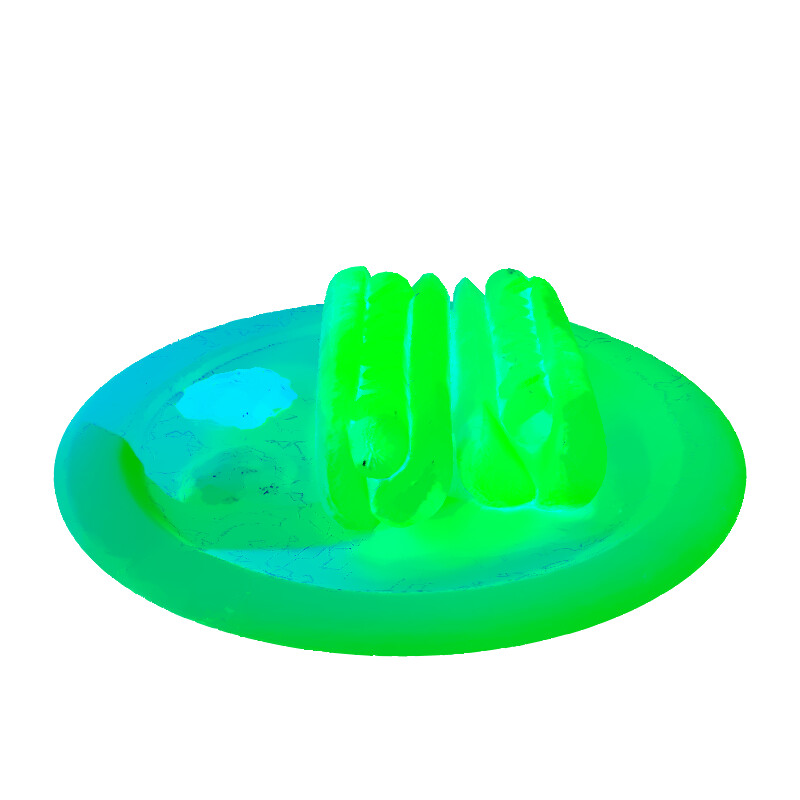}} &
			\raisebox{-0.5\height}{\includegraphics[width=\myfigsize]{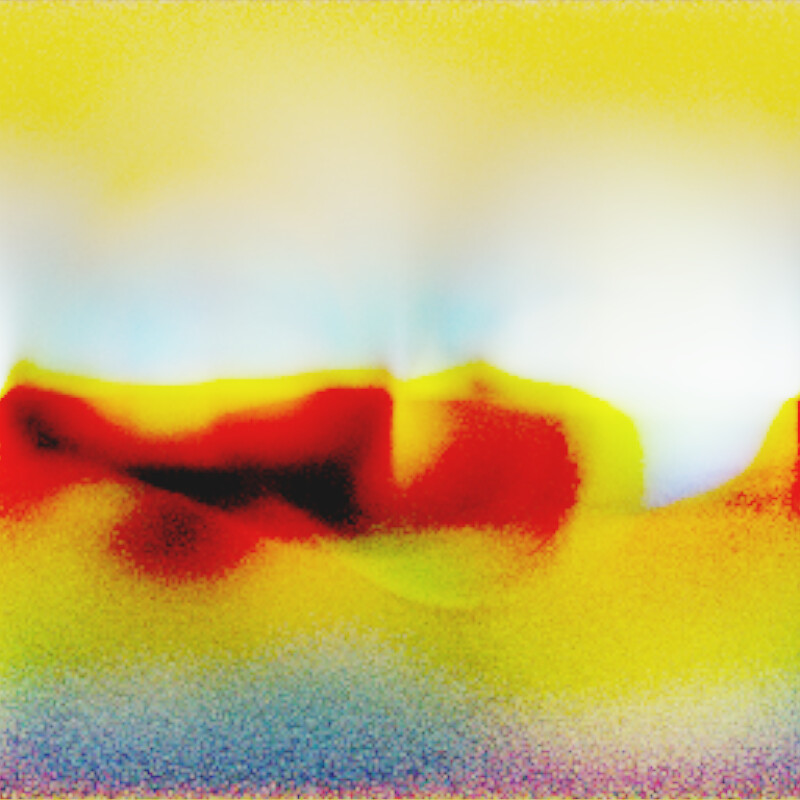}} \\

			\rotatebox[origin=c]{90}{w/ reg} 
			\raisebox{-0.5\height}{\includegraphics[width=\myfigsize]{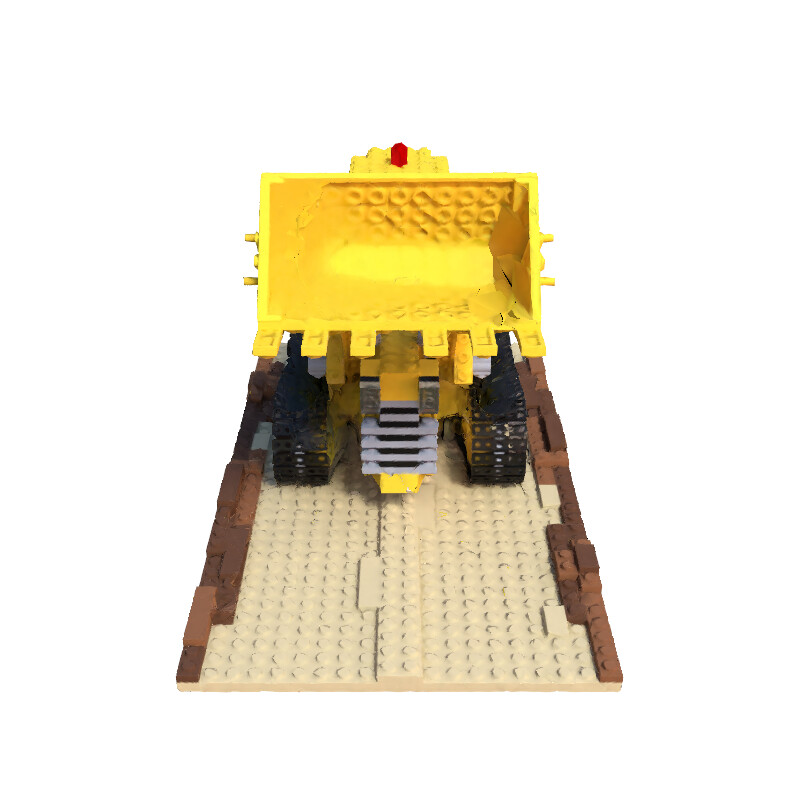}} &
			\raisebox{-0.5\height}{\includegraphics[width=\myfigsize]{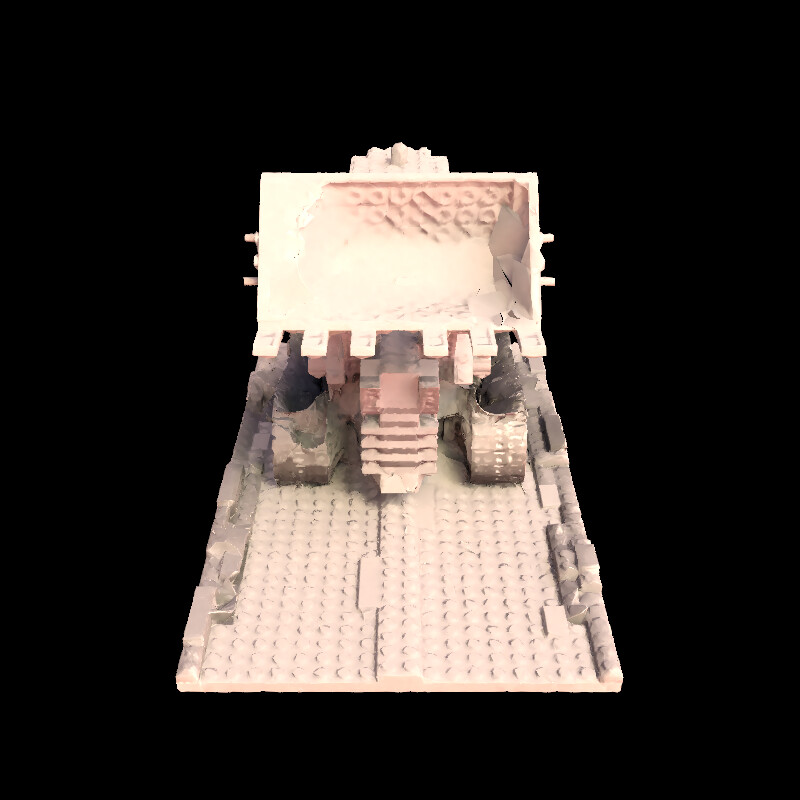}} &
			\raisebox{-0.5\height}{\includegraphics[width=\myfigsize]{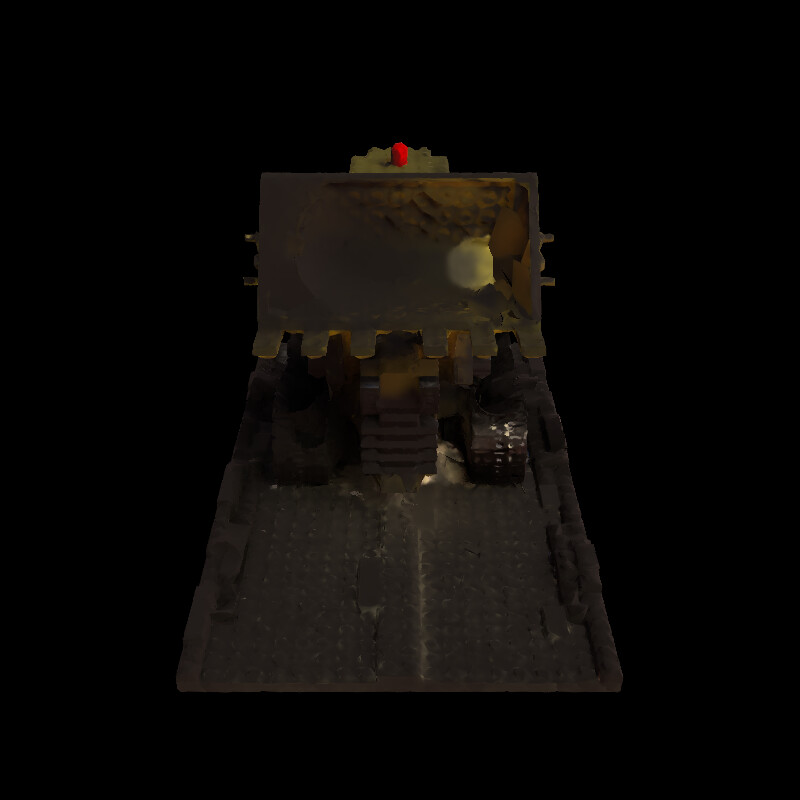}} &
			\raisebox{-0.5\height}{\includegraphics[width=\myfigsize]{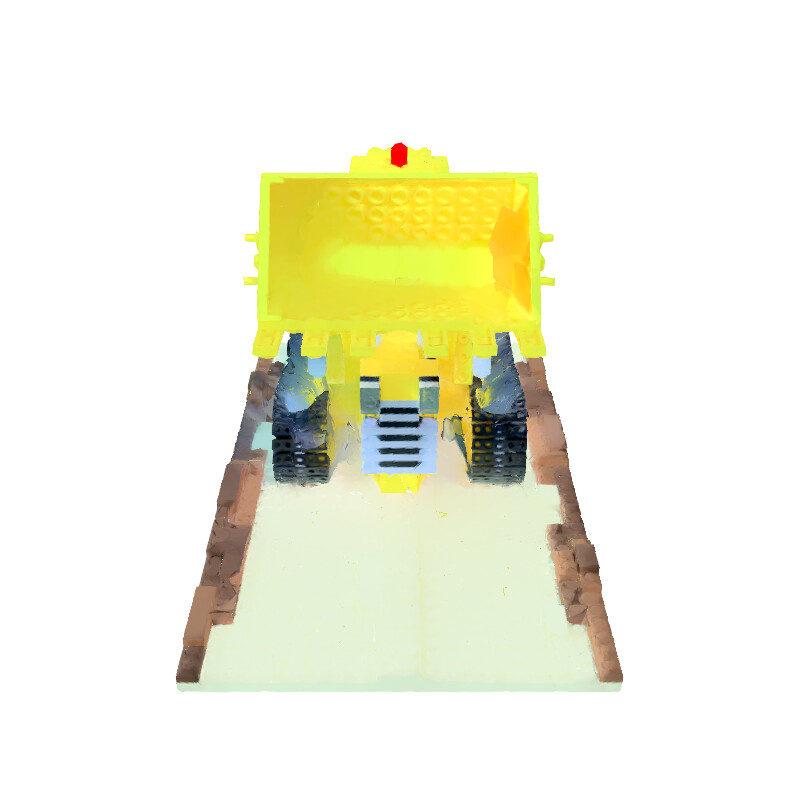}} &
			\raisebox{-0.5\height}{\includegraphics[width=\myfigsize]{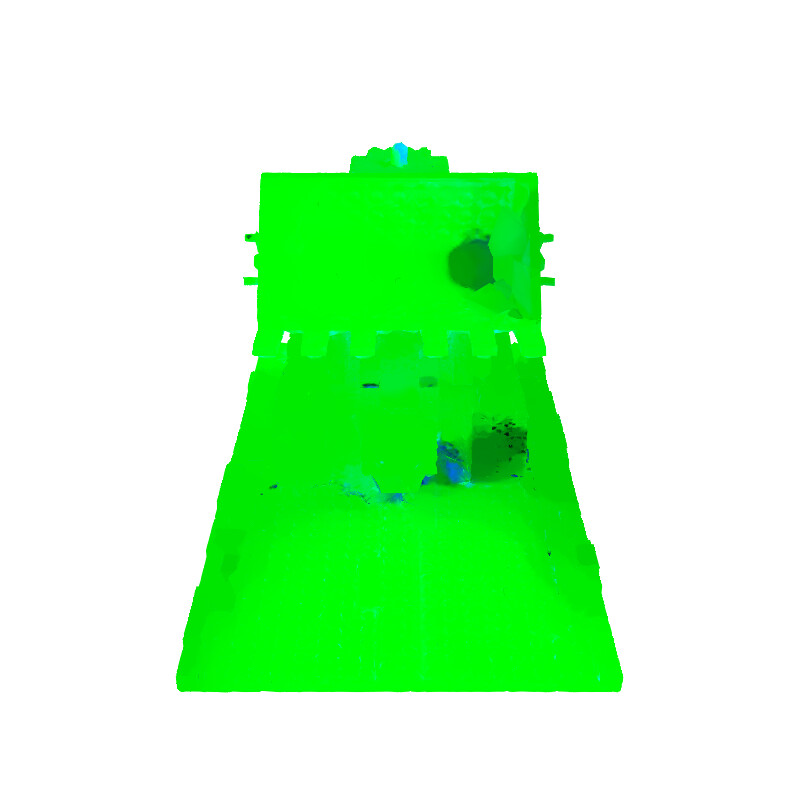}} &
			\raisebox{-0.5\height}{\includegraphics[width=\myfigsize]{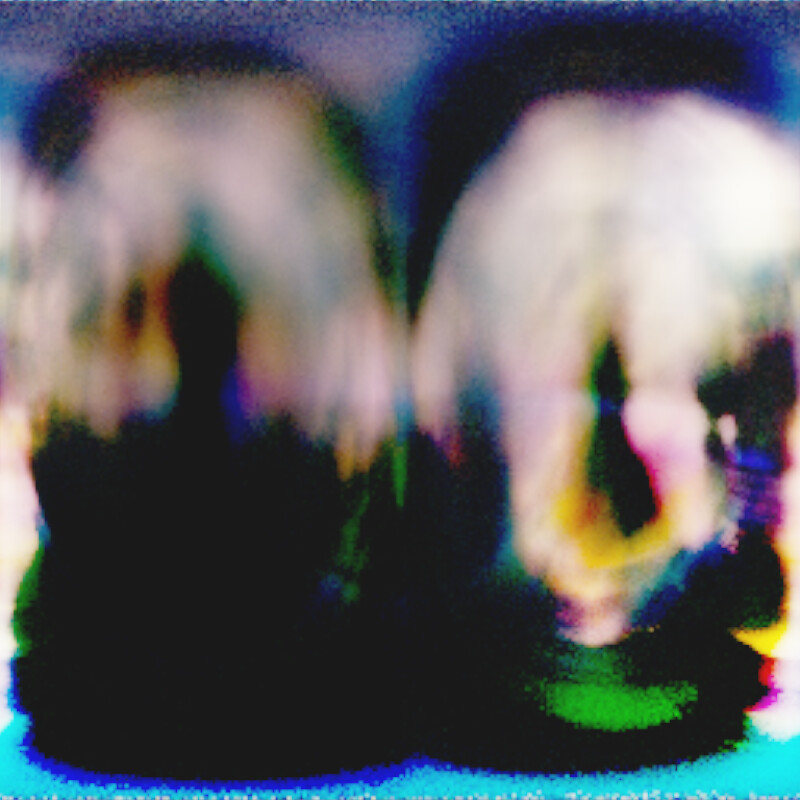}} \\

			\rotatebox[origin=c]{90}{w/o reg} 
			\raisebox{-0.5\height}{\includegraphics[width=\myfigsize]{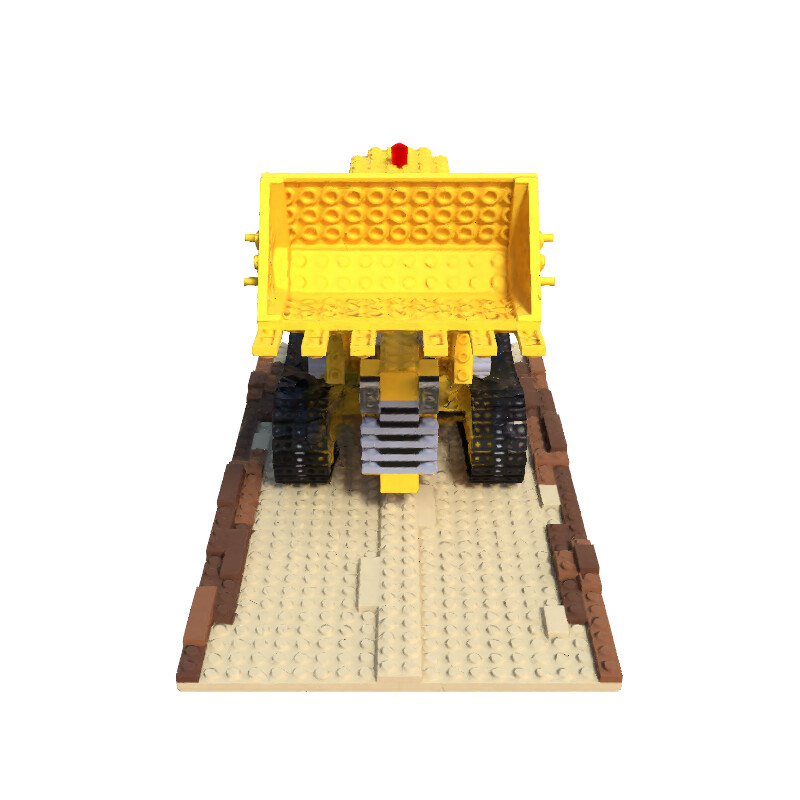}} &
			\raisebox{-0.5\height}{\includegraphics[width=\myfigsize]{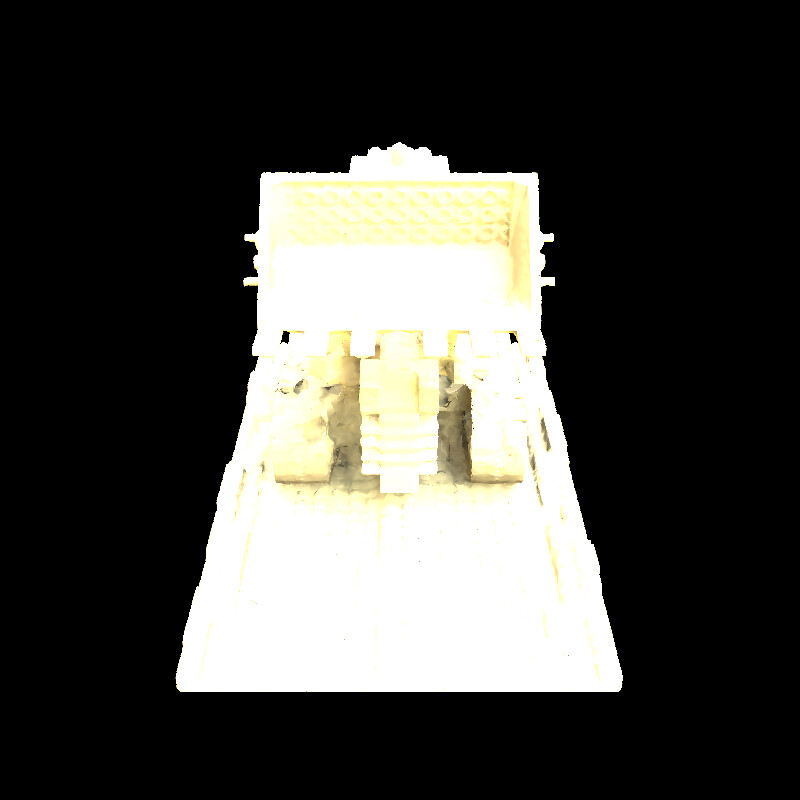}} &
			\raisebox{-0.5\height}{\includegraphics[width=\myfigsize]{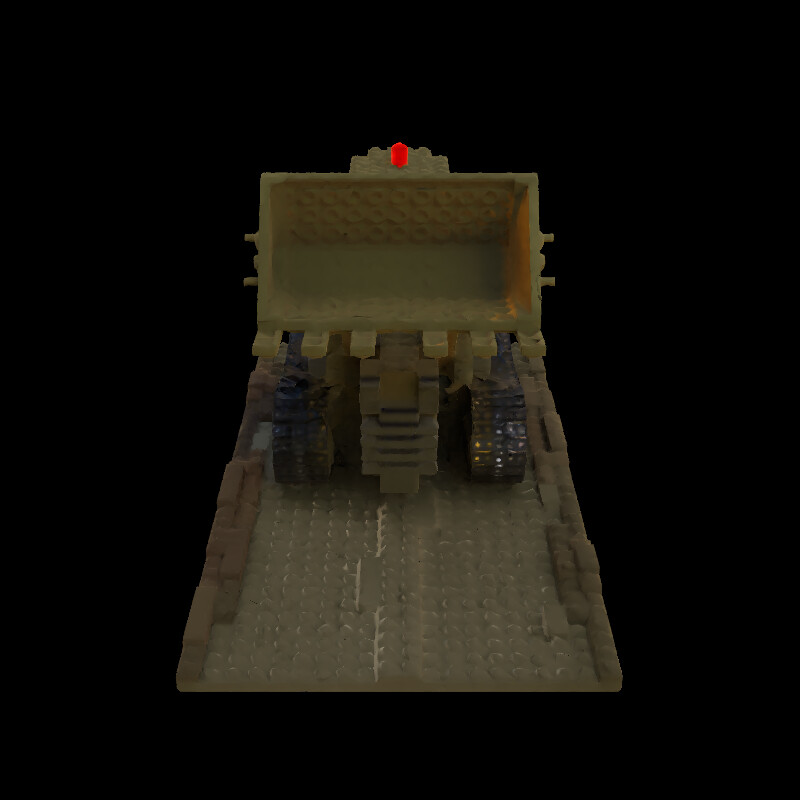}} &
			\raisebox{-0.5\height}{\includegraphics[width=\myfigsize]{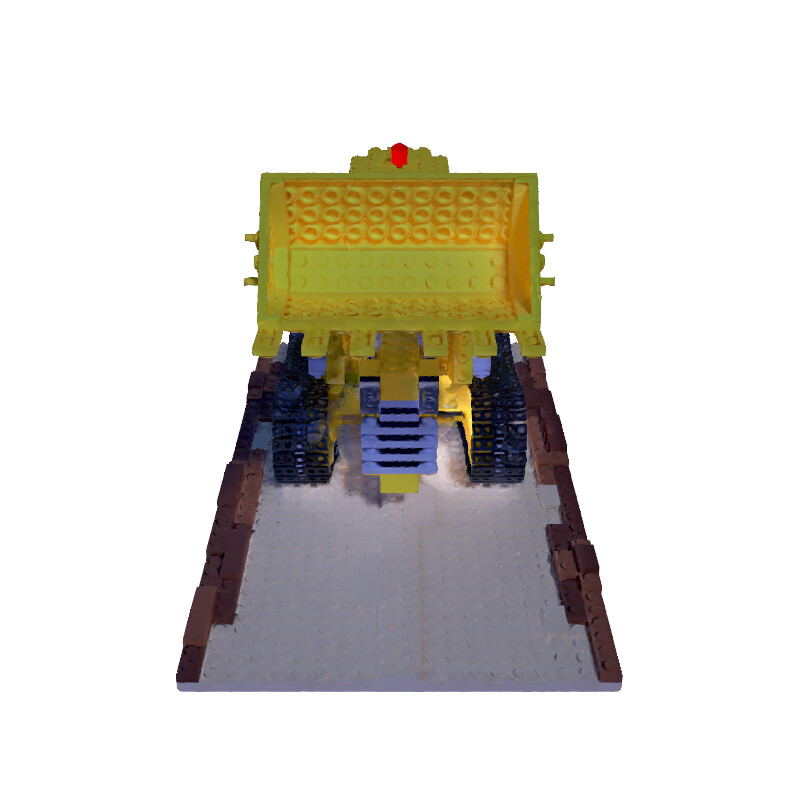}} &
			\raisebox{-0.5\height}{\includegraphics[width=\myfigsize]{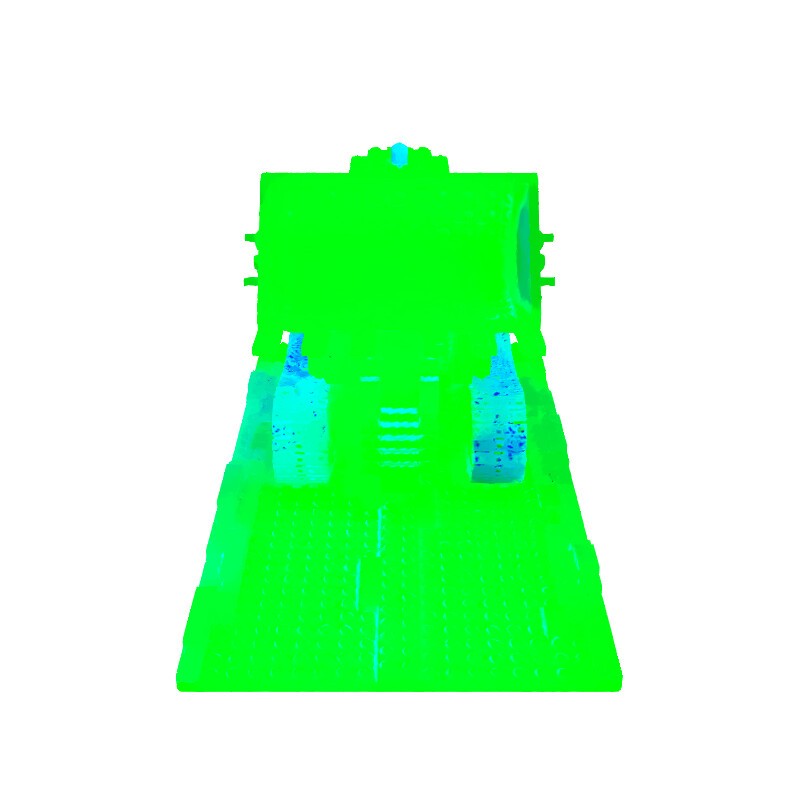}} &
			\raisebox{-0.5\height}{\includegraphics[width=\myfigsize]{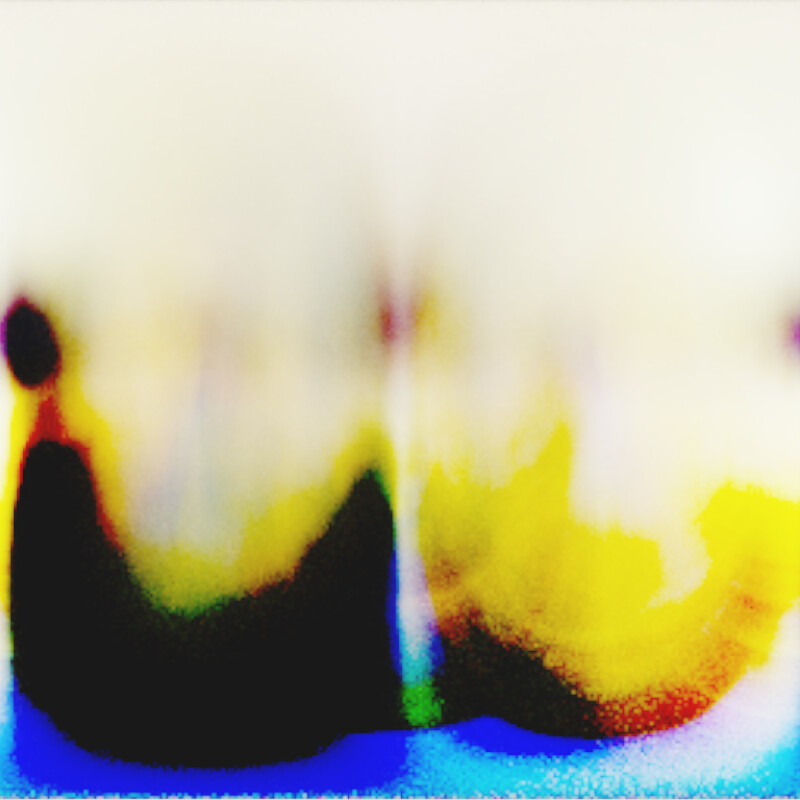}} \\

			\rotatebox[origin=c]{90}{w/ reg} 
			\raisebox{-0.5\height}{\includegraphics[width=\myfigsize]{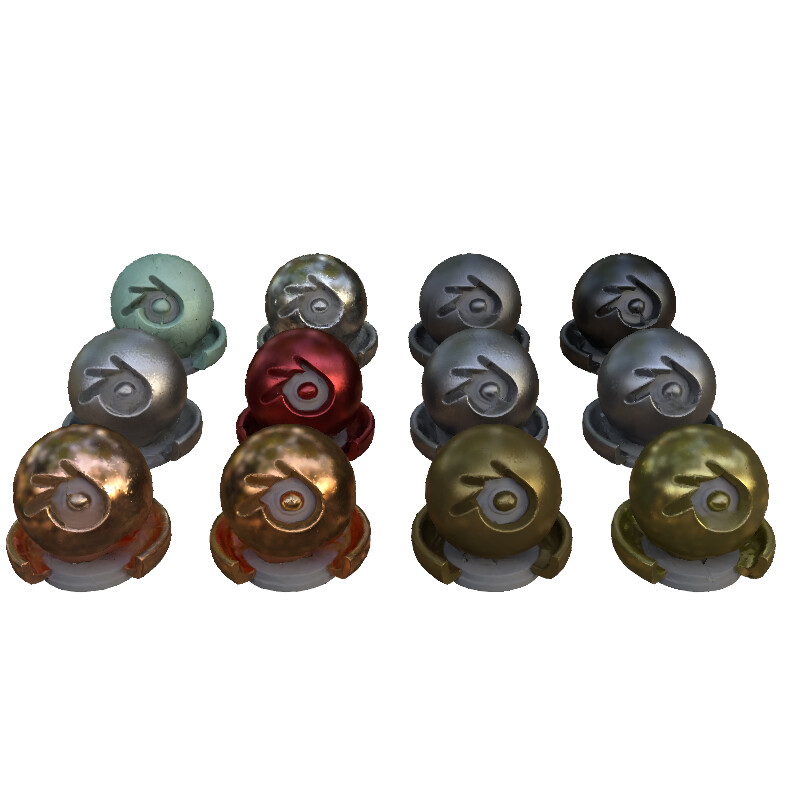}} &
			\raisebox{-0.5\height}{\includegraphics[width=\myfigsize]{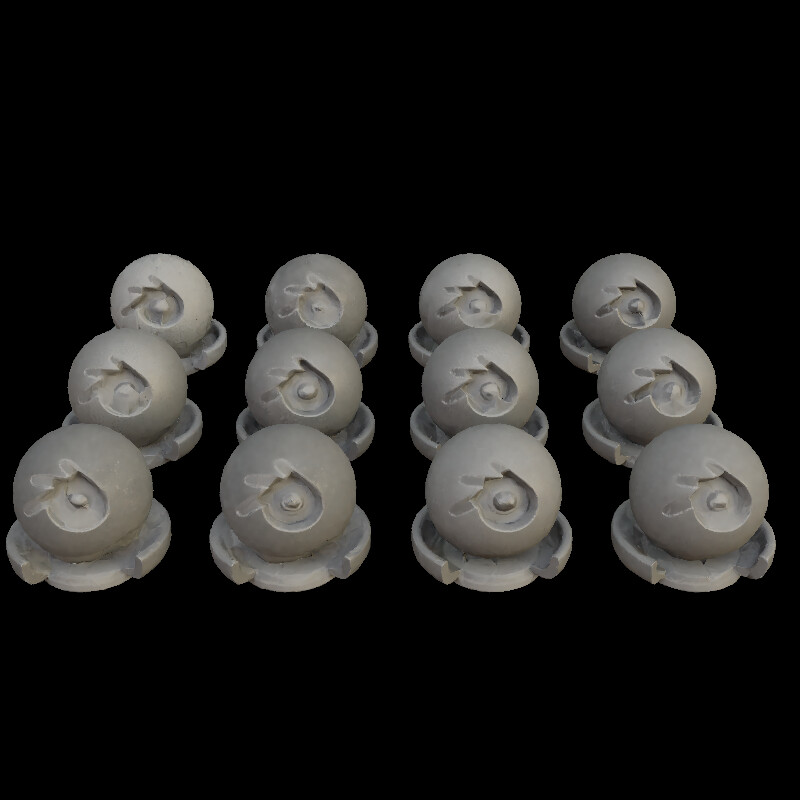}} &
			\raisebox{-0.5\height}{\includegraphics[width=\myfigsize]{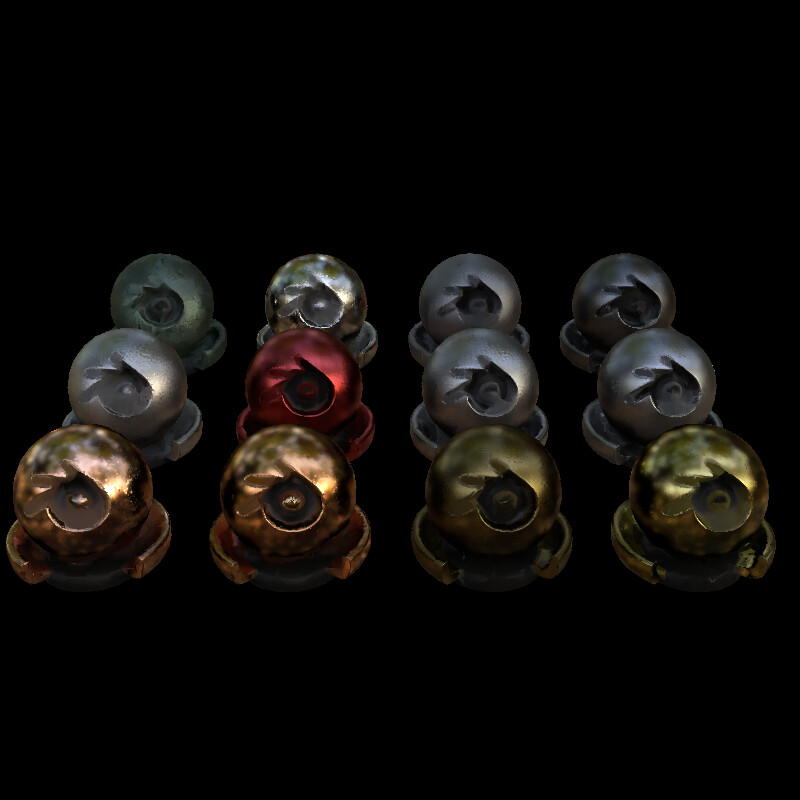}} &
			\raisebox{-0.5\height}{\includegraphics[width=\myfigsize]{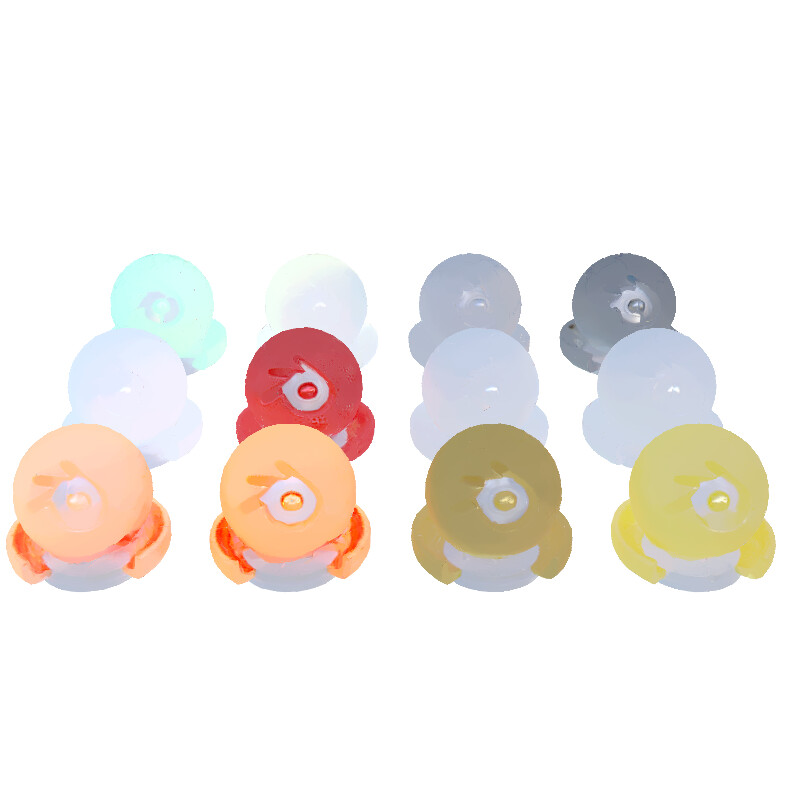}} &
			\raisebox{-0.5\height}{\includegraphics[width=\myfigsize]{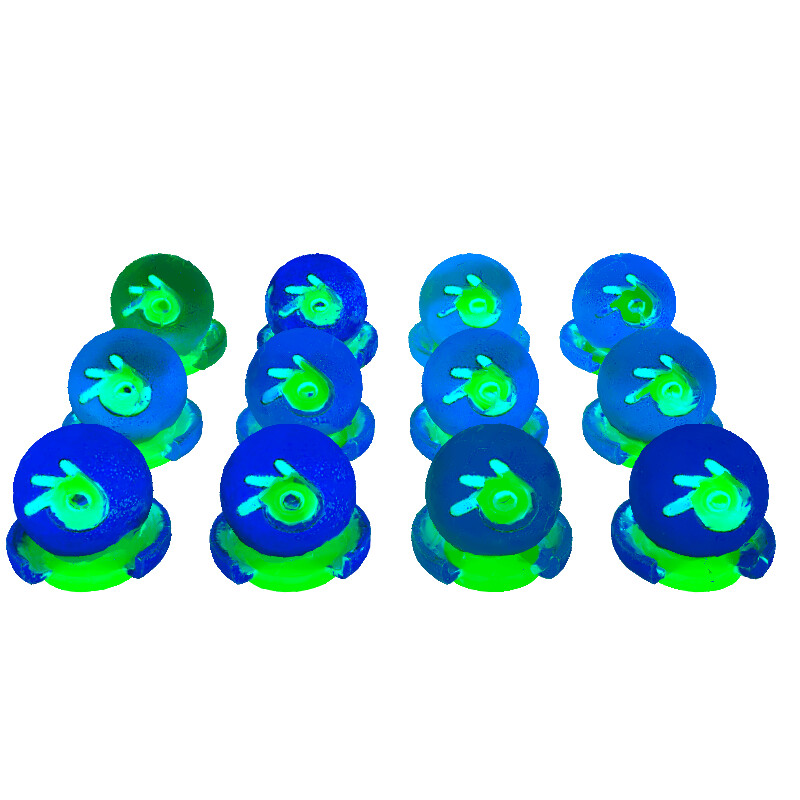}} &
			\raisebox{-0.5\height}{\includegraphics[width=\myfigsize]{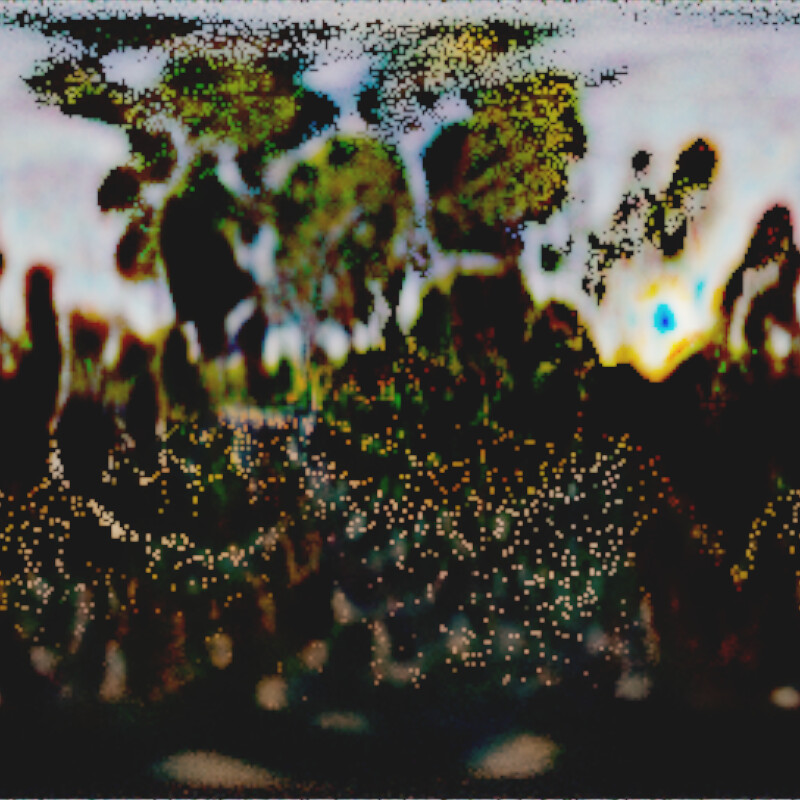}} \\

			\rotatebox[origin=c]{90}{w/o reg} 
			\raisebox{-0.5\height}{\includegraphics[width=\myfigsize]{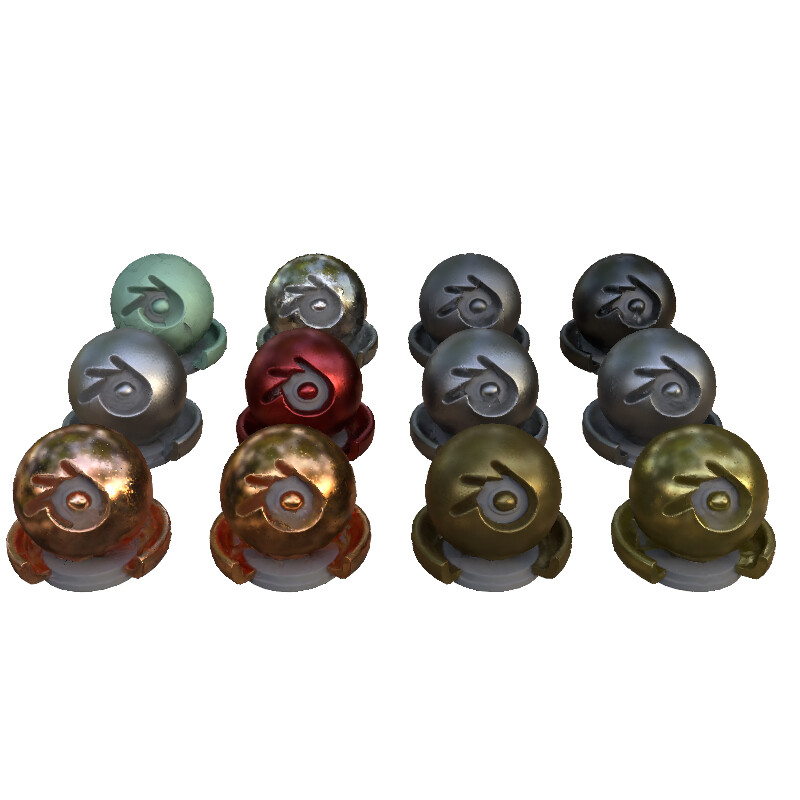}} &
			\raisebox{-0.5\height}{\includegraphics[width=\myfigsize]{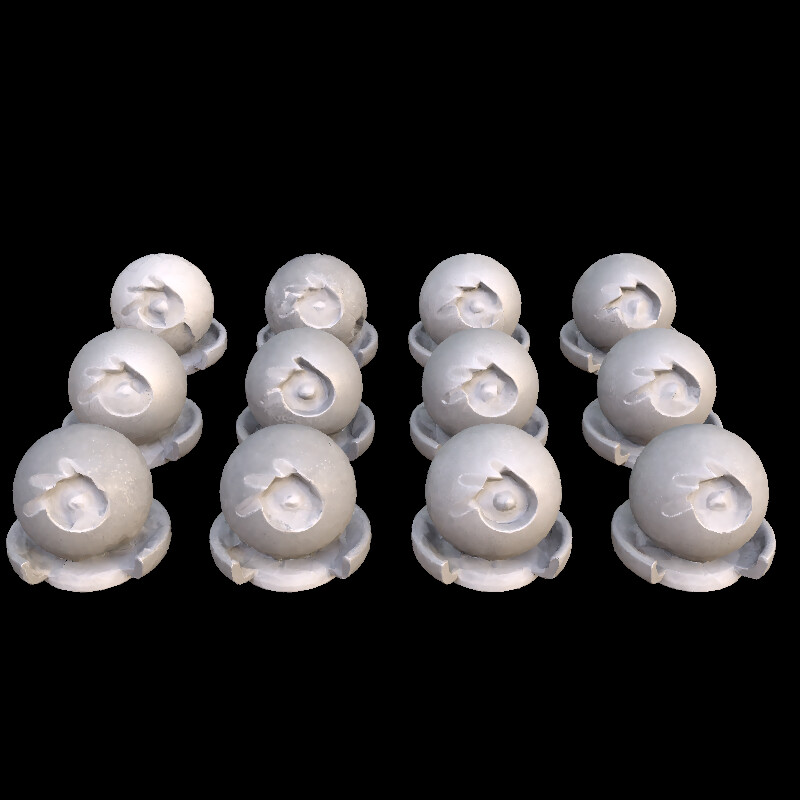}} &
			\raisebox{-0.5\height}{\includegraphics[width=\myfigsize]{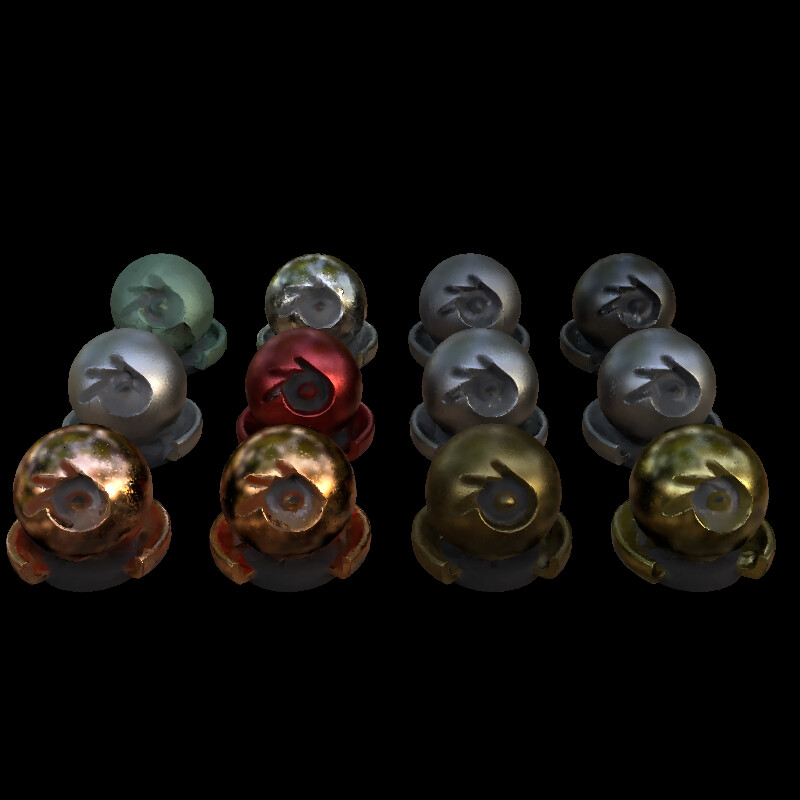}} &
			\raisebox{-0.5\height}{\includegraphics[width=\myfigsize]{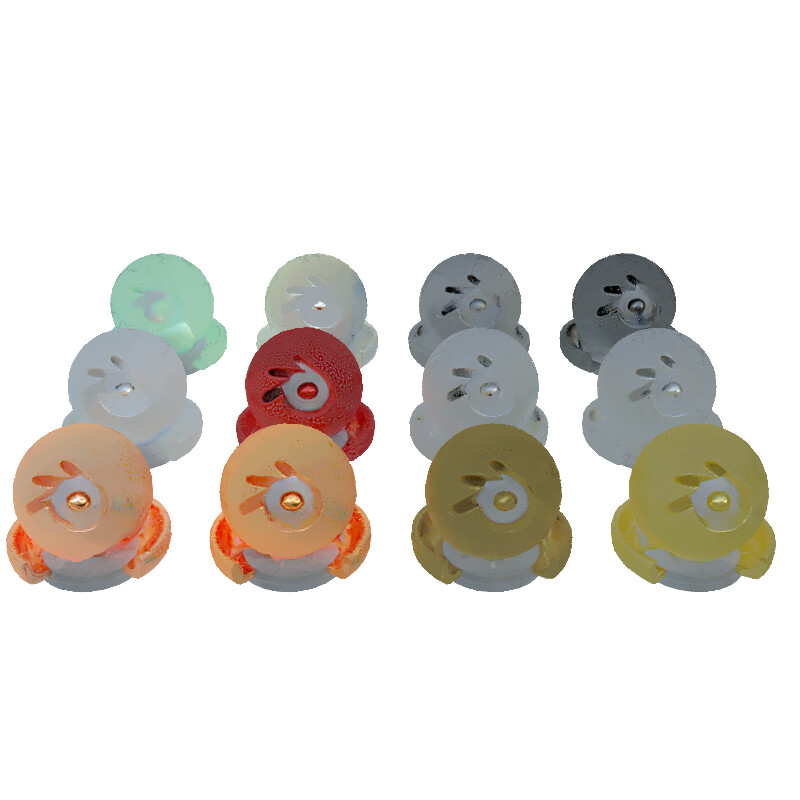}} &
			\raisebox{-0.5\height}{\includegraphics[width=\myfigsize]{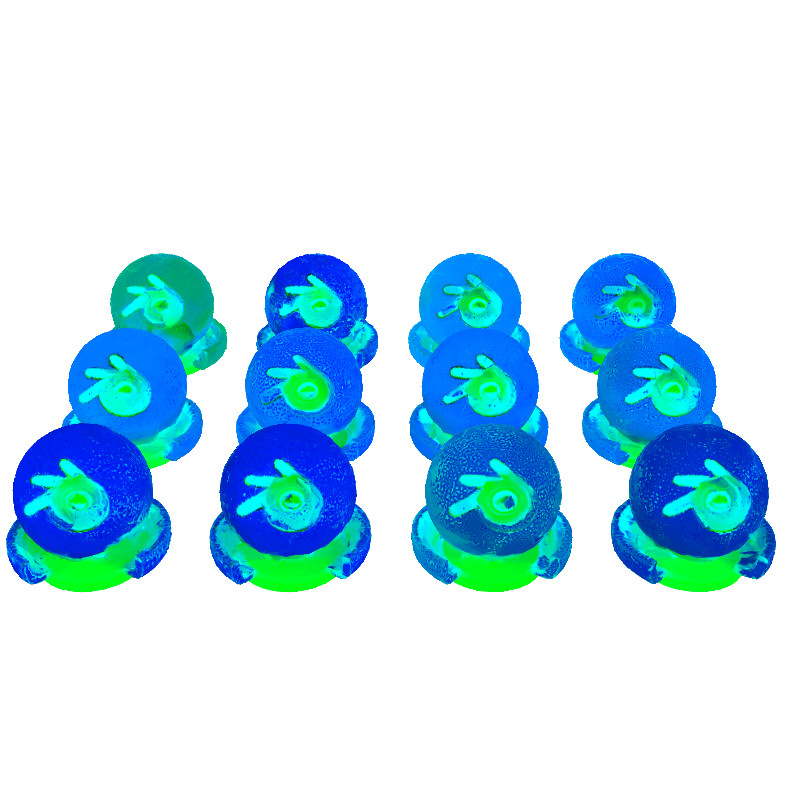}} &
			\raisebox{-0.5\height}{\includegraphics[width=\myfigsize]{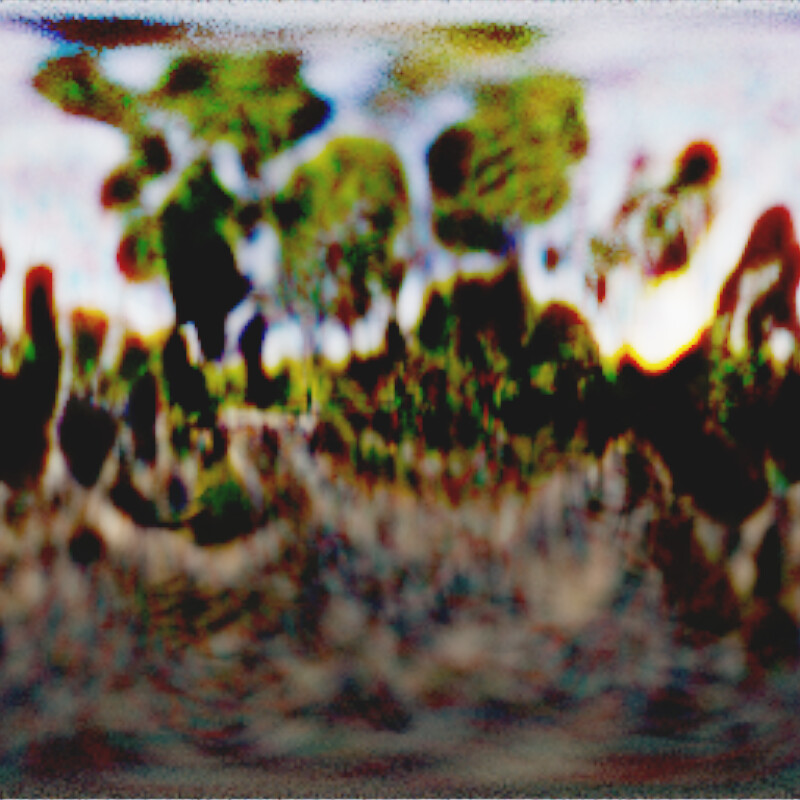}} \\

			\rotatebox[origin=c]{90}{w/ reg} 
			\raisebox{-0.5\height}{\includegraphics[width=\myfigsize]{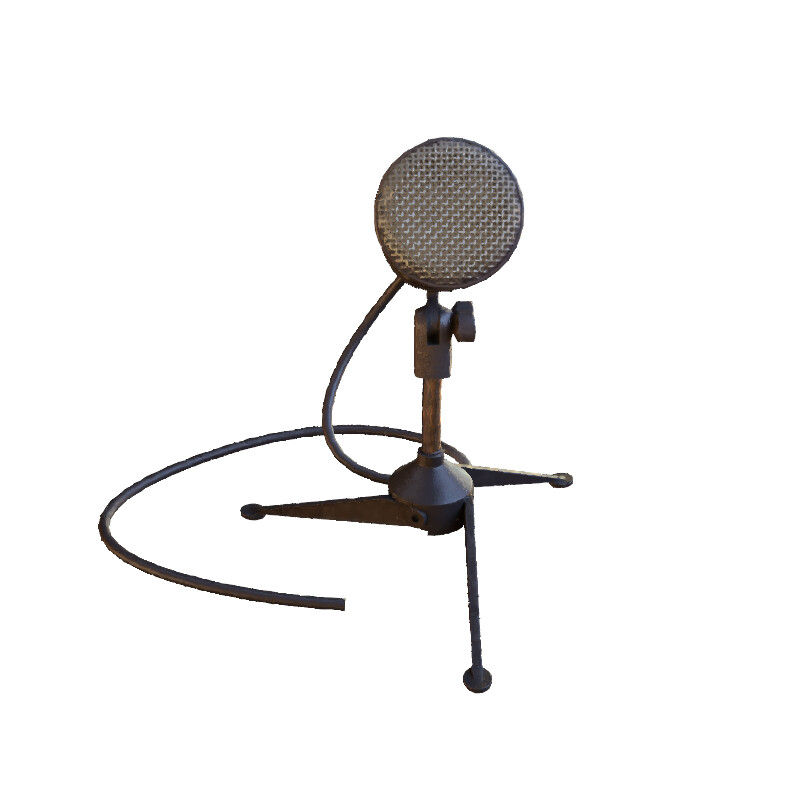}} &
			\raisebox{-0.5\height}{\includegraphics[width=\myfigsize]{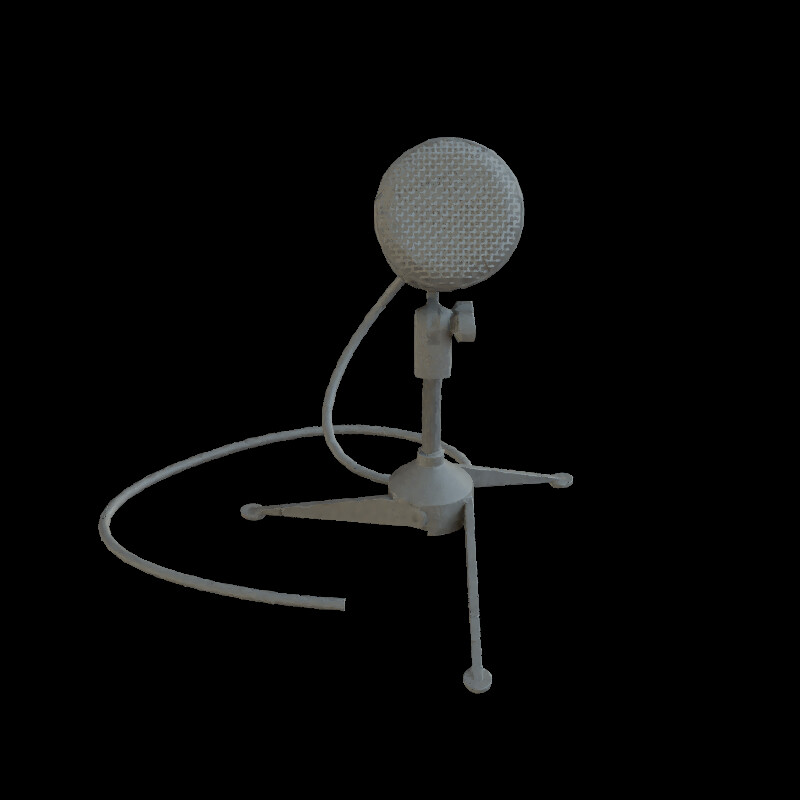}} &
			\raisebox{-0.5\height}{\includegraphics[width=\myfigsize]{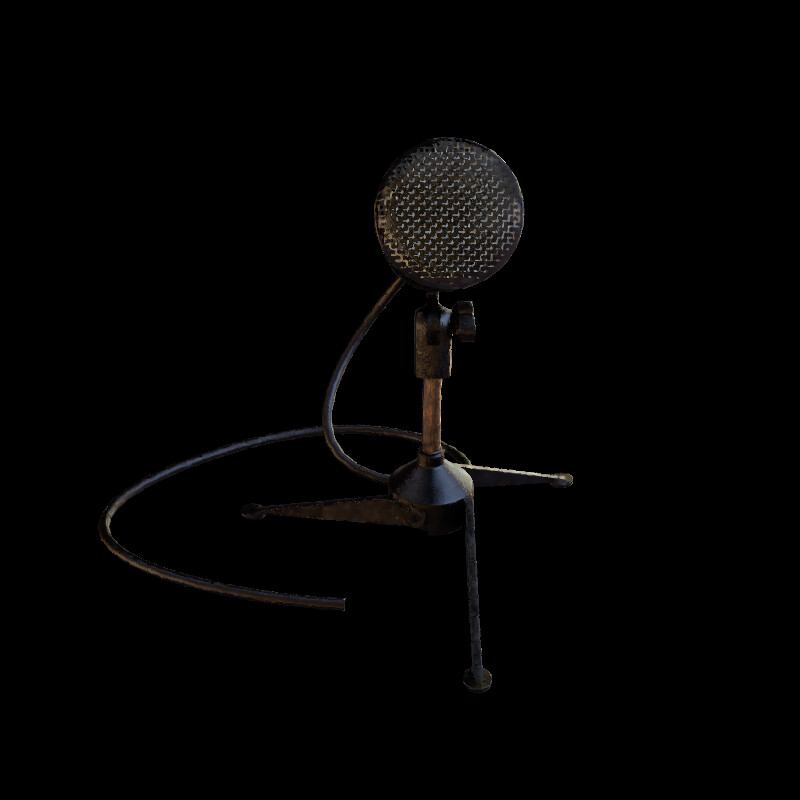}} &
			\raisebox{-0.5\height}{\includegraphics[width=\myfigsize]{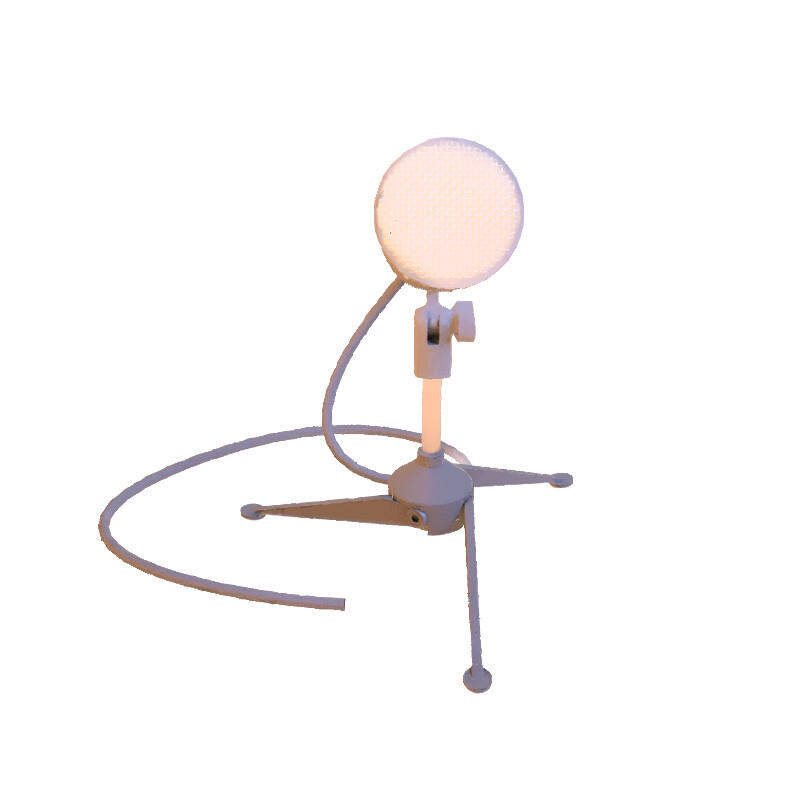}} &
			\raisebox{-0.5\height}{\includegraphics[width=\myfigsize]{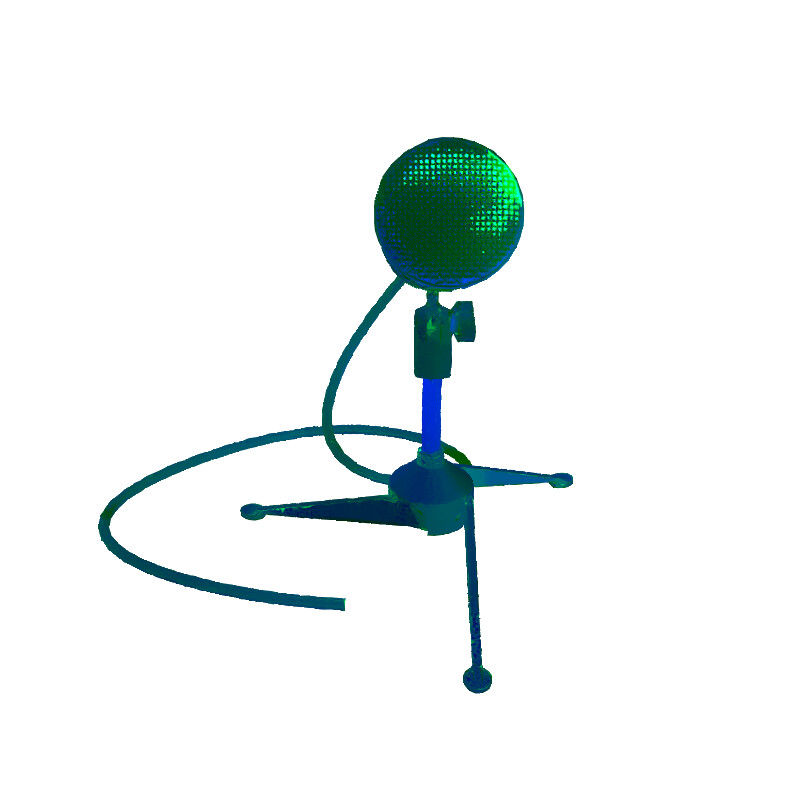}} &
			\raisebox{-0.5\height}{\includegraphics[width=\myfigsize]{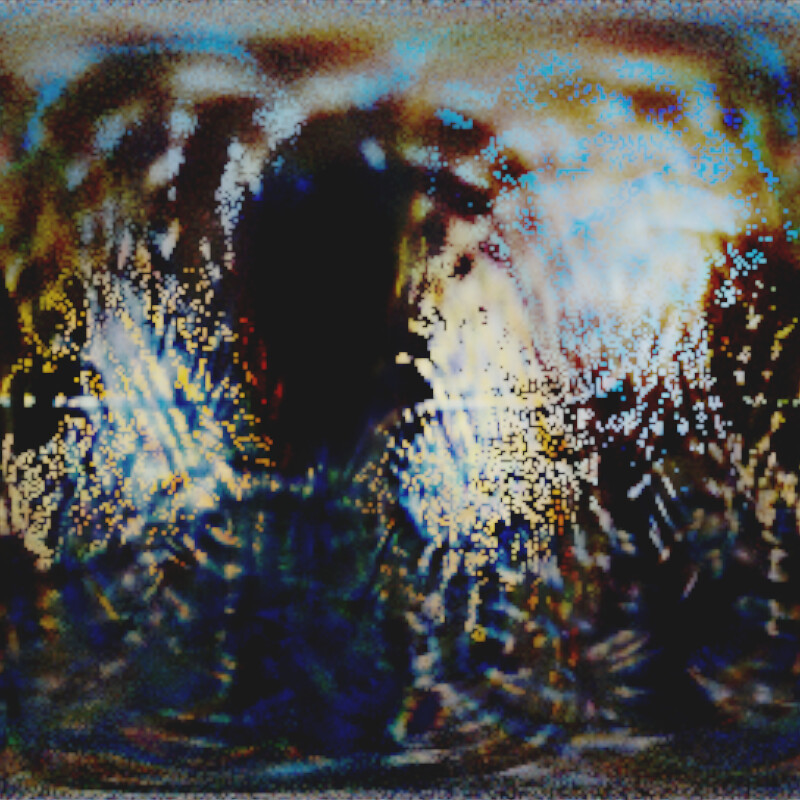}} \\

			\rotatebox[origin=c]{90}{w/o reg} 
			\raisebox{-0.5\height}{\includegraphics[width=\myfigsize]{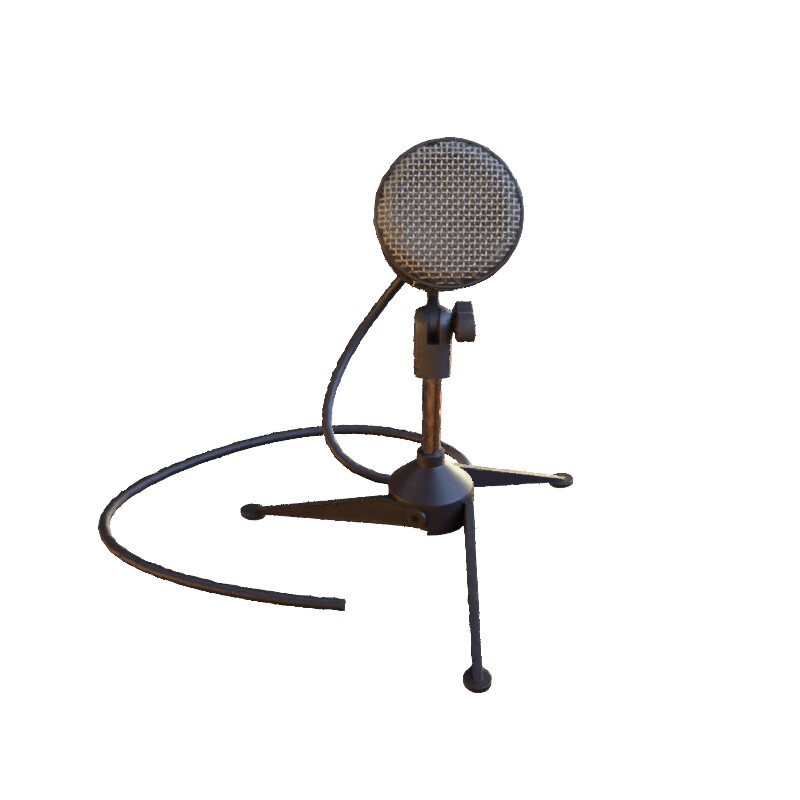}} &
			\raisebox{-0.5\height}{\includegraphics[width=\myfigsize]{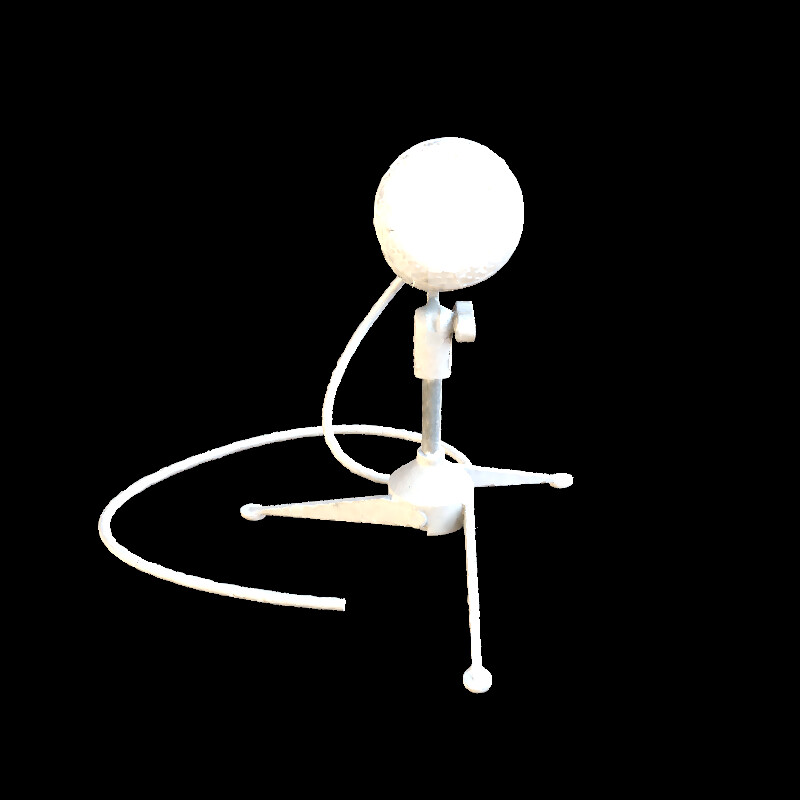}} &
			\raisebox{-0.5\height}{\includegraphics[width=\myfigsize]{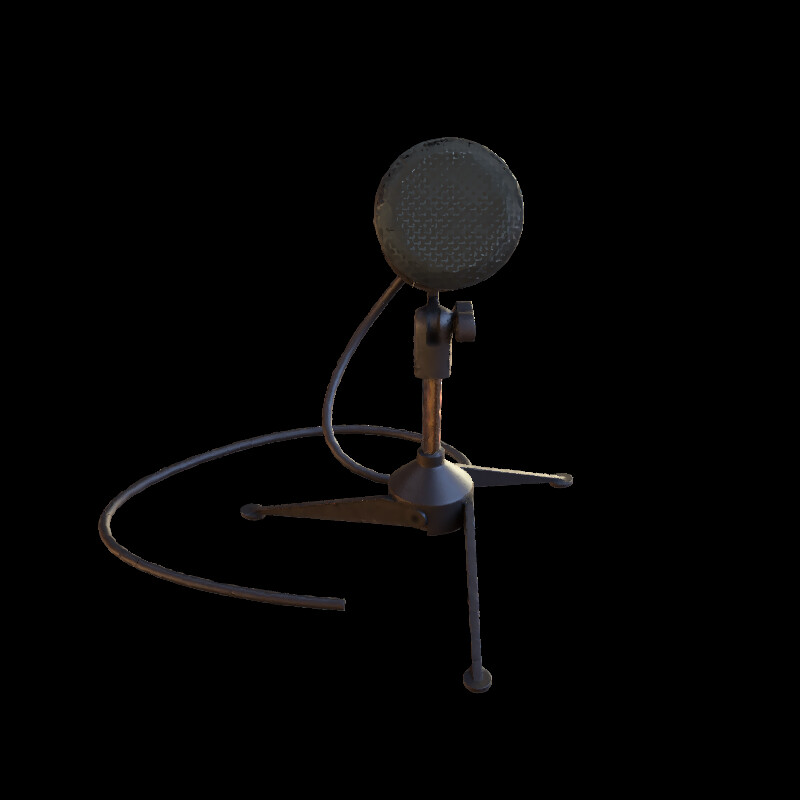}} &
			\raisebox{-0.5\height}{\includegraphics[width=\myfigsize]{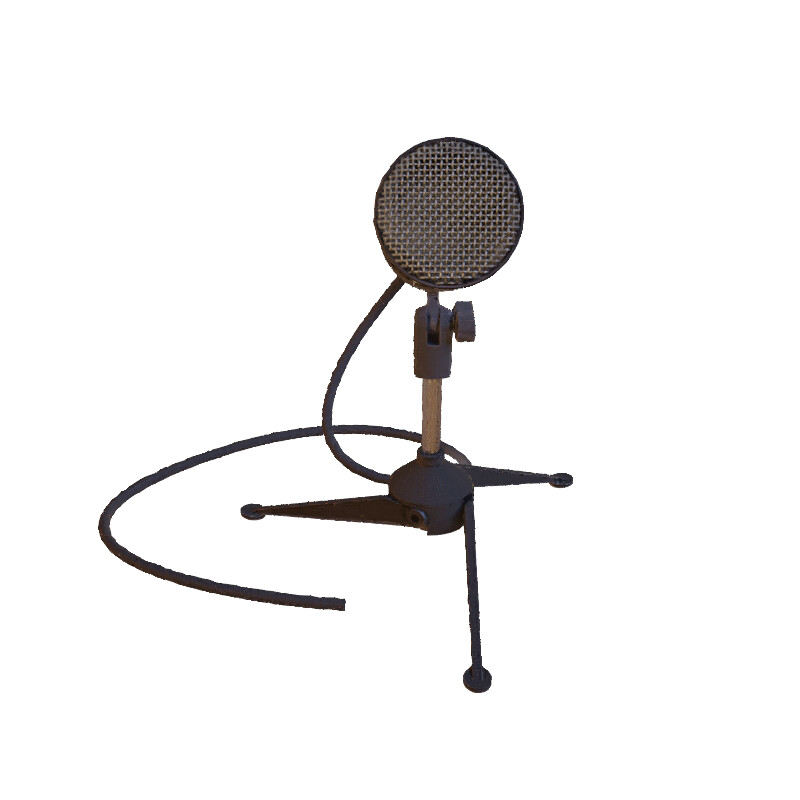}} &
			\raisebox{-0.5\height}{\includegraphics[width=\myfigsize]{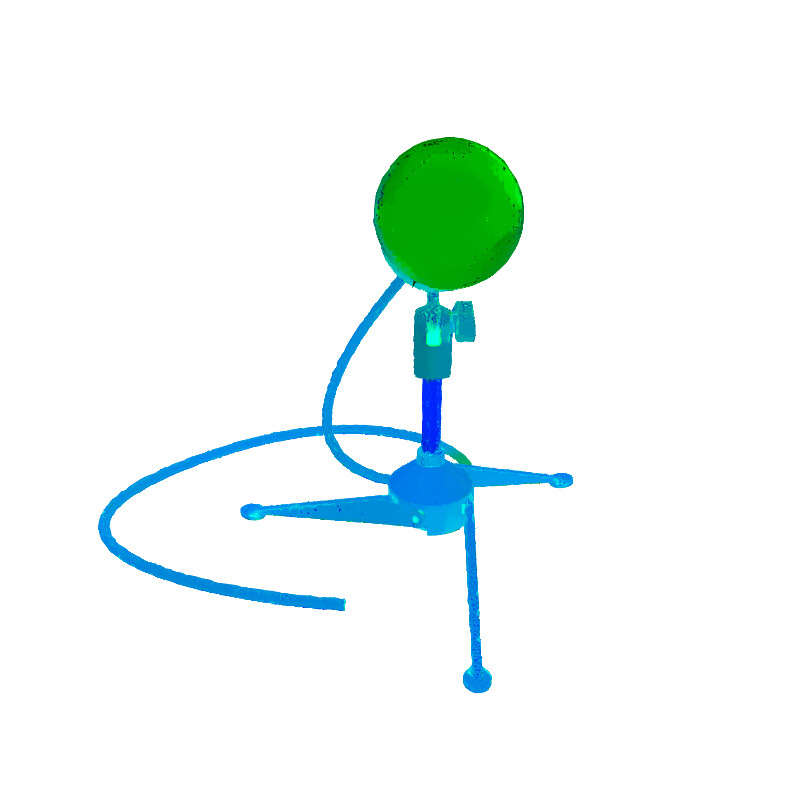}} &
			\raisebox{-0.5\height}{\includegraphics[width=\myfigsize]{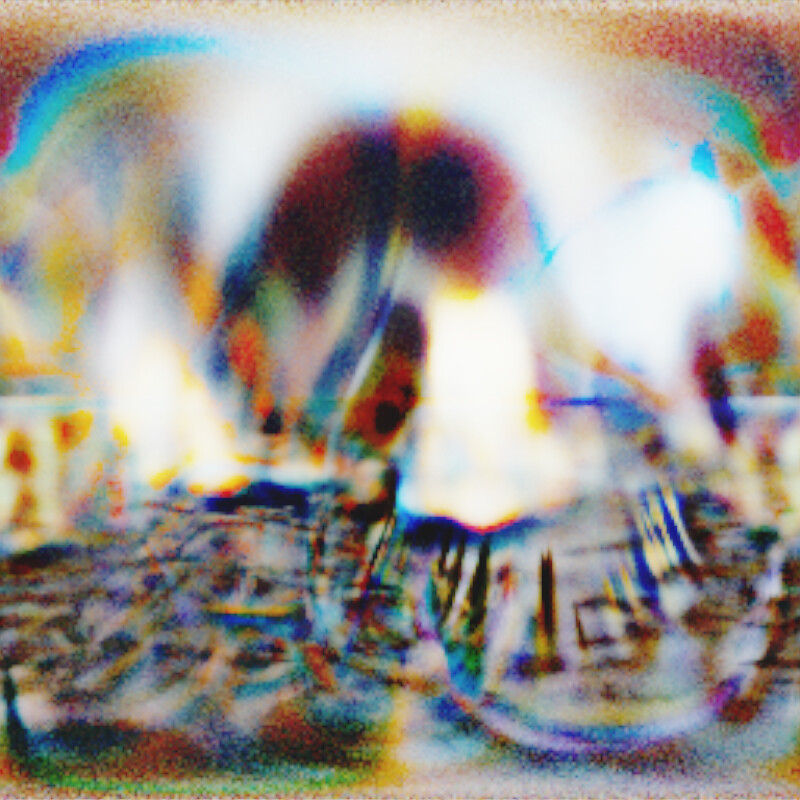}} \\

		\end{tabular}
		\caption{ Breakdown of shading terms with and without our light regularizer. Note that, for complex meshes (Hotdog and Lego), 
			shadows are more accurately baked into lighting terms, effectively delighting the $\kd$ textures. For scenes with simple visibility 
			(Chair, Materials, Microphone), the regularizer arguably contrast enhance the light probes too much, but we still note 
			smoother $\kd$. }
		\label{fig:regablation}
	\end{figure}
}

\maketitle

\begin{abstract}
  Recent advances in differentiable rendering have enabled high-quality reconstruction of 3D scenes from multi-view images. 
  Most methods rely on simple rendering algorithms: pre-filtered direct lighting or learned representations 
  of irradiance. We show that a more realistic shading model, incorporating ray tracing and Monte Carlo integration, 
  substantially improves decomposition into shape, materials \& lighting.
  Unfortunately, Monte Carlo integration provides estimates with significant noise, even at large sample counts,
  which makes gradient-based inverse rendering very challenging.  
  To address this, we incorporate multiple importance sampling and denoising in a novel inverse rendering pipeline. 
  This improves convergence and enables gradient-based optimization at low sample counts.
  We present an efficient method to jointly reconstruct geometry (explicit triangle meshes), materials, and lighting,
  which substantially improves material and light separation compared to previous work. We argue that denoising 
  can become an integral part of high quality inverse rendering pipelines. 
\end{abstract}

\let\thefootnote\relax\footnotetext{Project page: \color{magenta}\url{https://nvlabs.github.io/nvdiffrecmc/}}

\section{Introduction}
Differentiable rendering shows great promise for accurate multi-view 3D reconstruction from 
image observations. 
NeRF~\cite{Mildenhall2020} use differentiable volume rendering to create high quality view interpolation 
through neural, density-based light-fields. Surface-based
methods apply signed distance fields~\cite{Niemeyer2020CVPR,Oechsle2021,yariv2020multiview,Wang2021neus,Zhang2020physg} 
or triangle meshes~\cite{munkberg2021nvdiffrec, Shen2021} to capture high quality geometry.
Recent work~\cite{Boss2021,munkberg2021nvdiffrec, Zhang2021nerfactor} further 
decompose these representations into geometry, material, and environment light.

Most aforementioned methods rely on appearance baked into neural light fields or apply 
simple shading models. Typically, direct lighting without shadows is considered, combined with pre-filtered
representations of environment lighting~\cite{munkberg2021nvdiffrec,Zhang2020physg}. 
Some methods account for shadowing and indirect illumination~\cite{Boss2021,Zhang2021nerfactor,Zhang2022}, 
but often lock the geometry optimization when sampling the shadow term, or rely on learned representations
of irradiance. While results are impressive, the deviations from physically-based shading models makes
it harder for these methods to plausibly disentangle shape, material and lighting, as shown  
in Figure~\ref{fig:roller}.

\figRoller
In theory, it is straightforward to replace the rendering engines of these 3D reconstruction methods
with photorealistic differentiable renderers~\cite{Li2018, Loubet2019Reparameterizing, NimierDavid2019, Nimier2020, Zeltner2021,Zhang2020}
and optimize in a setting with more accurate light simulation, including global illumination effects. 
In practice, however, the noise in multi-bounce Monte Carlo rendering makes gradient-based optimization
challenging. Very high sample counts are required, which result in intractable iteration times.

In this paper, we bridge the gap between current multi-view 3D reconstruction and physically-based
differentiable rendering, and demonstrate high-quality reconstructions at competitive runtime performance. 
We attack the challenging case of extracting \emph{explicit triangle meshes}, PBR materials and environment
lighting from a set of multi-view images, in a format directly compatible with current DCC tools and game engines.
For improved visual fidelity, we compute direct illumination using Monte Carlo integration with ray tracing, 
and add several techniques to combat the increased noise levels. By carefully trading variance for bias,
we enable efficient gradient-based optimization in a physically-based inverse rendering pipeline.
Compared to previous 3D reconstruction methods, our formulation primarily improves material and light separation. 

Concretely, we reduce variance by combining multiple importance sampling~\cite{Veach1995} 
and differentiable denoisers. We evaluate both neural denoisers~\cite{Bako2017,Chaitanya2017,Hasselgren2020}
and cross-bilateral filters~\cite{Schied2017} in our pipeline. 
Furthermore, we decompose the rendering equation into albedo, demodulated diffuse lighting and specular lighting,
which enable precise regularization to improve light and material separation.
 
\section{Previous Work}
\paragraph{Neural methods for multi-view reconstruction} These methods fall in 
two categories: \emph{implicit} or \emph{explicit} scene representations. NeRF~\cite{Mildenhall2020} and follow-ups~\cite{martin2021nerf,niemeyer2021giraffe,pumarola2021d,kaizhang2020,wang2021nerfmm,garbin2021fastnerf,mueller2022instant,Reiser2021,yu2021plenoctrees,Wizadwongsa2021NeX}, use volumetric representations and compute radiance by ray marching through a neurally encoded 5D light field.
While achieving impressive results on novel view synthesis, geometric quality suffers from the ambiguity of volume 
rendering~\cite{kaizhang2020}. Surface-based rendering methods~\cite{Niemeyer2020CVPR, Oechsle2021, yariv2020multiview, Wang2021neus} 
optimizing the underlying surface directly using implicit differentiation, or gradually morph from a volumetric representation
into a surface representation. 
Methods with explicit representation estimate 3D meshes from images, where most approaches assume a given
mesh topology~\cite{softras,Chen2019dibrender,chen2021dibrpp}, 
but recent work also include topology optimization~\cite{Liao2018,Gao2020,Shen2021, munkberg2021nvdiffrec}.

\paragraph{BRDF and lighting estimation} To estimate surface radiometric properties from images, 
previous work on BTF and SVBRDF estimation rely on special viewing configurations, lighting patterns or complex capturing 
setups~\cite{hendrik2003planned,Gardner2003,Ghosh2009,Guarnera2016,Weinmann2015,bi2020neural,boss2020two,Schmitt2020CVPR,haindl_filip13visual}.
Recent methods exploit neural networks to predict BRDFs from images~\cite{Gao2019,Guo2020,li2020inverse,li2018learning,Merlin2021,Luan2021}.
Differentiable rendering methods~\cite{softras,Chen2019dibrender,StyleGAN3D,chen2021dibrpp,Hasselgren2021} learn to predict geometry, 
SVBRDF and, in some cases, lighting via photometric loss. 

Most related to our work are neural 3D reconstruction methods with \emph{intrinsic decomposition} 
of shape, materials, and lighting from images~\cite{Boss2021,boss2021neuralpil,munkberg2021nvdiffrec,Zhang2020physg,Zhang2021nerfactor,Zhang2022}.
Illumination is represented using mixtures of spherical Gaussians~\cite{Boss2021,munkberg2021nvdiffrec,Zhang2020physg,Zhang2022}, 
pre-filtered approximations~\cite{boss2021neuralpil,munkberg2021nvdiffrec}, or low resolution environment maps~\cite{Zhang2021nerfactor}. 
When the shadowing term is accounted for~\cite{Zhang2021nerfactor}, optimization is split into 
two passes where geometry is locked before the shadow term is sampled. Other approaches represent 
indirect illumination with neural networks~\cite{Wang2021learning,Zhang2022}. 

\paragraph{Image denoisers} Denoisers are essential tools in both real-time- and production renderers. 
Traditionally, variants of cross-bilateral filters are used~\cite{Zwicker2015}, 
which require scene-specific manual adjustments. More recently, neural denoisers~\cite{Bako2017, Chaitanya2017, Hasselgren2020} 
trained on large datasets have shown impressive quality without the need for manual tuning, and are now incorporated 
in most production renderers. We directly incorporate differentiable versions of these denoisers in our pipeline.
We are currently unaware of image denoisers applied in differentiable rendering, but we see a lot of potential
for denoisers in physically-based inverse rendering going forward.

\section{System}
We target the challenging task of joint optimization of shape, material and environment lighting from a 
set of multi-view images with known foreground segmentation masks and camera poses.
Our goal is to use physically-based rendering techniques to improve the intrinsic decomposition of lighting and materials, producing 
assets that can be relit, edited, animated, or used in simulation.
As a proof-of-concept, we extend a recent approach, \textsc{nvdiffrec}~\cite{munkberg2021nvdiffrec}, 
which targets the same optimization task (shape, materials and environment lighting).
Notably, they directly optimize a \emph{triangular} 3D model, which has obvious benefits: 
it is easy to import and modify the reconstructed models in existing DCC tools, and a triangular 
representation can exploit hardware-accelerated differentiable rasterization~\cite{Laine2020}. 
In our setting, triangular 3D models also means we can leverage hardware-accelerated 
ray-tracing for efficient shadow tests. \textsc{nvdiffrec} reports competitive results on view interpolation, 
material reconstruction, and relighting, and we will use their pipeline as a baseline
in our evaluations. 

\figSystem

Our system is summarized in 
Figure~\ref{fig:system}. A triangular mesh with arbitrary topology is optimized from a set of 
images through 2D supervision. Geometry is represented by a signed distance field
defined on a three-dimensional grid and reduced to a triangular surface mesh through deep marching tetrahedra
(DMTet)~\cite{Shen2021}. Next, the extracted surface mesh is rendered in a differentiable renderer,
using the physically-based (PBR) material model from Disney~\cite{Burley12}.
This material model combines a diffuse term with an isotropic, specular GGX lobe~\cite{Walter2007}. 
A tangent space normal map is also included to capture high frequency shading detail.
Finally, the rendered image is evaluated against a reference image using a photometric loss.
In contrast to \textsc{nvdiffrec}, which uses a simple renderer with deferred shading and the split-sum approximation
for direct lighting (without shadows), we instead leverage a renderer which evaluates the direct lighting integral 
using Monte Carlo integration and ray tracing (shadow rays). We represent the scene lighting using a high dynamic
range light probe stored as a floating point texture, typically at a resolution of 256$\times$256 texels. 
Finally, to combat the inherent variance that comes with Monte Carlo integration, we leverage
differentiable image denoising and multiple importance sampling.  

\paragraph{Optimization task}
Let $\phi$ denote the optimization parameters (shape, spatially varying materials and light probe).
For a given camera pose, $c$, our differentiable renderer produces an 
image $I_{\phi}(c)$. Given that we use Monte Carlo integration during 
rendering, this image inherently includes noise, and we apply a differentiable 
image denoiser, $D_{\theta}$, with parameters, $\theta$, to reduce the variance, 
$I_{\phi}^{\mathrm{denoised}}(c) = D_{\theta}(I_{\phi}(c))$.  
The reference image $I_{\mathrm{ref}}(c)$ is a view from the same camera.
Given a photometric loss function $L$, we minimize the empirical risk
\begin{equation}
	\underset{\phi, \theta}{\mathrm{argmin}}\ \mathbb{E}_{c}\big[L\big(D_\theta(I_{\phi}(c)), I_{\mathrm{ref}}(c)\big)\big]
	\label{eq:optimization}
\end{equation}
using Adam~\cite{Kingma2014} based on gradients w.r.t.~the optimization parameters,
$\partial L/\partial\phi$, and $\partial L/\partial\theta$, which are obtained through differentiable rendering.
We use the same loss function as \textsc{nvdiffrec}.
An example of the optimization process is illustrated in Figure~\ref{fig:training}, 

\figTraining

\subsection{Direct Illumination}
The outgoing radiance $L(\omega_o)$ in direction $\omega_o$ can be expressed using the rendering equation~\cite{Kajiya1986} as:
\begin{equation}
	L(\omega_o) = \int_\Omega L_i(\omega_i)f(\omega_i,\omega_o) (\omega_i \cdot \mathbf{n}) d\omega_i.
    \label{eq:ibl}
\end{equation}
This is an integral of the product of the incident radiance, $L_i(\omega_i)$ from 
direction $\omega_i$ and the BSDF $ f(\omega_i, \omega_o)$. The integration domain is 
the hemisphere $\Omega$ around the surface normal, $\mathbf{n}$. 

Spherical Harmonics (SH)~\cite{Chen2019dibrender} or Spherical Gaussians (SG)~\cite{Boss2021, Zhang2020physg} 
are often used as efficient approximations of direct illumination, but only work well for low- to medium-frequency lighting. In contrast, the \emph{split sum} 
approximation~\cite{Karis2013,munkberg2021nvdiffrec} captures all-frequency image based lighting, but does not incorporate shadows.
Our goal is all-frequency lighting \emph{including} shadows, which we tackle by evaluating the rendering equation using Monte Carlo integration:
\begin{equation}
	L(\omega_o) \approx \frac{1}{N} \sum_{i=1}^{N} \frac{L_i(\omega_i)f(\omega_i,\omega_o) (\omega_i \cdot \mathbf{n})}{p(\omega_i)},
\end{equation}
with samples drawn from some distribution $p(\omega_i)$. Note that $L_i(\omega_i)$ includes a visibility test, which can be 
evaluated by tracing a shadow ray in direction $\omega_i$. 
Unfortunately, the variance levels in Monte Carlo integration with 
low number of samples makes gradient-based optimization hard, particularly with complex lighting. In Section~\ref{sec:variance} we propose 
several variance reduction techniques, which enable an inverse rendering pipeline that efficiently reconstructs complex geometry,
a wide range of lighting conditions and spatially-varying BSDFs.

\paragraph{Shadow gradients}
In single view optimization~\cite{Vicini2022}, shape from shadows~\cite{Tiwary2022}, or direct optimization of the
position/direction of analytical light sources, shadow ray visibility gradients~\cite{Loubet2019Reparameterizing,bangaru2020}
are highly beneficial. 
However, in our multi-view setting (50+ views), similar to Loubet et~al.~\cite{Loubet2019Reparameterizing}, 
we observed that gradients of diffuse scattering are negligible compared to the gradients of primary visibility. 
Hence, for performance reasons, in the experiment presented in this paper, 
the shadow ray visibility gradients are detached when evaluating the hemisphere integral, and shape optimization
is driven by primary visibility gradients, obtained from \texttt{nvdiffrast}~\cite{Laine2020}. 

\section{Variance Reduction}
\label{sec:variance}

We evaluate direct illumination with high frequency environment map lighting combined with a wide range of 
materials (diffuse, dielectrics, and metals). Strong directional sunlight, highly specular, mirror-like materials, 
and the visibility component can all introduce significant levels of noise. 
To enable optimization in an inverse rendering setting at practical sample counts,  
we carefully sample each of these contributing factors to obtain a signal with low variance.
Below, we describe how we combat noise by using multiple importance sampling and denoising.

\subsection{Multiple Importance Sampling}
We leverage \emph{multiple importance sampling}~\cite{Veach1995} (MIS), 
a framework to weigh a set of different sampling techniques to reduce variance
in Monte Carlo integration. 
Given a set of sampling techniques, each with a sampling distribution $p_i$, the 
Monte Carlo estimator for an integral $\int_\Omega g(x) dx$ given by MIS is
\begin{equation}
\sum_{i=1}^n \frac{1}{n_i} \sum_{j=1}^{n_i} w_i(X_{i,j})\frac{g(X_{i,j})}{p_i(X_{i,j})}, \ \ \ \ \  w_i(x) = \frac{n_i p_i(x)}{\sum_k n_k p_k(x)}.
\end{equation} 
The weighting functions $w_i(x)$ are chosen using the \emph{balance heuristic}. 
Please refer to Veach's thesis~\cite{Veach1995} or the excellent PBRT book~\cite{Pharr2010} for further details.  

In our case, we apply MIS with three sampling techniques: light importance sampling, 
$p_\mathrm{light}(\omega)$, using a piecewise-constant 2D distribution sampling technique~\cite{Pharr2010}, 
cosine sampling, $p_\mathrm{diffuse}(\omega)$, for the diffuse lobe, and 
GGX importance sampling~\cite{Heitz2018GGX}, $p_\mathrm{specular}(\omega)$, for the specular lobe.
Unlike in forward rendering with known materials and lights, our material and light parameters are optimization variables.
Thus, the sampling distributions, $p_i$, are recomputed in each optimization iteration.
Following the taxonomy of differentiable Monte Carlo estimators of Zeltner et~al.~\cite{Zeltner2021},
our importance sampling is \emph{detached}, i.e., we do not back-propagate gradients to the scene parameters in the 
sampling step, only in the material evaluation. Please
refer to Zeltner et~al. for a careful analysis of Monte Carlo estimators for differentiable light transport. 

MIS is unbiased, but chaining multiple iterations of our algorithm exhibits \emph{bias}, as the current, importance sampled, 
iteration dictates the sampling distributions used in the next iteration.
Light probe intensities, for example, are optimized based on an importance sampled image, 
which are then used to construct the sampling distribution for the subsequent pass. For unbiased rendering, 
the sampling distribution must be estimated using a second set of uncorrelated samples instead.
Furthermore, we explicitly re-use the random seed from the forward pass during gradient backpropagation 
to scatter gradients to the exact same set of parameters that contributed to the forward rendering.
This approach is clearly biased~\cite{vicini2021path}, but we empirically note that this is very 
effective in reducing variance, and in our setting this variance-bias trade-off works in our favor. 

\subsection{Denoising}
For differentiable rendering, the benefits of denoising are twofold. First, it improves the image quality of the rendering
in the forward pass, reducing the optimization error and the gradient noise introduced in the image loss. Second, as gradients 
back-propagate through the denoiser's spatial filter kernel, gradient sharing between neighboring pixels is enabled. 
To see this, let's consider a simple denoiser, $O = X \circledast F$, where the noisy rendered image $X$ is filtered 
by a low-pass filter, $F$ ($\circledast$ represents an image space 2D convolution). Given a loss gradient, $\frac{\partial L}{\partial O}$, 
the gradients propagated back to the renderer 
are $\frac{\partial L}{\partial X} = \frac{\partial L}{\partial O} \circledast F^T$, which applies the same low-pass filter in 
case the filter is rotationally symmetric ($F = F^T$).
In other words, the renderer sees \emph{filtered} loss gradients in a local spatial footprint.
While denoisers inherently trade a reduction in variance for increased bias, we empirically note that denoising 
significantly helps convergence at lower sample counts, and help to reconstruct higher frequency environment lighting.
We show an illustrative example in Figure~\ref{fig:porsche_denoising}.

\figSeparation

Following previous work in denoising for production rendering~\cite{Bako2017}, Figure~\ref{fig:light_sep} shows how we separate lighting into 
diffuse, $\mathbf{c}_d$, and specular, $\mathbf{c}_s$ terms. 
This lets us denoise each term separately, creating denoised buffers, 
$D_{\theta}(\mathbf{c}_d)$, and $ D_{\theta}(\mathbf{c}_s)$. More importantly, we can use \emph{demodulated} diffuse lighting, 
which means that the lighting term has not yet been multiplied by material diffuse albedo, $\mathbf{k}_d$. 
In forward rendering, this is important as it decorrelates the noisy lighting from material textures,
thus selectively denoising the noisy Monte-Carlo estimates and avoiding to low-pass filter high-frequent texture information.
In inverse rendering, we can additionally use it to improve material 
and light decomposition by adding regularization on the lighting terms, as disucussed in Section~\ref{sec:priors}.
We compose the final image as $\mathbf{c} = \mathbf{k}_d \cdot D_{\theta}(\mathbf{c}_d) + D_{\theta}(\mathbf{c}_s)$.
We currently do not demodulate specular lighting because of the view dependent Fresnel term, 
but expect this to be improved in future work.

\figAblationDenoisePlots

\paragraph{Cross-bilateral filters}
Cross-bilateral filters are commonly used to remove noise in rendered images~\cite{Zwicker2015}. 
To evaluate this family of denoisers in our inverse pipeline, 
we adapted Spatio-temporal Variance-Guided Filtering~\cite{Schied2017} (SVGF),
which is a popular cross-bilateral filter using surface normals and per-pixel depth as edge-stopping guides.
The denoiser is applied to \emph{demodulated} diffuse lighting, to avoid smearing texture detail.
We disabled the temporal component of SVGF, as we focus on single frame rendering, 
and implemented the filter as a differentiable module to allow for loss gradients to propagate back to
our scene parameters.

\paragraph{Neural denoisers}
As a representative neural denoiser, we deploy the Open Image Denoiser (OIDN)~\cite{Afra21},
which is a U-Net~\cite{Ronneberger2015} pre-trained on a large corpus of rendered images.
The denoiser is applied to the rendered image before computing the image space loss.
As the network model is fully convolutional, it is trivially differentiable, and we can propagate gradients from the 
image space loss, through the denoiser back to the renderer. 

In Figure~\ref{fig:ablation-denoise-plots}, we provide an overview on the effect of denoisers during 3D scene reconstruction.
We observe that denoising is especially helpful at low sample counts, where we obtain similar reconstruction results at 8~spp with denoising, 
compared to 32~spp without a denoiser. At higher sample counts, however, the benefit from denoising diminishes, as variance in the Monte-Carlo estimates 
decreases.
Given that denoisers enable significantly faster iteration times, 
we consider it a valuable tool for saving computational
resources when fine-tuning model parameters for subsequent runs with high sample counts.
Additionally, we empirically found that using a denoiser during reconstruction yields higher quality light probes, as can be seen in
Figure~\ref{fig:porsche_denoising}, both at low and high sample counts.

We can also jointly optimize the denoiser parameters, $\theta$, with the 3D scene reconstruction task. i.e., 
the denoiser is \emph{fine-tuned} for the current scene.
Unfortunately, this approach has undesirable side-effects:
Features tend to get baked into the denoiser network weights instead of the materials or light probe.
This is especially apparent with the OIDN~\cite{Afra21} denoiser, which produced color shifts due to lack of regularization on the output.
We got notably better results with the hierarchical kernel prediction architecture from Hasselgren et~al.~\cite{Hasselgren2020}, which is more constrained. 
However, the results still lagged behind denoisers with locked weights. We refer to the supplemental material for details.

\section{Priors}
\label{sec:priors}
In our setting: 3D reconstruction from multi-view images with constant lighting, regularization is essential in order to disentangle
lighting and materials. Following previous work~\cite{Zhang2021nerfactor, munkberg2021nvdiffrec}, we apply
smoothness priors for albedo, specular, and normal map textures. Taking the albedo as an example, if $\kd\left(\mathbf{x}\right)$ 
denotes the diffuse albedo at world space position, $\mathbf{x}$,
and $\mathbf{\epsilon}$ is a small random displacement vector, we define the smoothness prior for albedo as:
\begin{equation}
	L_{\kd} = \sum_{\mathbf{x}_\text{surf}} \left|\kd\left(\mathbf{x}_\text{surf}\right) - \kd\left(\mathbf{x}_\text{surf} + \mathbf{\epsilon}\right)\right|,
\end{equation}
where $\mathbf{x}_\text{surf}$ are the world space positions at the primary hit point on the object. 
We note that the smoothness prior is not sufficient to disentangle material parameters and light, especially for scenes with 
high frequency lighting and sharp shadows. Optimization tends to bake shadows into the albedo 
texture (easy) rather than reconstruct a high intensity, small area in the environment map (hard). 
To enable high quality relighting, we explicitly want to enforce shading detail represented by lighting, and only bake remaining details into 
the material textures. We propose a novel regularizer term that is surprisingly effective. 
We compute a monochrome image loss between the demodulated lighting terms and the reference image: 
\begin{equation}
	L_\text{light} = |\mathrm{Y}\left(\mathbf{c}_d + \mathbf{c}_s\right) - \mathrm{V}\left(I_{\mathrm{ref}}\right)|.
\end{equation}
Here, $Y\left(\mathbf{x}\right) = \left(\mathbf{x}_r + \mathbf{x}_g + \mathbf{x}_b\right) / 3$ is a simple 
luminance operator, and $V\left(\mathbf{x}\right) = \max\left(\mathbf{x}_r, \mathbf{x}_g, \mathbf{x}_b\right)$ is the HSV value
component. The rationale for using HSV-value for the reference image is that the $\max$ operation approximates demodulation, e.g., 
a red and white pixel have identical values.  We assume that the demodulated lighting is mostly monochrome, in which case 
$Y\left(\mathbf{x}\right) \sim V\left(\mathbf{x}\right)$, and given that we need to propagate gradients to
$\mathbf{c}_d$ and $\mathbf{c}_s$, $Y$ avoids discontinuities. 
This regularizer is limited by our inability to demodulate the reference image. The HSV-value ignores chrominance, but we cannot 
separate a shadow from a darker material. This has not been a problem in our tests, but could interfere with optimization if 
the regularizer is given too much weight. Please refer to the supplemental materials for complete regularizer details.

\section{Experiments}
\label{sec:results}

\begin{table}[b]
	\caption{
		Summarized relighting results for NeRFactor (CC-BY-3.0), NeRF (CC-BY-3.0) and our synthetic datasets. The NeRFactor dataset contains four scenes, each scene  
		has eight validation views and eight different light probes (256 validation images). 
		For the NeRF dataset (which contain higher frequency lighting), 
		we use the \texttt{Chair}, \texttt{Hotdog}, \texttt{Lego}, \texttt{Materials} and \texttt{Mic} scenes, with eight validation views and 
		four light probe configurations (160 validation images). Our dataset contains a variation of high and low frequency lighting with 
		geometrically complex objects. The image metric scores are arithmetic means over all images.
	}
	\label{tab:relighting}
	\centering
	{
	\setlength{\tabcolsep}{1mm}	
	\begin{tabular}{lccccccccc}
		\toprule
		& \multicolumn{3}{c}{\textbf{NeRFactor synthetic}} & \multicolumn{3}{c}{\textbf{Nerf synthetic}}  & \multicolumn{3}{c}{\textbf{Our synthetic}} \\
		& \multicolumn{1}{c}{PSNR$\uparrow$} & \multicolumn{1}{c}{SSIM$\uparrow$} & LPIPS$\downarrow$ & 
		\multicolumn{1}{c}{PSNR$\uparrow$} & \multicolumn{1}{c}{SSIM$\uparrow$} & LPIPS$\downarrow$ & 
		\multicolumn{1}{c}{PSNR$\uparrow$} & \multicolumn{1}{c}{SSIM$\uparrow$} & LPIPS$\downarrow$ \\
		\midrule
		Our                & 26.0~dB & 0.924 & 0.060 & 26.5~dB & 0.932 & 0.055 & 27.1~dB & 0.950 & 0.027\\
		\textsc{nvdiffrec} & 24.8~dB & 0.910 & 0.063 & 23.3~dB & 0.889 & 0.076 & 23.7~dB & 0.925 & 0.049\\
		\textsc{NeRFactor} & 22.2~dB & 0.896 & 0.087 &   -     &  -    &   -   & - & - & - \\
		\bottomrule
	\end{tabular}
	}
\end{table}

\figRelight

In our experiments, we use \textsc{nvdiffrec}~\cite{munkberg2021nvdiffrec} as a baseline, 
and refer to their work for thorough comparisons against related work. 
We focus the evaluation on the quality of material and light separation. 
At test time, all view interpolation results are generated without denoising at 2k~spp. 
All relighting results are rendered in Blender Cycles at 64~spp with
denoising~\cite{Afra21}.
Table~\ref{tab:relighting} shows a quantitative comparison with \textsc{nvdiffrec} and 
NeRFactor~\cite{Zhang2021nerfactor} on the NeRFactor relighting setup.
Note that there is an indeterminate scale factor between material reflectance (e.g., albedo) and the light 
intensity. This is accounted for by scaling each image to match the average luminance of the reference for the 
corresponding scene. The same methodology is applied for all algorithms in our comparisons.  We outperform 
previous work, providing better material reconstruction. Figure~\ref{fig:relight_nerf_collage} shows visual examples.

\figSceneEdit

We additionally perform relighting on the synthetic NeRF dataset, which is substantially more challenging than the NeRFactor variant,
due to high frequency lighting and global illumination effects. \textsc{nvdiffrec} produces severe artifacts, as exemplified by the
Hotdog scene in Figure~\ref{fig:relight_nerf_collage}. Table~\ref{tab:relighting} 
shows a significant increase in image quality for our approach. The visual examples show that our results are plausible, though not without artifacts.
Finally, we constructed a novel synthetic dataset with three scenes with highly complex geometry to stress-test the system. Each scene
contains 200 training views and 200 novel views for evaluation.
The three scenes are shown in Figures~\ref{fig:roller}~,~\ref{fig:training}~,and~\ref{fig:porsche_denoising}. 
Quantitatively we outperform previous work by a larger margin, and Figure~\ref{fig:roller} shows very little shading in the albedo textures.

\figPhotogrammetry

In Figures~\ref{fig:scene_edit}~and~\ref{fig:nsvf}, we apply our method to datasets with real photos. These sets are more difficult, due to 
inaccuracies in foreground segmentation masks and camera poses. We extract triangle meshes that 
can be trivially edited in 3D modeling software, and in Figure~\ref{fig:scene_edit} we use Blender to perform material 
editing and relighting with convincing results. 
Figure~\ref{fig:nsvf} shows a breakdown of geometry, material parameters, and environment light. 

\figPorscheDenoising

To study the impact of denoising, we optimized the \texttt{Porsche} scene w/ and w/o denoising. 
As shown in Figure~\ref{fig:porsche_denoising}, denoising improves both visual quality and environment lighting detail 
at equal sample counts. The noise levels varies throughout the scene, and we note that denoising is particularly 
helpful in regions with complex lighting or occlusion, such as the specular highlight and cockpit.

\begin{table}[tb]
	\caption{
		View interpolation results. 
		For reference, the \textsc{NeRFactor} scores are 26.9~dB PSNR and SSIM of 0.930 on the NerFactor synthetic dataset, 
		and Mip-NeRF has 35.0~dB PSNR and SSIM 0.978 on the Nerf synthetic dataset.
		The image metric scores are arithmetic means over all test images.
	}
	\label{tab:view_interpolation}
	\centering
	{
		\setlength{\tabcolsep}{1mm}	
		\begin{tabular}{lcccccccc}
			\toprule
			& \multicolumn{2}{c}{\textbf{NeRFactor synthetic}} 
			& \multicolumn{2}{c}{\textbf{Nerf synthetic}}  
			& \multicolumn{2}{c}{\textbf{Our synthetic}} 
			& \multicolumn{2}{c}{\textbf{Real-world}} \\
			& \multicolumn{1}{c}{PSNR$\uparrow$} & \multicolumn{1}{c}{SSIM$\uparrow$} 
			& \multicolumn{1}{c}{PSNR$\uparrow$} & \multicolumn{1}{c}{SSIM$\uparrow$}
			& \multicolumn{1}{c}{PSNR$\uparrow$} & \multicolumn{1}{c}{SSIM$\uparrow$}
			& \multicolumn{1}{c}{PSNR$\uparrow$} & \multicolumn{1}{c}{SSIM$\uparrow$} 
			\\
			\midrule
			Our                & 29.6~dB & 0.951 & 28.4~dB & 0.938 & 25.6~dB & 0.934 & 25.29~dB & 0.899 \\
			\textsc{nvdiffrec} & 31.7~dB & 0.967 & 30.4~dB & 0.958 & 25.8~dB & 0.944 & 26.58~dB & 0.918 \\
			\bottomrule
		\end{tabular}
	}
\end{table}

Neural light-fields, e.g., Mip-NeRF~\cite{barron2021mipnerf} excel at view interpolation. 
We enforce material/light separation through additional regularization, which slightly degrades 
view interpolation results, as shown in Table~\ref{tab:view_interpolation}. 
Our scores are slightly below \textsc{nvdiffrec}, but, as shown above, we provide considerably 
better material and lighting separation.

\paragraph{Compute resources} 
Tracing rays to evaluate the direct illumination is considerably more expensive than 
pre-filtered environment light approaches. We leverage hardware-accelerated ray intersections,
but note that our implementation is far from fully optimized.
Our method scales linearly with sample count, which gives us a simple way to trade quality 
for performance. With a batch size 8 at a rendering resolution of 512$\times$512, we get the following iteration times 
on a single NVIDIA A6000 GPU.
\begin{center} 
\begin{tabular}{lcccccc}
    \toprule
    & \small{\textsc{nvdiffrec}} & \small{Our 2~spp} & \small{Our 8~spp} & \small{Our 32~spp} & \small{Our 128~spp} & \small{Our 288~spp} \\ 
    \midrule
    Iteration time  & 340~ms & 280~ms & 285~ms & 300~ms & 360~ms & 450~ms \\ 
    \bottomrule
\end{tabular}
\end{center}
Our denoising strategies enable optimization at low sample counts. 
Unless otherwise mentioned, for the results presented in the paper, we use high quality settings 
of 128+ rays per pixel with 5000$\times$2 iterations (second pass with fixed topology and 2D textures),  
which takes $\sim$4 hours (A6000).

\section{Conclusions}
\label{sec:conclusions}

Our restriction to direct illumination shows up in scenes with global illumination. 
Similarly, we do not handle specular chains (geometry seen through a glass window). 
For this, we need to integrate multi-bounce path tracing, which is a clear avenue for future work, but comes
with additional challenges in increased noise-levels, visibility gradients through specular chains, and drastically increased
iteration times. 
Our renderer is intentionally biased to improve optimization times, but unbiased rendering could expect to generate better results for very
high sample counts. Other limitations include lack of efficient regularization of material specular parameters and reliance on a foreground
segmentation mask. Our approach is computationally intense, requiring a high-end GPU for optimization runs.

To summarize, we have shown that differentiable Monte-Carlo rendering combined with variance-reduction techniques is practical and applicable
to multi-view 3D object reconstruction of explicit triangular 3D models. Our physically-based renderer clearly improves material and light
reconstruction over previous work. By leveraging hardware accelerated ray-tracing and differentiable image denoisers, we remain competitive 
to previous work in terms of optimization time.

{\small
\bibliographystyle{ieee_fullname}
\bibliography{paper}
}

\pagebreak
\appendix

\section{Optimization and Regularization}
\label{sec:optimization}

\paragraph{Image Loss} Our renderer uses physically-based shading and produces images with high dynamic range.
Therefore, the objective function must be robust to the full range of floating-point values.
Following recent work in differentiable rendering~\cite{Hasselgren2021,munkberg2021nvdiffrec}, 
our image space loss, $L_\mathrm{image}$, computes the $L_1$ norm on tonemapped colors. As tone map operator, $T$,
we transform linear radiance values, $x$, according to $T(x) = \Gamma(\log(x + 1))$,
where $\Gamma(x)$ is the sRGB transfer function~\cite{srgb96}:
\begin{eqnarray}
	\Gamma(x) &=&  
	\begin{cases}
		12.92x & x \leq 0.0031308 \\
		(1+a)x^{1/2.4} -a & x > 0.0031308
	\end{cases} \\ 
	a &=& 0.055.  \nonumber
\end{eqnarray}

\paragraph{Regularizers} While it is desirable to minimize regularization, in our setting with multi-view images with 
constant lighting, we have to rely on several priors to guide optimization towards a result with a good separation
of geometry, materials, and lighting. As mentioned in the paper, 
we rely on smoothness regularizers for albedo, $\kd$, specular parameters, $\korm$ and geometric surface normal, 
$\mathbf{n}$, following:
\begin{equation}
	L_{\kd} = \frac{1}{\left|\mathbf{x}_\text{surf}\right|}\sum_{\mathbf{x}_\text{surf}} \left|\kd\left(\mathbf{x}_\text{surf}\right) - \kd\left(\mathbf{x}_\text{surf} + \bm{\epsilon}\right)\right|,
\end{equation}
where $\mathbf{x}_\text{surf}$ is a world space position on the surface of the object
and $\bm{\epsilon} \sim \mathcal{N}(0,\sigma\!\!=\!\!0.01)$ is a small random displacement vector.
Regularizing the geometric surface normal (before normal map perturbation) is novel compared to \textsc{nvdiffrec} 
and helps enforce smoother geometry, particularly in early training.

We note that normal mapping (surface normals perturbed through a texture lookup) also benefits from 
regularization. While normal mapping is a powerful tool for simulating local micro-geometry, the decorrelation
of the geometry and surface normal can be problematic. In some optimization runs, we observe that 
with normal mapping enabled, the environment light is (incorrectly) leveraged as a color dictionary,
where the normal perturbation re-orients the normals of a geometric element to look up a desired color. 
To prevent this behavior, we use the following regularizer.
Given a normal perturbation $\mathbf{n}'$ in tangent space, our loss is defined as
\begin{equation}
	L_{\mathbf{n}'} = \frac{1}{\left|\mathbf{x}_\text{surf}\right|}\sum_{\mathbf{x}_\text{surf}} 
	1 - 
	\underbrace{
		\frac{\mathbf{n}'\left(\mathbf{x}_\text{surf}\right) + \mathbf{n}'\left(\mathbf{x}_\text{surf} + \bm{\epsilon}\right)}
		{\left|\mathbf{n}'\left(\mathbf{x}_\text{surf}\right) + \mathbf{n}'\left(\mathbf{x}_\text{surf} + \bm{\epsilon}\right)\right|}
	}_{\textrm{Half-angle vector}}
	\cdot
	(0,0,1).
	\label{eq:normalsmooth}	
\end{equation}
Intuitively, we enforce that normal perturbations, modeling micro-geometry, randomly selected in a small local area,
have an expected value of the unperturbed tangent space surface normal, $(0,0,1)$. 

\figLightReg

As mentioned in the paper, we additionally regularize based on monochrome image loss between the demodulated lighting terms 
and the reference image: 
\begin{equation}
	L_\text{light} = |Y\left(T(\mathbf{c}_d + \mathbf{c}_s)\right) - V\left(T(I_{\mathrm{ref}})\right)|_1.
\end{equation}
Here, $T$ is the tonemap operator described in the previous paragraph.
$\mathbf{c}_d$ is demodulated diffuse lighting, $\mathbf{c}_s$ is specular lighting, and $I_{\mathrm{ref}}$ is the target 
reference image. Monochrome images are computed through the simple luminance operator, 
$Y\left(\mathbf{x}\right) = \left(\mathbf{x}_r + \mathbf{x}_g + \mathbf{x}_b\right) / 3$, 
and HSV-value, $V\left(\mathbf{x}\right) = \max\left(\mathbf{x}_r, \mathbf{x}_g, \mathbf{x}_b\right)$. While the 
regularizer is limited by our inability to demodulate the reference image, we show in 
Figure~\ref{fig:lighting_reg} that it greatly increases lighting detail, which is particularly effective in datasets 
with high frequency lighting.  Figure~\ref{fig:regablation} shows the impact of the regularizer on a larger set of scenes.

We compute the final loss as a weighted combination of the image loss and regularizer terms:
\begin{equation}
	L = L_\text{image} 
	+ \underbrace{\lambda_{\kd}}_{=0.1}L_{\kd} 
	+ \underbrace{\lambda_{\korm}}_{=0.05}L_{\korm} 
	+ \underbrace{\lambda_{\mathbf{n}}}_{=0.025}L_{\mathbf{n}} 
	+ \underbrace{\lambda_{\mathbf{n}'}}_{=0.25}L_{\mathbf{n}'} 
	+ \underbrace{\lambda_\text{light}}_{=0.15}L_\text{light},
\end{equation}
where we show the default weights used for most results (unless otherwise noted).

\paragraph{Optimization details} We use Adam~\cite{Kingma2014} (default settings) to optimize parameters for geometry, material, 
and lighting. Referring to DMTet~\cite{Shen2021}, geometry is parameterized as SDF values on a three-dimensional grid, with a perturbation 
vector per grid-vertex. Material and lighting are encoded in textures, or neural network (e.g., an MLP with positional encoding) 
encoded high-frequency functions. We typically use different learning rates for geometry, material, and lighting, with lighting having 
the highest learning rate and material the lowest. We closely follow the publically available \textsc{nvdiffrec} code base and refer to that for 
details. In Figure~\ref{fig:probes}, we visualize two examples of optimizing the light probe, and note that 
we can capture high-frequency lighting details for specular objects.

\figProbes

We note that the early phases of optimization (starting from random geometry with large topology changes) 
are crucial in producing a good final result. 
The shadow test poses a challenge since  changes in geometry may drastically change the lighting intensity in a previously shadowed 
area, causing the optimization process to get stuck in bad local minima. This is shown in Figure~\ref{fig:shadow_ramp}, where the 
optimizer fails to carve out the shape of the object. 
We combat this by incrementally blending in the shadow term as optimization progresses. 

Recall from the main paper that we compute the color according to the rendering equation: 
\begin{equation}
	L(\omega_o) = \int_\Omega L_i(\omega_i)f(\omega_i,\omega_o) (\omega_i \cdot \mathbf{n}) d\omega_i,
	\label{eq:ibl}
\end{equation}
where $L_i(\omega_i) = L_i'(\omega_i)H(\omega_i)$ can be separated into a lighting term, $L_i'(\omega_i)$, and visibility term, $H(\omega_i)$. 
Rather than using binary visibility, we introduce a light 
leakage term, $\tau$, and linearly fade in the shadow contributions over the first 1750 iterations:
\begin{eqnarray}
	H(\omega_i, \tau) &=&  
	\begin{cases}
		1-\tau & \text{if intersect\_ray}(\omega_i)\\
		1 & \text{otherwise}
	\end{cases}. 
\end{eqnarray}

\figShadowRamp

In Figure~\ref{fig:shadow_ramp} we note that gradually blending in the shadow term has a large impact on early convergence. 
In particular, since we start from a random topology, carving out empty space or adding geometry may
have a large impact on overall shading, causing spiky and noisy gradients. 

The denoiser may also interfere with early topology optimization (blurred visibility gradients). 
This is particularly prominent when the denoiser parameters are trained along with scene parameters. 
Therefore, we similarly, ramp up the spatial footprint, $\sigma$, in the case of a bilateral 
denoiser. For neural denoisers, which have no easily configurable filter width, we instead linearly blend between the noisy 
and denoised images to create a smooth progression.
\section{Denoising}
\label{sec:denoiser}

\paragraph{Bilateral Denoiser}
As a representative bilateral denoiser, we adapt the spatial component of 
Spatiotemporal Variance-guided Filtering~\cite{Schied2017} (SVGF), Our filter is given by
\begin{equation}
	\hat{\mathbf{c}}\left(p\right) = 
	\frac{
		\sum_{q \in \Omega} \mathbf{c}\left(q\right) \cdot w\left(p, q\right)
	}{
		\sum_{q \in \Omega} w\left(p, q\right)
	},
\end{equation}
where $\mathbf{c}$ are the pixel colors, $w\left(p, q\right)$ are the bilateral weights between pixels $p$ and $q$, and $\Omega$ denotes 
the filter footprint. The bilateral weight is split into three components as follows
\begin{equation}
	w\left(p, q\right) = 
	\underbrace{
		\vphantom{\max(0, \mathbf{n}\left(p\right) \cdot \mathbf{n}\left(q\right))^{\sigma_\mathbf{n}}}
		e^{-\frac{|p - q|^2}{2\sigma^2}}
	}_{w}
	\underbrace{
		\vphantom{\max(0, \mathbf{n}\left(p\right) \cdot \mathbf{n}\left(q\right))^{\sigma_\mathbf{n}}}
		e^{-\frac{\left|z\left(p\right) - z\left(q\right)\right|}{\sigma_{z}\left|\nabla z\left(p\right) \cdot (p - q)\right|}}
	}_{w_{z}}
	\underbrace{
		\max(0, \mathbf{n}\left(p\right) \cdot \mathbf{n}\left(q\right))^{\sigma_\mathbf{n}}
	}_{w_\mathbf{n}}.
\end{equation}
The spatial component, $w$, is a Gaussian filter with its footprint controlled by $\sigma$. 
We linearly increase $\sigma$ from 1e-4 to 2.0 over the first 1750 iterations. 
The depth component, $w_z$ implements an edge 
stopping filter based on image space depth. It is a soft test, comparing the depth differences between pixels 
$z(p)$ and $z(q)$ with a linear prediction based on the local depth gradient $\nabla z(p)$.  
The final component, $w_\mathbf{n}$, is computed as the scalar product over surface normals $\mathbf{n}(p)$ and $\mathbf{n}(q)$. 
Following SVGF, we use $\sigma_z=1$ and $\sigma_\mathbf{n} = 128$.
We also omitted the \'A-Trous wavelet filtering of SVGF, which is primarily a run-time performance optimization,
and evaluate the filter densely in the local footprint. We propagate gradients back to 
the renderer, i.e., the noisy input colors, but no gradients to the filter weights.

\figAblationDenoisePlotsLiveTrain
\figAblationDenoiseLight
\figAblationDenoiseGeom

\paragraph{Neural denoisers} For OIDN, we use the pre-trained weights from the official code~\cite{Afra21}, and 
wrap the model (a direct-prediction U-Net) in a PyTorch module to allow for gradient propagation to the input colors. 
We also propagate gradients to the network weights, in order to fine-tune the denoiser per-scene,
but unfortunately, the direct-prediction architecture of OIDN made live-training highly unstable in our setting.

For the kernel-predicting neural denoiser, we leverage the architecture from Hasselgren et~al.~\cite{Hasselgren2020}, 
but omit the adaptive sampling part and  the recurrent feedback loop. The network weights were initialized to random 
values using Xavier initialization. The hierarchical kernel prediction in this architecture constrains the denoiser, 
and enables live-trained neural denoising in our setting. That said, in our evaluations, the live-trained variant
produce worse reconstructions than the pre-trained denoisers.

We study the impact of different denoising algorithms on scene reconstruction, including 
jointly optimizing the denoising network, in Figure~\ref{fig:ablation-denoise-plots-live}.
Unfortunately, when fine-tuning the denoiser to the specific scene as part of optimization, we note a tendency 
for features to get baked into the denoiser's network weights instead of the desired scene parameters (overfitting), 
negatively impacting the reconstruction quality. 
We showcase the isolated effect of denoising on light probe optimization in Figure~\ref{fig:ablation-denoise-light}.
Figure~\ref{fig:ablation-denoise-geom} illustrates the effect of denoising when jointly optimizing geometry, material and lighting.

\figShadowGradients
\figShadowGradientsPorsche

\section{Visibility Gradients from Shadows}

In Figure~\ref{fig:shadow_gradients} we show the impact of the visibility gradients~\cite{Loubet2019Reparameterizing,bangaru2020}
for the shadow test on a few targeted reconstruction examples. In this ablation, we disabled denoising, and applied a 
full shadow term directly (instead of linearly ramping up the shadow contribution, as discussed in Section~\ref{sec:optimization}). 
While clearly beneficial in the simpler examples, e.g., single-view reconstruction and 
targeted optimization (finding the light position), for multi-view reconstruction with 
joint shape, material, and environment light optimization, the benefits of the shadow visibility gradients is less  
clear. In our experiments, gradients from the shadow rays in the evaluation of direct lighting in the hemisphere,
typically have negligible impact compared to gradients from primary visibility.
Below, we report the view-interpolation scores over the validation sets for three scenes.
\begin{center} 
	\begin{tabularx}{\textwidth}{lYYY@{\hskip 20mm}YYY}
		\toprule
		& \multicolumn{3}{c}{PSNR$\uparrow$} & \multicolumn{3}{c}{SSIM$\uparrow$} \\
		& \small{Chair} & \small{Lego} & \small{Porsche} & \small{Chair} & \small{Lego} & \small{Porsche} \\
		\midrule
		w/  gradients  & 29.15		  & 27.03		 & 27.83           &  0.948        &  0.918       &    0.941        \\
		w/o gradients  & 29.06		  & 26.83		 & 28.21           &  0.948        &  0.918       &    0.945        \\  
		\bottomrule
	\end{tabularx}
\end{center}
In contrast, we observed that the shadow ray gradients sometimes increase noise levels and degrade reconstruction quality.
We show a visual example in Figure~\ref{fig:shadow_gradients_porsche}.

\section{Sampling}

\figAblationMIS

In Figure~\ref{tab:ablation-MIS}, we examine the effect of importance sampling strategies on reconstruction quality.
We compare cosine sampling, BSDF importance sampling~\cite{Heitz2018GGX}, light probe importance sampling (using piecewise-constant 2D distribution sampling~\cite{Pharr2010}), 
and multiple importance sampling~\cite{Veach1995} (MIS) using the balance heuristic. 
MIS consistently generates high quality reconstruction for a wide range of materials.

\figAblationCorrelationPlot

For unbiased results, the forward and backward passes of the renderer should use independent (\emph{decorrelated}) random samples.
However, we observe drastically reduced noise levels, hence improved convergence rates, when using \emph{correlated} samples,
i.e., re-using the same random seeds in the forward and backward pass, especially for low sample count optimization runs.
This effect mainly stems from gradients flowing back to the exact same set of parameters which contributed to the forward rendering, 
instead of randomly selecting another set of parameters, which naturally increases variance.
In Figure~\ref{fig:ablation-correlation-plot}, we compare the two approaches over a range of sample counts.
\figMeshesAO

\section{Geometry}

In Table~\ref{tab:chamfer} we relate our method to Table 8 (supplemental) of the \textsc{nvdiffrec} paper~\cite{munkberg2021nvdiffrec}, 
and note similar geometric quality as \textsc{nvdiffrec}.
The Lego scene is an outlier caused by our normal smoothness regularizer from Eq.~\ref{eq:normalsmooth}. The scene is particularly challenging for geometric smoothing
since it contains very complex geometry full of holes and sharp edges. Please refer to Figure~\ref{fig:ao-meshes} for a visual comparison.

\begin{table}
	\centering
	\caption
	{
		Chamfer L1 scores on the extracted meshes. Lower score is better.
	}
	\begin{tabularx}{\textwidth}{lYYYYY}
		\toprule
		& \small{Chair} & \small{Hotdog} & \small{Lego} & \small{Materials} & \small{Mic} \\
		\midrule
		PhySG              & 0.1341 & 0.2420 & 0.2592 &    N/A & 0.2712 \\
		NeRF (w/o mask)    & 0.0185 & 4.6010 & 0.0184 & 0.0057 & 0.0124 \\
		NeRF (w/ mask)     & 0.0435 & 0.0436 & 0.0201 & 0.0082 & 0.0122 \\
		\textsc{nvdiffrec} & 0.0574 & 0.0272 & 0.0267 & 0.0180 & 0.0098 \\
		Our                & 0.0566 & 0.0297 & 0.0583 & 0.0162 & 0.0151 \\
		\bottomrule
	\end{tabularx}
	\label{tab:chamfer}
\end{table}
\section{Relighting and View Interpolation} 

We present relighting results with per scene breakdown for all synthetic datasets in Table~\ref{tab:all_relighting}. 
Figures~\ref{fig:relight_nerfactor_hotdog},~\ref{fig:relight_ficus_hotdog},~and~\ref{fig:relight_lego_hotdog} show visual examples of the NeRFactor synthetic dataset. 
This dataset is intentionally designed to simplify light and material separation (low frequency lighting, no prominent 
shadowing, direct lighting only).
Figures~\ref{fig:relight_nerf_hotdog},~\ref{fig:relight_nerf_materials},~and~\ref{fig:relight_nerf_mic} show visual examples from the NeRF synthetic dataset. This 
dataset was designed view-interpolation, and contains particularly hard scenes with 
all-frequency lighting. Table~\ref{tab:all_view_interp} presents per-scene view interpolation scores for the synthetic datasets.

\tabRelightingBreakdown
\tabViewBreakdown

\figNerfactorRelightProbe{figures/supplemental/relighting/nerfactor_hotdog/}{2}{1cm}{2cm}{Hotdog}{fig:relight_nerfactor_hotdog}
\figNerfactorRelight{figures/supplemental/relighting/nerfactor_ficus/}{1}{2cm}{1.5cm}{Ficus}{fig:relight_ficus_hotdog}
\figNerfactorRelight{figures/supplemental/relighting/nerfactor_lego/}{1}{2cm}{1.5cm}{Lego}{fig:relight_lego_hotdog}
\figNerfRelightProbe{figures/supplemental/relighting/nerf_hotdog/}{2}{3cm}{3cm}{Hotdog}{fig:relight_nerf_hotdog}
\figNerfRelight{figures/supplemental/relighting/nerf_materials/}{2}{3cm}{5cm}{Materials}{fig:relight_nerf_materials}
\figNerfRelight{figures/supplemental/relighting/nerf_mic/}{3}{1cm}{1cm}{Microphone}{fig:relight_nerf_mic}

\figNewreg
\section{Scene Credits} 

Mori Knob from Yasotoshi Mori (CC BY-3.0). 
Bob model (CC0) by Keenan Crane. 
Rollercoaster and Porsche scenes from LDraw resources (CC BY-2.0) by Philippe Hurbain. 
Apollo and Damicornis models courtesy of the Smithsonian 3D repository~\cite{Smithsonian2020} (CC0). 
The Family scene is part of the Tanks\&Temples dataset~\cite{Knapitsch2017} (CC BY-NC-SA 3.0), 
the Character scene is part of the BlendedMVS dataset~\cite{Yao2020blendedmvs} (CC BY-4.0) 
and the Gold Cape scene is part of the NeRD dataset~\cite{Boss2021} (CC BY-NC-SA 4.0).
The NeRF~\cite{Mildenhall2020} and NeRFactor~\cite{Zhang2021nerfactor} datasets (CC BY-3.0) contain renders from modified blender models located on blendswap.com:
chair by 1DInc (CC0), drums by bryanajones (CC-BY),  ficus by Herberhold (CC0), hotdog by erickfree (CC0), 
lego by Heinzelnisse (CC-BY-NC), materials by elbrujodelatribu (CC0), mic by up3d.de (CC0), 
ship by gregzaal (CC-BY-SA). Light probes from Poly Haven~\cite{polyhaven}: Aerodynamics workshop and Boiler room by Oliksiy Yakovlyev, 
Dreifaltkeitsberg by Andreas Mischok, and Music hall by Sergej Majboroda (all CC0). The probes provided in the NeRFactor dataset 
are modified from the probes (CC0) shipped with Blender.

\end{document}